\documentclass[a4paper,12pt]{article}

\usepackage{psfrag}
\usepackage{mathrsfs} 

\usepackage{graphicx}
\usepackage{subfigure}
\usepackage[margin=10pt,font=small,labelfont=bf,
labelsep=period]{caption}
\usepackage{amsmath}
\usepackage{amssymb}
\usepackage{dsfont} 
\usepackage[bottom]{footmisc} 
\usepackage{indentfirst} 
\usepackage{upref} 
\usepackage{cite} 
\numberwithin{equation}{section}

\newcommand{\ddx}{\textrm{d}^d x\:}
\newcommand{\Tr}{\text{Tr}\:}

\renewcommand{\baselinestretch}{1.5}
\begin{document}
\begin{titlepage}
\renewcommand{\baselinestretch}{1.1}
\title{\begin{flushright}
\normalsize{MZ-TH/12-10}
\bigskip
\vspace{1cm}
\end{flushright}
The ``tetrad only'' theory space: Nonperturbative renormalization flow and 
Asymptotic Safety}
\date{}
\author{U. Harst and M. Reuter\\
{\small Institute of Physics, University of Mainz}\\[-0.2cm]
{\small Staudingerweg 7, D-55099 Mainz, Germany}}
\maketitle\thispagestyle{empty}

\begin{abstract} 
We set up a nonperturbative gravitational coarse graining flow and the 
corresponding functional renormalization group equation on the as to yet 
unexplored ``tetrad only'' theory space. It comprises action functionals which 
depend on the tetrad field (along with the related background and ghost fields) 
and are invariant under the semi-direct product of spacetime diffeomorphisms 
and local Lorentz transformations. This theory space differs from that of 
Quantum Einstein Gravity (QEG) in that the tetrad rather than the metric 
constitutes the fundamental variable and because of the additional symmetry 
requirement of local Lorentz invariance. It also differs from ``Quantum 
Einstein Cartan Gravity'' (QECG) investigated recently since the spin 
connection is not an independent field variable now. We explicitly compute the 
renormalization group flow on this theory space within the tetrad version of 
the Einstein-Hilbert truncation. A detailed comparison with analog results in 
QEG and QECG is performed in order to assess the impact the choice of a 
fundamental field variable has on the renormalization behavior of the 
gravitational average action, and the possibility of an asymptotically safe 
infinite cutoff limit is investigated. Implications for nonperturbative studies 
of fermionic matter coupled to quantum gravity are also discussed. It turns out 
that, in the context of functional flow equations, the ``hybrid calculations'' 
proposed in the literature (using the tetrad for fermionic diagrams only, and 
the metric in all others) are unlikely to be quantitatively reliable. Moreover 
we find that, unlike in perturbation theory, the non-propagating Faddeev-Popov 
ghosts related to the local Lorentz transformations may not be discarded but 
rather contribute quite significantly to the beta functions of Newton's 
constant and the cosmological constant.
\end{abstract}

\end{titlepage}
\newpage 
\section{Introduction}
In classical General Relativity there exists a remarkably rich variety of 
different variational principles which give rise to Einstein's equation, or 
equations equivalent to it but expressed in terms of different field variables. 
The best known examples are the Einstein-Hilbert action expressed in terms of 
the metric, $S_{\text{EH}}[g_{\mu\nu}]$, or the tetrad, respectively, 
$S_{\text{EH}}[e^a_{\ \mu}]$. The latter action functional is obtained by 
inserting the representation of the metric in terms of vielbeins into the 
former: $g_{\mu\nu}=\eta_{ab} e^a_{\ \mu} e^b_{\ \nu}$. 

Another classically equivalent formulation, at least in absence of spinning 
matter, is provided by the first order Hilbert-Palatini action 
$S_{\text{HP}}[e^a_{\ \mu}, \omega^{ab}_{\ \ \mu}]$ which, besides the tetrad, 
depends on the spin connection $\omega^{ab}_{\ \ \mu}$ assuming values in the 
Lie algebra of ${\sf O(}1,3)$. Variation of $S_{\text{HP}}$ with respect to 
$\omega^{ab}_{\ \ \mu}$ leads, in vacuo, to an equation of motion which 
expresses that this connection has vanishing torsion. It can be solved 
algebraically as $\omega=\omega(e)$ which, when inserted into $S_{\text{HP}}$, 
brings us back to $S_{\text{EH}}[e]\equiv S_{\text{HP}}[e,\omega(e)]$. 

Another equivalent formulation is based upon the self-dual Hilbert-Palatini 
action $S_{\text{HP}}^{\text{sd}}[e^a_{\ \mu}, \omega^{(+)\, ab}_{\ \ \ \mu}]$ 
which depends only on the (complex, in the Lorentzian case) self-dual 
projection of the spin connection, $\omega^{(+)\,ab}_{\ \ \ \mu}$ 
\cite{A,R,T,Kiefer}. This action in turn is closely related to the Plebanski 
action \cite{Pleb}, containing additional 2-form fields, and to the 
Capovilla-Dell-Jacobson action \cite{CDJ} which involves essentially only a 
self-dual connection. Similarly, Krasnov's diffeomorphism invariant Yang-Mills 
theories \cite{Krasnov} allow for a ``pure connection'' reformulation of 
General Relativity as well as deformations thereof.

The above variational principles are Lagrangian in nature; the fields employed 
provide a parametrization of configuration space. The corresponding Legendre 
transformation yields a Hamiltonian description in which the ``carrier fields'' 
of the gravitational interaction parametrize a phase-space now. In this way the 
ADM-Hamiltonian \cite{ADM} and Ashtekar's Hamiltonian \cite{Ash-Ham}, for 
instance, make their appearance.

Regarding the ongoing search for a quantum (field) theory of gravity this 
multitude of classical formalisms offers many equally plausible possibilities 
to explore. A priori it is not clear which one of the above hamiltonian 
systems, if any, is linked to the as to yet unknown fundamental quantum theory 
in the simplest or most easy to guess way.

In the traditional approach of ``quantizing'' a known classical system more 
input than the field equations, such as the Lagrangian is needed, so that 
classically equivalent theories might possibly give rise to inequivalent 
quantum theories. Among those, at most one can be ``correct'', in the sense of 
being realized in Nature. Of course, given the limitations of the available 
observational and experimental data it is not clear whether the most natural 
and/or simplest presentation of the classical limit emerging from this 
``correct'' theory is in the above list, or even close to it. Up to now, 
because of the many well known conceptual and technical problems \cite{Kiefer} 
we are not in a position to discriminate the various classical gravity theories 
on the basis of the {\it quantum} properties they imply.

Rather than trying to quantize a given classical dynamical system, there is 
another strategy one can adopt in order to search for a quantum theory 
consistent with the observed classical limit, the Asymptotic Safety program 
\cite{wein,mr,oliver,frank1,NJP}. One of its advantages as compared to a 
``quantization'' is that it depends on the classical input data to a lesser 
extent. The idea is to fix a certain theory space of action functionals, a 
coarse graining flow of it, and then search for a renormalization group (RG) 
fixed point (FP) on it at which the infinite ultraviolet (UV) cutoff limit can be 
taken in a ``safe'' way. 

While originally motivated by the possibility of sidestepping the problem of 
perturbative nonrenormalizability, this search strategy in principle can {\it 
predict} the theory's fundamental action. The only input needed is {\it theory 
space}. Once it is chosen one can ``turn the crank'' and, in case a suitable 
fixed point is found, construct a UV-regularized functional integral 
representation of the resulting theory \cite{elisa1}. Only at this very last 
stage we can identify the hamiltonian system which, implicitly, was quantized 
by taking the continuum limit at the respective fixed point.

To characterize a theory space $\mathcal{T}$ we must pick a certain set of 
fields, collectively denoted $\Phi$, a space of action functionals $A[\Phi]$, 
and a group ${\bf G}$ of symmetry transformations under which they are required 
to be invariant. In this sense, the above classical gravity theories motivate 
us to explore, for instance, the case where $\Phi\equiv g_{\mu \nu}$ is the 
metric and ${\bf G}$ the diffeomorphism group, or, as in Einstein-Cartan 
theory, $\Phi\equiv (e^a_{\ \mu}, \omega^{ab}_{\ \ \mu})$ where ${\bf G}$ is 
the semidirect product of local Lorentz transformations and spacetime 
diffeomorphisms. 

We emphasize that these spaces and symmetries are the only ``inspiration'' 
drawn from the classical examples. Their dynamics, i.\,e. the specific 
classical action they postulate, plays no special r{\^o}le in the Asymptotic 
Safety program. It is just one special point in the pertinent theory space, and 
usually not the sought for fixed point of the RG flow.

Most of the work on Asymptotic Safety has been done in ``Einstein'' gravity 
which, by definition, is based upon a theory space $\mathcal{T}_{\text{E}}$ of 
functionals\footnote{The dots stand for the background fields and Faddeev-Popov 
ghosts to be introduced later.} $A[g_{\mu\nu},\cdots]$ invariant under ${\bf 
G}={\sf Diff}({\cal M})$, the diffeomorphisms of the spacetime manifold ${\cal 
M}$. Recently also first investigations of the ``Einstein-Cartan'' choice
\begin{equation}
 \mathcal{T}_{\text{EC}}=
 \big\{
   A[e^a_{\ \mu}, \omega^{ab}_{\ \ \mu},\cdots]; 
   {\bf G}={\sf Diff}({\cal M})\ltimes {\sf O(}4)_{\rm loc}
 \big\}
\end{equation}
were published \cite{e-omega}. Here ${\sf O(}4)$ plays the r{\^o}le of the 
Euclidean Lorentz group\footnote{We shall consider the case of Euclidean 
signature throughout.}, and ${\sf O(}4)_{\rm loc}$ denotes the group of the 
corresponding local gauge transformations.

The present paper instead is devoted to the ``tetrad only'' theory space 
pertaining to a $d$ dimensional spacetime ${\cal M}$:
\begin{equation}
 \mathcal{T}_{\text{tet}}= 
 \big\{ 
   A[e^a_{\ \mu}, \cdots]; 
   {\bf G} ={\sf Diff}({\cal M})\ltimes {\sf O(}d)_{\rm loc}
 \big\}.
\end{equation}
With actions depending on the vielbein only, this space is intermediate between 
$\mathcal{T}_{\text{E}}$ and $\mathcal{T}_{\text{EC}}$: Coming from the 
``Einstein'' side it generalizes ${\cal T}_{\text{E}}$ by declaring $e^a_{\ 
\mu}$ the fundamental field and the metric a composite thereof, $g_{\mu\nu} = 
\eta_{ab} e^a_{\ \mu} e^b_{\ \nu}$. Conversely, coming from the 
``Einstein-Cartan'' side, every $A[e, \omega,\cdots]\in 
\mathcal{T}_{\text{EC}}$ implies a certain $A[e,\cdots]\in 
\mathcal{T}_{\text{tet}}$ upon inserting $\omega=\omega_{\text{LC}}(e)$, where 
$\omega_{\text{LC}}(e)$ is the torsion-free (Levi-Civita) connection the 
vielbein gives rise to.

There are various independent motivations for this investigation.

\noindent{\bf (A)} The first functional RG based results obtained on the 
Einstein-Cartan theory space $\mathcal{T}_{\text{EC}}$, in a truncation with a 
scale dependent Hilbert-Palatini action (including a running Immirzi term), 
show certain characteristic differences in comparison with the familiar case of 
$\mathcal{T}_{\text{E}}$ truncated with a running Einstein-Hilbert action; in 
particular, the ${\cal T}_{\text{EC}}$ results show a stronger RG scheme and 
gauge fixing dependence than the older ones on the ``Einstein'' case 
\cite{e-omega}. It would be interesting to know whether these differences are 
mainly due to the use of the different truncations, different field variables, 
or both. In the present paper we shall change only the field variable (and the 
group {\bf G} correspondingly), but not the truncation, and so it should be 
possible to disentangle the two sources of deviations to some extent.

We note here that, like most settings of quantum field theory, the flow 
equation of the average action is {\it not} invariant under diffeomorphisms in 
field space, $\Phi \mapsto \Phi'(\Phi)$. Thus, at intermediate steps, as long 
as one does not compute observables, there is no reason to expect any field 
parametrization independence. Moreover, and perhaps this is even more 
important, the gauge fixing and ghost sectors are quite different for ${\sf 
Diff}({\cal M})$ and ${\sf Diff}({\cal M})\ltimes {\sf O(}d)_{\rm loc}$, 
respectively. Therefore the $\beta$-functions for the running Newton constant 
$G_k$ or cosmological constant $\bar{\lambda}_k$, for instance, may well depend 
on whether the functional renormalization group equation (FRGE) is formulated 
in terms of the metric or tetrad. Similar remarks apply also to a recent study 
of the perturbative RG running of $G_k$ and the Immirzi parameter 
\cite{bene-speziale}.

\noindent{\bf (B)} On theory spaces involving fermions coupled to gravity 
introducing vielbeins is compulsory. Besides the pure gravity couplings, such 
as $G_k,\ \bar\lambda_k$, etc. the average action will then depend on 
additional couplings related to the matter field monomials. If we collectively 
denote these couplings by $u_{\text{grav}}$ and $u_{\text{mat}}$, respectively, 
their $\beta$-functions are of the form
\begin{align}
 \beta_{\text{grav}} &= 
   \beta^{\text{grav}}_{\text{grav}}(u_{\text{grav}}) 
     +\beta^{\text{mat}}_{\text{grav}}(u_{\text{grav}},u_{\text{mat}})\\
 \beta_{\text{mat}} &= 
   \beta^{\text{mat}}_{\text{mat}}(u_{\text{mat}})
     +\beta^{\text{grav}}_{\text{mat}}(u_{\text{grav}},u_{\text{mat}})
   \label{betamatter}
\end{align}
Diagrammatically speaking, the two parts $\beta^{\text{grav}}_{\text{grav}}$ 
and $\beta^{\text{mat}}_{\text{grav}}$ of the pure gravity $\beta$-functions 
stem from the graviton and matter loops, respectively. Conversely, the running 
of the matter couplings has a part due to pure matter loops, 
$\beta^{\text{mat}}_{\text{mat}}$, plus mixed matter-gravity contributions, 
$\beta^{\text{grav}}_{\text{mat}}$.

In order to get a first impression of the impact the fermions have on the 
gravitational RG flow one might neglect the running of the matter couplings, 
and try to compute $\beta_{\text{grav}}$ only. While the evaluation of 
$\beta^{\text{mat}} _{\text{grav}}$ from the fermion loops clearly requires a 
vielbein and a spin connection, the pure gravity part 
$\beta^{\text{grav}}_{\text{grav}}$ does not obviously do so. From a pragmatic 
point of view it is therefore tempting to take the 
$\beta^{\text{grav}}_{\text{grav}}$ part from a (much simpler, and already 
available) computation in the metric formalism. The invariants $I[g_{\mu\nu}]$ 
occurring in the latter one would interpret as $I[e]\equiv 
I[g_{\mu\nu}=\eta_{ab} e^{a}_{\ \mu}e^{b}_{\ \nu}]$. For some (but not all) 
field monomials in $\mathcal{T}_{\text{tet}}$ this establishes a correspondence 
to monomials in $\mathcal{T}_{\text{E}}$, and one can try to identify their 
running prefactors; for instance, $\bar{\lambda}_k\int \text{d}^d x\, 
\sqrt{g}\in {\cal T}_{\text{E}}\ \leftrightarrow \ \bar{\lambda}_k \int 
\text{d}^d x\, e\in {\cal T}_{\text{tet}}$, when $g$ and $e$ denote the 
determinants of $g_{\mu\nu}$ and $e^a_{\ \mu}$, respectively.

Thus it seems that only the fermion loops, $\beta^\text{mat}_\text{grav}$, need 
to be calculated. This requires fixing a Lorentz gauge in order to associate a 
unique $e^a_{\ \mu}$ to a given $g_{\mu\nu}$, and for $\omega^{ab}_{\ \ \mu}$ 
one might take the unique Levi-Civita connection associated to this vielbein, 
$\omega^{\ \ \ ab}_{\text{LC} \ \ \mu}(e)$.

We shall refer to this procedure as a {\it hybrid calculation}. Clearly it can 
be meaningful at most within a truncation of $\mathcal{T}_{\text{E}}$ and 
$\mathcal{T}_{\text{tet}}$ that allows an identification of monomials; an 
example is the Einstein-Hilbert action regarded as a functional of $g_{\mu\nu}$ 
and $e^a_{\ \mu}$, respectively, with the same two couplings $G_k$ and 
$\bar{\lambda}_k$ occurring in both cases. At the exact level there exists 
certainly no such one-to-one correspondence between action monomials in 
$\mathcal{T}_{\text{E}}$ and $\mathcal{T}_{\text{tet}}$. Nevertheless, if it 
was possible to establish the ``hybrid'' scheme as a reliable approximation, 
this would be of considerable importance for the feasibility of practical 
calculations.

As to yet, all investigations for the gravity + fermions theory space, in 
particular in the Asymptotic Safety context, are, in fact, hybrid computations 
of this form \cite{percacci-perini,vacca, Eichhorn-Gies}. They combine the 
metric-formalism $\beta$-functions for $G_k$ and $\bar{\lambda}_k$ in the 
Einstein-Hilbert truncation with certain matter contributions 
$\beta^{\text{mat}}_{\text{grav}}$, solve for $u_{\text{grav}}(k)$, and insert 
the result into \eqref{betamatter} to obtain the running of the matter 
couplings. In ref. \cite{Eichhorn-Gies} the gravity corrections to certain 
4-fermion couplings $u_{\text{mat}}$ were studied in this way.

A necessary condition for the consistency of the hybrid approach is that the 
pure gravity part $\beta^{\text{grav}}_{\text{grav}}$ does not change much when 
we switch from $g_{\mu\nu}$ to $e^{a}_{\ \mu}$ as the fundamental field 
variable in the Einstein-Hilbert truncation. In the present paper we shall be 
able to explicitly test whether or not this is actually the case. It will be 
one of our main results that the hybrid scheme is very hard, if not impossible 
to justify, at least at the quantitative level. We shall demonstrate in detail 
that if one aims at some degree of numerical precision, one should consistently 
work with the vielbein and its corresponding ghost system already at the pure 
gravity level.

\noindent{\bf (C)} Picking the vielbein as the fundamental field variable 
requires fixing a ${\sf O(}d)_{\rm loc}$ gauge. In perturbation theory, a 
popular choice is the Deser-van Nieuwenhuizen algebraic gauge fixing condition 
where the antisymmetric part of the $d \times d$ matrix $e^a_{\ \mu}$ is 
required to vanish \cite{DvN,Joek}. As ${\sf O(}d)$ has $\frac{1}{2}\,d\,(d-1)$ 
parameters, this reduces the $d^2$ independent components of $e^a_{\ \mu}$ to 
$\frac{1}{2} d(d+1)$, which is precisely the number of independent fields 
$g_{\mu\nu}$ has in $d$ dimensions.

It was shown that, for this gauge, and in perturbation theory, no Faddeev-Popov 
ghosts need to be introduced for the ${\sf O(}d)_{\rm loc}$ factor of {\bf G}, 
and that it allows to explicitly express vielbein fluctuations purely in terms 
of metric fluctuations \cite{Woodard}. Therefore the point of view was 
advocated that even in presence of fermions the vielbein can be eliminated in 
favor of the metric.

While this method was proven to be correct in a well defined perturbative 
context, recently it has been proposed to use this same procedure, in 
particular the omission of the ${\sf O(}d)_{\rm loc}$ ghosts, also in the 
context of a nonperturbative flow equation for the gravity--fermion system 
\cite{vacca,Eichhorn-Gies}. If applicable, it would provide a very economic 
framework for hybrid computations of the type sketched above.

However, as we are going to discuss in detail there are reasons to doubt that 
the perturbative arguments justifying the omission of the ${\sf O(}d)_{\rm 
loc}$ ghosts carry over to the nonperturbative setting of the FRGE. In fact, in 
perturbation theory the ghosts are omitted since their inverse propagator 
contains no derivatives, they are non-propagating, leading to a trivial 
Faddeev-Popov determinant. In the FRGE, instead, a straightforward evaluation 
of the functional traces cuts off all field modes in a uniform fashion, no 
matter if their kinetic term contains 2, or more, or no derivatives at all.

In the present paper we shall explicitly evaluate the contributions to 
$\beta_{\text{grav}}$ from the non-propagating ${\sf O(}d)_{\rm loc}$ ghosts 
pertaining to the symmetric vielbein gauge, and we shall analyze whether they 
really can be discarded in setting up the flow equation for the average action.

Fortunately, the particular fermionic $\beta$-functions computed in 
\cite{Eichhorn-Gies} happen to be independent on whether the ${\sf 
O}(d)_{\text{loc}}$ ghosts are retained or not. However, in future extensions 
of such studies it will be important to know how to treat them correctly.

The remaining sections of this article are organized as follows. In Section 2 
we summarize various preliminaries on the gravitational average action and its 
FRGE which will be needed later on. In Section 3 we focus on the ``tetrad 
only'' theory space $\mathcal{T}_{\text{tet}}$ in the Einstein-Hilbert 
truncation, and calculate the corresponding $\beta$-functions. The resulting RG 
flow is analyzed with numerical methods in Section 4 then. Our results, in 
particular on the issues (A)--(C) raised above, are summarized in Section 5.

\section{The average action approach to quantum gravity}
Introducing the scale-dependent effective average action $\Gamma_k$ it has been 
possible to construct a functional RG flow for quantum gravity \cite{mr}. This 
``running action'' can be considered the generating functional of the 1\,PI 
correlation functions that take into account quantum fluctuation of all scales 
between the UV and an infrared cutoff scale $k$. For $k\rightarrow \infty$ it 
is closely related to the bare action $S$ and for vanishing cutoff it coincides 
with the usual effective action $\Gamma=\Gamma_{k=0}$. Its scale-dependence is 
governed by an exact renormalization group equation:
\begin{equation}\label{ERGE1}
 \partial_t \Gamma_k = 
   \frac{1}{2}\,\text{STr}\:
   \Bigg[
     \frac{\partial_t \widehat{R}_k}{\Gamma^{(2)}_k+\widehat{R}_k}
   \Bigg].
\end{equation}
Here $t\equiv \ln k$, and $\Gamma^{(2)}_k$ denotes the matrix of the second 
functional derivative of $\Gamma_k$ with respect to the dynamical fields. 
Furthermore, $\widehat{R}_k$ is an operator that implements the infrared cutoff 
in the path integral by replacing the bare action $S$ with $S+\Delta_k S$ where 
$\Delta_k S$ is quadratic in the fluctuations, $\Delta_k S \propto \int \varphi 
\widehat{R}_k\, \varphi$. Finally, the supertrace in \eqref{ERGE1} comprises a 
trace over all internal indices as well as an integral/sum over all modes of 
$\varphi$; for fermionic fields it contains an additional minus sign 
\cite{avact}.

As $\Gamma_k$ is a generic point in ``theory space'', i.\,e. a functional of a 
given set of fields restricted only by the required symmetries, solving this 
exact equation is usually a formidable task. For this reason one has to resort 
to truncations of theory space in order to find approximate solutions to eq. 
\eqref{ERGE1}. This is done by expanding $\Gamma_k$ in a basis of integrated 
field monomials $I_\alpha$, i.\,e. $\Gamma_k[\,\cdot\,] =\sum_\alpha 
c_\alpha(k) I_\alpha[\,\cdot\,]$ and restricting the sum to a finite number of 
terms. The scale-dependence of $\Gamma_k$ is then described by a finite number 
of running couplings $c_\alpha(k)$. If we project the RHS of \eqref{ERGE1} onto 
this subspace of theory space the functional equation reduces to a coupled 
system of ordinary differential equations in these couplings.

If we describe pure gravity with the metric as field variable, the simplest 
truncation is the Einstein-Hilbert truncation with only two running couplings: 
Newton's constant $G_k$ and the cosmological constant $\bar{\lambda}_k$. As 
gravity is a gauge theory we also have to add a gauge fixing and a ghost term 
to the truncation ansatz; its running shall be ignored in our approximation. 
Our ansatz for $\Gamma_k$ can therefore be decomposed into a ``bosonic part'' 
$\breve{\Gamma}_k$ and the classical ghost contribution $S_{\text{gh}}$:
\begin{equation}
 \Gamma_k=
 \Gamma^{\text{EH}}_{k}+\Gamma_{\text{gf}}+S_{\text{gh}}
 \equiv
 \breve{\Gamma}_k+S_{\text{gh}}
\end{equation}
Using this decomposition the FRGE $\eqref{ERGE1}$ can be written in the 
following form:
\begin{equation}\label{FRGE}
 \partial_t \Gamma_k= 
 \frac{1}{2}\Tr 
   \left[ \frac{\partial_t \widehat{R}_k}{\breve{\Gamma}^{(2)}_k + 
   \widehat{R}_k}
   \right]
   - \Tr \left[ 
    \frac{\partial_t \widehat{R}^{\text{gh}}_k}{S^{(2)}_{\text{gh}} + 
    \widehat{R}^{\text{gh}}_k}
   \right].
\end{equation}

The gravitational average action heavily relies on the background field method 
\cite{DeWitt-books}. The field chosen to represent gravity is split arbitrarily 
into a background part and a fluctuation: $\phi=\bar\phi + \bar \varphi$. 
$\Gamma_k$ is constructed as a background gauge invariant functional of both 
fields, $\Gamma_k[\phi, \bar\phi]$, i.\,e. it is invariant under a simultaneous 
action of ${\bf G}$ on both $\phi$ and $\bar\phi$. As we only deal with a 
so-called single metric truncation \cite{elisa2} in this paper, we will set the 
fluctuations to zero after the second derivative $\Gamma^{(2)}$ with respect to 
the fluctuations has been taken. At the end we therefore arrive at a system of 
differential equations for the running couplings parametrizing 
$\Gamma_k[\bar\phi]=\Gamma_k[\bar\phi, \bar\phi]$.

For example, in the ``tetrad only'' case the average action is a curve $k 
\mapsto \Gamma_k$ in the theory space $\mathcal{T}_{\text{tet}}$ which, to be 
precise now, consists of ${\sf Diff}({\cal M})\ltimes {\sf O(}d)_{\rm loc}$ 
invariant functionals of the type $A[e^a_{\ \mu},\bar{e}^a_{\ \mu}, C^\mu, 
\bar{C}_\mu, \Sigma^{ab}, \bar{\Sigma}_{ab}]$; besides the vielbein and its 
background, they depend on the diffeomorphism ghosts $(C^\mu, \bar{C}_\mu)$ and 
${\sf O(}d)_{\rm loc}$ ghosts $(\Sigma^{ab},\bar{\Sigma}_{ab})$. Instead of 
$e^a_{\ \mu}$ we shall often consider the vielbein fluctuation 
$\bar{\varepsilon}^a_{\ \mu}\equiv e^a_{\ \mu}- \bar{e}^a_{\ \mu}$ the 
independent argument of the action.

\section{Tetrad theory space in Einstein-Hilbert truncation}
At this point we take two decisions. One of them refers to the deeper level of 
the exact theory, the other to the practical (computational) level of concrete 
approximations.

First, we fix the theory space to be the ``tetrad only'' one, 
$\mathcal{T}_{\text{tet}}$, so that all actions to be considered depend only on 
$e^a_{\ \mu}$, along with the corresponding background and ghost fields.

Second, to be able to perform practical calculations we decide to truncate 
$\mathcal{T}_{\text{tet}}$ by an ansatz for $\Gamma_k$ which is essentially a 
$k$-dependent version of the Einstein-Hilbert action reexpressed in terms of 
the tetrad, $S_{\text{EH}}[g(e)]$.

\subsection{The FRGE on $\mathcal{T}_{\text{tet}}$}
In this subsection we derive the RG flow of tetrad gravity in the 
Einstein-Hilbert truncation 
\begin{equation}
 \breve{\Gamma}_k[e,\bar{e}]=
  -\frac{1}{16 \pi G_k} \int \ddx \sqrt{g(e)} \,
  \bigg(
    R(g(e))- 2 \bar{\lambda}_k
  \bigg)
  +\Gamma_{\text{gf}}[e,\bar{e}].
\end{equation}
This action involves two running couplings, the cosmological constant 
$\bar{\lambda}_k$ and Newton's constant $G_k$; the latter is frequently 
expressed in terms of the dimensionless function $Z_{Nk}$ according to 
$G_k\equiv Z^{-1}_{Nk}\,\bar G$ with a constant $\bar{G}$.

To be as general as possible we re-express the metric in terms of the new field 
variable $e^a_{\ \mu}$ in the following way:
\begin{equation}\label{DefXi}
 g_{\mu\nu}= \xi^{-1} e^a_{\ \mu} e^b_{\ \nu} \eta_{a b}.
\end{equation}
This representation resembles the usual vielbein decomposition of the metric, 
except for the additional free parameter $\xi>0$. For this reason we will refer 
to the field $e^a_{\ \mu}$ as a generalized vielbein for a given $g_{\mu\nu}$. 
Treating $e^a_{\ \mu}$ as the independent variable we assume that the basis 
1-forms $e^a=e^a_{\ \mu} \text{d} x^\mu$ indeed form a non-degenerate co-frame. 
The parameter $\xi$ is merely a mathematical tool that enables us to study a 
continuous class of field redefinitions at a time.

As for the usual vielbein this generalized decomposition of the metric is not 
unique, but there exists an ${\sf O(}d)$ manifold of vielbein fields 
corresponding to the same metric. We will treat this arbitrariness as an 
additional gauge freedom, such that the total group of gauge transformations is 
given by ${\bf G} = {\sf Diff}({\cal M}) \ltimes {\sf O(}d)_{\rm loc}$. 
Compared to the metric formulation we therefore have to add a second gauge 
fixing term; the corresponding background gauge invariant ghost-action can be 
constructed using the formalism introduced in \cite{je-uli}.

If we decompose both the metric $g_{\mu\nu}\equiv 
\bar{g}_{\mu\nu}+\bar{h}_{\mu\nu}$ and the vielbein $e^a_{\ \mu}\equiv 
\bar{e}^a_{\ \mu}+\bar{\varepsilon}^a_{\ \mu}$ into background fields and 
fluctuations, we find
\begin{equation}
\bar{g}_{\mu \nu}+ \bar{h}_{\mu \nu} 
  = 
  g_{\mu \nu} 
  = 
  \xi^{-1}(\bar{e}^a_{\ \mu}+\bar{\varepsilon}^a_{\ \mu})
          (\bar{e}^b_{\ \nu}+\bar{\varepsilon}^b_{\ \nu})\eta_{a b}
  =
  \xi^{-1}\bar{e}^a_{\ \mu}\bar{e}^b_{\ \nu}\eta_{a b} 
  + 2\xi^{-1} \bar{\varepsilon}_{(\mu \nu)} 
  + \mathcal{O}(\bar{\varepsilon}^2).
\end{equation}
Here and in the following we use the background vielbein $\bar{e}^a_{\ \mu}$ to 
change the type of the first (i.\,e., frame) index of the vielbein fluctuation: 
$\bar{\varepsilon}_{\mu \nu}=\eta_{a b} \bar{e}^a_{\ \mu} 
\bar{\varepsilon}^b_{\ \nu}$. We see that the symmetric part of the vielbein 
fluctuations, $\bar{\varepsilon}_{(\mu \nu)}$, is proportional to the metric 
fluctuations $\bar{h}_{\mu\nu}$ in lowest order, while we can relate the 
additional $d(d-1)/2$ gauge degrees of freedom carried by $e^a_{\ \mu}$ to the 
antisymmetric part of the fluctuations, $\bar{\varepsilon}_{[\mu \nu]}$. 

This observation motivates the following choice of gauge conditions. For the 
diffeomorphisms we choose the usual harmonic gauge fixing function for metric 
fluctuations, replacing $\bar{h}_{\mu \nu} \mapsto 2 \,\xi^{-1} 
\bar{\varepsilon}_{(\mu \nu)}$, with $\kappa\equiv (32 \pi \bar{G})^{-1/2}$:
\begin{equation}
 F_\mu 
 =
 2 \sqrt{2} \kappa\,\xi^{-1}\Big( \bar{D}^\nu \bar{\varepsilon}_{(\mu \nu)}
  -\frac{1}{2}\bar{D}_\mu \bar{\varepsilon}^{\nu}_{\ \nu}\Big).
\end{equation}
The ${\sf O(}d)$ transformations are gauge fixed using
\begin{equation}
 G^{ab}
 =
 2 \, \xi^{-\frac{1}{2}}\, \bar{g}^{\mu \nu} 
 \bar{\varepsilon}^{[a}_{\ \mu} \bar{e}^{b]}_{\ \nu}
 =
 2\, \xi^{-\frac{1}{2}} \bar{\varepsilon}^{[ab]},
\end{equation}
corresponding to a suppression of the antisymmetric vielbein fluctuations.

With these gauge conditions the gauge fixing term in the effective average 
action assumes the usual form, involving parameters $\alpha_D$ and $\alpha_L$:
\begin{equation}
 \Gamma_{\text{gf},k}[e,\bar{e}]
  = 
 \frac{1}{2 \alpha_D} \int \ddx \sqrt{\bar{g}} \bar{g}^{\mu \nu} F_{\mu} F_\nu 
  + \frac{1}{2 \alpha_L} \int \ddx \sqrt{\bar g}\, G^{ab}G_{ab}.
\end{equation}
In the following we fix the diffeomorphism gauge parameter $\alpha_D$ to 
$\alpha_D= 1/Z_{N k}$ which leads to the same cancelation in the kinetic 
operator as in metric gravity \cite{mr}.

In order to obtain a {\it background {\bf G}-invariant} ghost action with 
respect to both ${\sf O(}d)_{\text{loc}}$ transformations and diffeomorphisms, 
we can make use of the Faddeev-Popov construction only if we first 
reparametrize the gauge transformations in such a way, that the new generators 
of diffeomorphisms and ${\sf O(}d)$ transformations commute. This corresponds 
to an ${\sf O(}d)$ covariantization of the Lie derivative. Following this 
procedure, described in detail in \cite{e-omega,je-uli}, while treating the 
ghost sector classically (i.\,e. we can set $e=\bar{e}$ already at the level 
of the ghost action) we arrive at 
\begin{equation}
 S_{\text{gh}}[C,\bar C,\Sigma',\bar \Sigma';\bar e]= 
-\!\!\int\! \ddx \bar{e} 
  \begin{pmatrix}
   \bar{C}^{\mu}\\
   \!
   \bar{\Sigma}{}'{}^{\mu \nu}\!
  \end{pmatrix}^{\!\!\!\!T}
  \begin{pmatrix}
   \sqrt{2}\xi^{-1}\big(\delta^\mu_{\ \rho} \bar{D}^2 
   \!+\! \bar{R}^\mu_{\ \rho} \big) & 0\\
   2 \, \xi^{-\frac{1}{2}} \bar{\mu}\, \delta^\mu_{\ \rho} \bar{D}^\nu & 
   \!\!\!\!\!2\xi^{-\frac{1}{2}}\bar{\mu}^2 \delta^{[\mu}_\rho 
   \delta^{\nu]}_\sigma 
  \end{pmatrix}
  \begin{pmatrix}
   C^{\rho}\\
   \!{\Sigma'}^{\rho \sigma}\!
  \end{pmatrix}
\end{equation}
Here $\bar{C}^{\mu},\, C^{\mu}$ represent the diffeomorphism ghosts and 
$\bar{\Sigma}^{\mu\nu},\, \Sigma^{\mu\nu}$ the ${\sf O(}d)$ ghost fields.

As the infinitesimal transformation under diffeomorphisms contains a 
derivative, while the corresponding ${\sf O(}d)$ transformation does not, the 
diffeomorphism ghosts have a canonical mass dimension of one unit less 
compared to the ${\sf O(}d)$ ghosts. In order to obtain a Hessian operator of 
a well-defined mass dimension we have rescaled the fields 
$\bar{\Sigma}^{\mu\nu}=\bar{\mu}\bar{\Sigma}'{}^{\mu\nu}$, 
$\Sigma^{\mu\nu}=\bar{\mu}\Sigma'{}^{\mu\nu}$ with an arbitrary mass parameter 
$\bar{\mu}$; consequently the Hessian operator obtains a mass dimension of 2.

\subsection{Structure of the vielbein sector}
After having presented the details of our truncation we can now pass on to the 
evaluation of the FRGE \eqref{FRGE} in this truncation. On the LHS of the 
equation, after setting $\bar{e}=e$, we obtain the same result as in the 
metric version of the Einstein-Hilbert truncation \cite{mr}:
\begin{equation}
 \partial_t \Gamma_k[e,e]
  =
  2 \kappa^2 \int \ddx \sqrt{g(e)} 
  \big[-R(g(e)) \partial_t Z_{N k} 
  + 2 \partial_t \big(Z_{Nk}\bar\lambda_k\big)\big]
\end{equation}
On the RHS of the FRGE, however, we find two types of additional contributions 
to the supertrace as compared to those already present in the metric 
description. While the second type of contributions is due to the extended 
gauge group of the theory, the first type is closely linked to the off-shell 
character of the FRGE. This can be seen as follows.

In order to obtain $\breve{\Gamma}^{(2)}$ we expand $\breve{\Gamma}_k$ to 
second order in the vielbein fluctuations and read off the operator from the 
quadratic term $\breve{\Gamma}^{\text{quad}}_k$. As $\Gamma_{\text{gf}}$ is 
already quadratic in the fluctuations we only have to expand 
$\Gamma_{\text{EH},k}$. For
\begin{equation}
 \Gamma^{\text{quad}}_{\text{EH}} =
  \frac{1}{2} \left.\delta^2_e \Gamma_{\text{EH}}\right|_{e=\bar{e}}
\end{equation}
we find
\begin{equation}
 \begin{split}\label{GammaQuad1}
 \Gamma^{\text{quad}}_{\text{EH}} 
  &= 
   \frac{1}{2}\cdot \frac{4}{\xi^2} 
   \int \ddx\!{}_1\ddx\!{}_2 
   \left. 
     \frac{\delta^2 \Gamma_{\text{EH}}}{\delta g_{\rho \sigma} (x_2) 
     \delta g_{\mu\nu}(x_1)}
   \right|_{g=\bar{g}} 
   \bar{\varepsilon}_{(\mu\nu)}(x_1)\bar{\varepsilon}_{(\rho\sigma)}(x_2)\\
  &+ \frac{1}{2}\cdot\frac{2}{\xi} \int \ddx 
   \left.
    \frac{\delta   \Gamma_{\text{EH}}}{\delta g_{\mu \nu}(x)}
   \right|_{g=\bar{g}} 
   \bar{\varepsilon}_{a(\nu}(x)\bar{\varepsilon}^a_{\ \mu)}(x)
 \end{split}
\end{equation}
Here we have used the chain rule for functional derivatives. Obviously, the 
first term on the RHS of \eqref{GammaQuad1} corresponds exactly to the one 
known from the metric calculation, while the second term is due to the field 
redefinition. We note that those two terms come with different powers of 
$\xi$, which enables us to keep track of their respective origin during the 
entire calculation and in the final result. This was in fact our main 
motivation for introducing this book-keeping device.

Note also that in \eqref{GammaQuad1} the term due to the field redefinition is 
proportional to the first variation $\delta \Gamma_{\text{EH}} / \delta 
g_{\mu\nu}$. So it would vanish if we were to go ``on shell'', i.\,e. to 
insert a special metric or vielbein which happens to be a stationary point of 
$\Gamma_{\text{EH}}$. We emphasize that in the process of computing 
$\beta$-functions this would be a severe mistake. To see this, consider an 
(exact) average action expanded as 
\begin{equation}
 \Gamma_k[\phi,\bar{\phi}]= \sum_\alpha c_\alpha(k) I_\alpha[\phi,\bar{\phi}],
\end{equation}
where $c_\alpha(k)$ denote the running couplings and the $I_\alpha$'s are 
${\bf G}$-invariant basis functionals (integrated field monomials, say) 
independent of $k$. When represented in this fashion one may think of 
$\Gamma_k$ as a ``generating function'' for the set of running couplings, 
$\{c_\alpha (k)\}$, which are ``projected out'' by expanding $\Gamma_k$ in the 
basis $\{I_\alpha[\,\cdot\,,\,\cdot\,]\}$. In this picture the fields $\phi$, 
$\bar{\phi}$ have a subordinate status only. They serve as arguments of the 
$I_\alpha$'s, and their only r{\^o}le is that of a dummy variable needed in 
order to define the basis functionals $I_\alpha$. Therefore, in order for the 
set $\{I_\alpha\}$ to remain {\it complete} it is in general not possible to 
narrow down the function space $\phi$, $\bar\phi$ are drawn from in any way, 
for instance by stationary point conditions or the like. In this sense, the 
average action and its associated FRGE are intrinsically ``off shell'' in 
nature.

At most at the level of truncations where the set $\{ I_\alpha\}$ is 
incomplete anyhow we may opt for special choices of the fields (e.\,g. 
satisfying convenient symmetry conditions) as long as the invariants in the 
truncation ansatz when calculated for these fields can still be distinguished 
from all other invariants and from each other. This is an often used 
computational trick that simplifies practical calculations without affecting 
the result in any way.

For the total quadratic part of the action $\breve{\Gamma}_k$ we obtain, with 
$\sqrt{\bar{g}}\equiv \bar{e}$,
\begin{align}\label{GammaQuad}
\breve{\Gamma}^{\text{quad}}_k[\bar{\varepsilon};\bar{e}]
 &= 
  \frac{4 Z_{N k} \kappa^2}{\xi^2} \int \ddx \sqrt{\bar g}\; 
  \bar{\varepsilon}_{(\mu \nu)}
  \big[ 
    -\! K^{\mu\nu}_{\ \ \rho \sigma} \bar{D}^2 
    + U^{\mu\nu}_{\ \ \rho \sigma}
  \big] \bar{\varepsilon}^{(\rho \sigma)}\\ 
 &+ \frac{2 Z_{Nk}\kappa^2}{\xi} \int \ddx \sqrt{\bar{g}} 
  \left( 
   \bar{R}^{\mu \nu} + \Lambda \bar{g}^{\mu\nu} 
   - \frac{\bar{R}}{2} \bar{g}^{\mu \nu}
  \right) \bar{\varepsilon}_{a (\nu} \bar{\varepsilon}^{a}_{\ \mu)}
  + \frac{1}{2 \alpha_L} \frac{4}{\xi} \int \ddx \sqrt{\bar{g}} 
  \bar{\varepsilon}^{[ab]}\bar{\varepsilon}_{[ab]}\nonumber 
\end{align}
where
\begin{equation}
 K^{\mu\nu}_{\ \ \rho \sigma}\equiv\frac{1}{4}
 \Big(
  \delta^{\mu}_{\rho}\delta^{\nu}_{\sigma}
  +\delta^{\mu}_{\sigma}\delta^{\nu}_{\rho}
  -\bar{g}^{\mu\nu}\bar{g}_{\rho \sigma}
 \Big)
\end{equation}
and
\begin{equation}
 U^{\mu\nu}_{\ \ \rho \sigma}\equiv\frac{1}{4}
 \big[
  \delta^{\mu}_{\rho}\delta^{\nu}_{\sigma}
  +\delta^{\mu}_{\sigma}\delta^{\nu}_{\rho}
  -\bar{g}^{\mu\nu}\bar{g}_{\rho \sigma}\big] 
  \big( \bar{R} - 2 \bar{\lambda}_k\big) 
  + \frac{1}{2} \big[ \bar{g}^{\mu\nu} \bar{R}_{\rho \sigma}
  +\bar{g}_{\rho \sigma} \bar{R}^{\mu \nu}\big]
  -\delta^{(\mu}_{(\rho} \bar{R}^{\nu)}_{\ \sigma)}
  -\bar{R}^{(\nu \ \mu)}_{\ (\rho \ \sigma)}
\end{equation}
We observe that the first term on the RHS of \eqref{GammaQuad} is exactly the 
contribution known from the metric computation \cite{mr}; in particular thanks 
to $\alpha_D = 1/Z_{Nk}$ all non-minimal terms in the differential operator 
canceled. The second and third terms in \eqref{GammaQuad} correspond to the 
already mentioned first and second type of new contributions, respectively.

In a next step we decompose the vielbein fluctuations $\bar{\varepsilon}_{\mu 
\nu}$ into their symmetric traceless part $\widehat{\varepsilon}_{\mu \nu}$, 
antisymmetric part $\tilde{\varepsilon}_{\mu \nu}$, and trace part 
$\phi_{\varepsilon}$, according to
\begin{equation}\label{Decomposition}
 \bar{\varepsilon}_{\mu \nu}
 = 
 \widehat{\varepsilon}_{\mu \nu}+ \tilde{\varepsilon}_{\mu \nu} 
 + \frac{1}{d}\,\bar{g}_{\mu\nu}\phi_{\varepsilon}
\end{equation}
with $\widehat{\varepsilon}_{\mu \nu}=\widehat{\varepsilon}_{\nu \mu}$, 
$\widehat{\varepsilon}^{\mu}_{\ \mu}=0$ and $\tilde{\varepsilon}_{\mu 
\nu}=\bar{\varepsilon}_{[\mu \nu]}$. In addition we specify the background 
spacetime to be a maximally symmetric Einstein space with
\begin{equation}\label{EinsteinBackg}
 \bar{R}_{\mu \nu \rho \sigma}
  =
  \frac{1}{d(d-1)}\big[ \bar{g}_{\mu \rho}\bar{g}_{\nu \sigma}
  -\bar{g}_{\mu \sigma}\bar{g}_{\nu \rho}\big] \bar{R} 
  \quad \text{and} \quad 
  \bar{R}_{\mu\nu} =\frac{1}{d}\, \bar{g}_{\mu \nu} \bar{R}.
\end{equation}
This spacetime is still sufficiently general to identify the contributions to 
the relevant invariants $\int\! \sqrt{\bar{g}}$ and $\int\! \sqrt{\bar{g}} 
\bar{R}$ unambiguously. Within the present truncation it is thus a permissable 
restriction of the function space of the metric; it does not affect the 
generality of the calculation and so is an example of the computational trick 
mentioned above.

Using the relations \eqref{Decomposition} and \eqref{EinsteinBackg} the 
quadratic part of the action reads
\begin{equation}
 \begin{split}\label{GammaQuaddec}
  \breve{\Gamma}^{\text{quad}}_k[\bar{\varepsilon};\bar{e}]
  &=
  \frac{Z_{Nk} \kappa^2}{2} \frac{4}{\xi^2} \int \ddx \sqrt{\bar{g}} 
  \Bigg\{ 
   \widehat{\varepsilon}_{\mu \nu} \Big[-\!\bar{D}^2 
   + (\xi-2)\bar{\lambda}_k + C_T(\xi)\bar{R}\Big] 
   \widehat{\varepsilon}^{\mu\nu}\\
  &\hspace{2cm} 
   +\tilde{\varepsilon}_{\mu\nu} \xi \bigg[\frac{1}{Z_{N k}\kappa^2 \alpha_L}+ 
   \bar{\lambda}_k - \frac{d-2}{2d}\bar{R}\bigg] \tilde{\varepsilon}^{\mu\nu}\\
  &\hspace{2cm} -\frac{d-2}{2d}\phi_{\varepsilon} 
   \bigg[ -\!\bar{D}^2- \bigg(2+\frac{2\xi}{d-2}\bigg)\bar{\lambda}_k
   + C_S(\xi)\bar{R}\bigg]\phi_\varepsilon\Bigg\}
 \end{split}
\end{equation}
with the constants
\begin{equation}
 C_T(\xi)\equiv \frac{d(d-3)+4}{d(d-1)}- \frac{d-2}{2d}\xi, \qquad 
 C_S(\xi)\equiv\frac{d-4 +\xi}{d}.
\end{equation}
Note that whereas the symmetric tensor $\widehat{\varepsilon}_{\mu \nu}$ has a 
standard positive definite kinetic term, its antisymmetric counterpart is 
non-propagating; the $\tilde{\varepsilon}_{\mu \nu}$-bilinear contains no 
derivatives at all, but only a (gauge dependent) mass term. Note also that in 
$d>2$ the trace part $\phi_{\varepsilon}$ has a ``wrong sign'' kinetic term, 
reflecting the well known conformal factor instability \cite{mr}.

Let us now fix the precise form of the cutoff operator $\widehat{R}_k$ in the 
various sectors of field space. Generically it has the structure
\begin{equation}
 \widehat{R}_k={\cal Z}_k k^2 R^{(0)}(-\bar{D}^2/k^2),
\end{equation}
where ${\cal Z}_k$ is a matrix in field space, and $R^{(0)}(u)$ is a 
dimensionless ``shape function'' that interpolates smoothly between 
$R^{(0)}(0)=1$ and $\lim_{u\rightarrow \infty}R^{(0)}(u)=0$. At least in 
simple matter field theories on a rigid background spacetime, there is a 
simple rule for finding a suitable $\mathcal{Z}_k$, and this rule has also 
been used in the metric calculation in \cite{mr}: If a certain field mode has 
a kinetic operator of the form $[-\bar{D}^2 + \cdots]$, the $\mathcal{Z}_k$ is 
fixed in such a way that in the sum $\Gamma_k+ \Delta_k S$ this operator gets 
replaced by $[-\bar{D}^2 + k^2 R^{(0)}(-\bar{D}^2/k^2)+ \cdots]$. 

In the case at hand it is straightforward to implement this rule for 
$\widehat{\varepsilon}_{\mu\nu}$ and $\phi_\varepsilon$. In the different 
sectors we choose 
\begin{equation}
 ({\cal Z}_k)_{\widehat{\varepsilon}\widehat{\varepsilon}}
 =
 2 Z_{N k} \xi^{-2} \kappa^2,\quad
 ({\cal Z}_k)_{\tilde{\varepsilon}\tilde{\varepsilon}}
 =
 2 \xi^{-1} Z_{N k} \kappa^2,\quad
 ({\cal Z}_k)_{\phi_{\varepsilon}\phi_{\varepsilon}}
 =
 -2 \xi^{-2} Z_{N k} \kappa^2\frac{d-2}{2d}.
\end{equation}
As for the antisymmetric tensor $\tilde{\varepsilon}_{\mu\nu}$, we fixed the 
corresponding $\mathcal{Z}_k$ in such a way that, taking the overall prefactor 
into account, the addition of $\widehat{R}_k$ to the inverse propagator 
replaces the square brackets in the $\tilde{\varepsilon}_{\mu\nu}$-bilinear of 
\eqref{GammaQuaddec} by
\begin{equation}
 \bigg[k^2 R^{(0)}(-\bar{D}^2/k^2)+\frac{1}{Z_{N k}\kappa^2 
 \alpha_L}+\bar{\lambda}_k - \frac{d-2}{2d}\bar{R}\bigg].
\end{equation}

Now we have specified all ingredients entering the supertrace on the RHS of 
\eqref{FRGE} in the different sectors. 

First of all we note that the contributions of the antisymmetric sector vanish 
in the limit of $\alpha_L\rightarrow 0$, as this part of the trace is given by
\begin{equation}
\begin{split}
  &\frac{1}{2}\Tr\!{}_{\tilde{\varepsilon}\tilde{\varepsilon}}
  \Bigg[
  \frac{\partial_t\big(Z_{Nk} k^2 R^{(0)}(-\bar{D}^2/k^2)\big)}{Z_{Nk}\big(k^2 
  R^{(0)}(-\bar{D}^2/k^2)+\bar{\lambda}_k + 1/(Z_{Nk}\kappa^2 \alpha_L) - 
  \bar{R} (d-2)/(2d)\big)}\Bigg]\\
  &=\frac{\alpha_L}{2} \Tr\!{}_{\tilde{\varepsilon}\tilde{\varepsilon}}   
  \Bigg[\frac{\partial_t\big(Z_{Nk} k^2  
  R^{(0)}(-\bar{D}^2/k^2)\big)}{Z_{Nk}\big(\alpha_L k^2  
  R^{(0)}(-\bar{D}^2/k^2)+\alpha_L \bar{\lambda}_k + 1/(Z_{Nk}\kappa^2 ) - 
  \alpha_L \bar{R} (d-2)/(2d)\big)}\Bigg]\\
 &\xrightarrow{\alpha_L\rightarrow 0} 0.
 \end{split}
\end{equation}
This behavior is easy to understand as the limit $\alpha_L\rightarrow 0$ 
corresponds to a sharp implementation of the ${\sf O(}d)$ gauge condition that 
introduces a delta functional $\delta[\tilde{\varepsilon}_{\mu \nu}]$ into the 
path integral. Since the domain of tensors with 
$\tilde{\varepsilon}_{\mu\nu}=0$ is invariant under the coarse graining 
operation it is obvious that the antisymmetric fluctuations should not 
contribute to any RG running in this limit. From now on we will choose the 
gauge $\alpha_L=0$ in order to simplify the discussion.

In this particularly simple gauge the quadratic form \eqref{GammaQuaddec} is 
structurally similar to the corresponding equation in the metric formalism, 
see eq. (4.12) in \cite{mr}. However, the prefactors of $\bar{\lambda}_k$ in 
the various terms of $\breve{\Gamma}^{\text{quad}}_k$ and the now 
$\xi$-dependent coefficients $C_S(\xi)$, $C_T(\xi)$ of the curvature scalar 
$\bar{R}$ are different and this will have a rather significant impact on the 
resulting RG flow. Replacing these constants appropriately in the original 
metric calculation we can obtain the ``bosonic'' contributions to the 
$\beta$-functions without a new calculation from those of \cite{mr}.

\subsection{Propagating and non-propagating ghosts}
Let us move on and discuss the ghost sector. Here we choose the cutoff 
operator to be
\begin{equation}
 \widehat{R}^{\text{gh}}_k  
  =
  \begin{pmatrix}
   \sqrt{2} \xi^{-1}\delta^{\mu}_{\ \rho} k^2 R^{(0)} (-\bar{D}^2/k^2) & 0\\
   0 &  Z^{\text{gh}}_{Lk}\delta^{[\mu}_{\rho} \delta^{\nu]}_{\sigma} 
   k^2 R^{(0)}(-\bar{D}^2/k^2)
\end{pmatrix}.
\end{equation}
In the diffeomorphism-ghost sector we have adjusted 
$\mathcal{Z}^{\text{gh}}_k$ to the kinetic term according to the above rule. 

In the ${\sf O(}d)$ ghost sector, however, there is no kinetic term; the 
ghosts do not propagate. Nevertheless, a consistent application of the FRGE 
requires us not to ignore, but to systematically integrate out these 
non-propagating modes in the same way as all the others, i.\,e. ordered, and 
eventually cut off according to their $\bar{D}^2$-eigenvalue. Therefore we 
introduce a cutoff-operator (with a prefactor unrelated to the couplings in 
$\Gamma_k$, denoted by $Z^{\text{gh}}_{Lk}$) in this sector as 
well.\footnote{Recall that ideally, at the exact level, the cutoff action 
$\Delta_k S$ would be independent of the running couplings present in 
$\Gamma_k$ \cite{avact}.}

In the gauge chosen, the inverse ghost propagator 
$S^{(2)}_{\text{gh}}+\widehat{R}^{\text{gh}}_k$ is a triangular matrix, such 
that the contributions of the different sectors to the trace decouple.

For any constant choice of $Z^{\text{gh}}_{Lk}=Z^{\text{gh}}_{L}$ we obtain 
contributions of the ${\sf O(}d)$ ghost sector of the form
\begin{equation}\label{GhostTrace}
 \Tr\Bigg[
 \frac{\partial_t (Z^{\text{gh}}_{L}k^2 R^{(0)})}{-M^2 +Z^{\text{gh}}_{L}k^2 
  R^{(0)}}\,\delta^{[\mu}_{\rho} \delta^{\nu]}_{\sigma}\Bigg]
 =
 \Tr \Bigg[ \frac{k^{-2}\partial_t 
 (k^2 R^{(0)})}{-\frac{M^2}{Z^{\text{gh}}_{L}k^2}  
 +R^{(0)}}\,\delta^{[\mu}_{\rho} \delta^{\nu]}_{\sigma}\Bigg]
\end{equation}
with the abbreviation $M^2\equiv2\bar{\mu}^2\xi^{-1/2}$. Introducing the 
dimensionless mass parameter $\mu \equiv\bar{\mu}/k$, and then neglecting any 
further running of $\mu$, we observe that the trace \eqref{GhostTrace} depends 
only on the $k$-independent dimensionless quantity 
\begin{equation}\label{DimlessQuant}
 -\frac{M^2}{Z^{\text{gh}}_{L}k^2}\equiv 
 -\frac{2 \mu^2}{Z^{\text{gh}}_{L}\xi^{1/2}}. 
\end{equation}
In order to avoid divergences due to a vanishing denominator in 
\eqref{GhostTrace} we have to choose a negative value for $Z_L^{\text{gh}}$, 
as known from the conformal sector. Since both parameters, $\mu$ and 
$Z^{\text{gh}}_L$, occur only in the combination \eqref{DimlessQuant} we can 
mimic any choice of $Z^{\text{gh}}_L<0$ by choosing a suitable $\mu$. (In 
particular $Z^{\text{gh}}_L\mapsto -1$, upon replacing $\mu^2\mapsto 
-\mu^2/Z^{\text{gh}}_L$.)

In the following we will discuss three distinguished choices of 
$Z^{\text{gh}}_L$:

\vspace{0.1cm}
\noindent{\bf (i)} $Z^{\text{gh}}_L=-1$: the cutoff term is unrelated to 
$\Gamma_k$, the ${\sf O(}d)$ ghost contribution will therefore depend on $\mu$ 
and $\xi$.

\vspace{0.1cm}
\noindent{\bf (ii)} $Z^{\text{gh}}_L=-M^2/k^2=-2\mu^2\xi^{-1/2}$: the cutoff 
is optimally adapted to the form of $\Gamma_k$ leading to a cancelation of the 
parameters $\mu$ and $\xi$. This procedure is closest to the above rule for 
usual kinetic term adaptation and we therefore expect the most reliable 
results for this choice.

\vspace{0.1cm}
\noindent{\bf (iii)} $Z^{\text{gh}}_L\rightarrow0$: no cutoff term introduced. 
This choice corresponds to neglecting the ${\sf O(}d)$ ghost modes completely, 
the trace \eqref{GhostTrace} vanishes.

\vspace{0.1cm}
As explained above, these three choices are equivalent to using 
$Z^{\text{gh}}_L=-1$ and setting $\mu^2$ equal to $\mu^2$, $\xi^{1/2}/2$, and 
$\mu\rightarrow\infty$, respectively. We shall refer to them as the {\it ghost 
adaptation schemes (i)--(iii)} from now on.

\subsection{The interpolating beta functions}
The remaining part of the calculation consists of projecting out the 
invariants $\int\!\sqrt{\bar{g}}$ and $\int\! \sqrt{\bar{g}} \bar R$ from the 
supertrace in order to find the $\beta$-functions for $G_k$ and 
$\bar{\lambda}_k$; it follows exactly the metric calculation in \cite{mr}.

If we turn over to dimensionless couplings
\begin{equation}
 g_k= \frac{k^{d-2}}{32 \pi Z_{Nk} \kappa^2}= k^{d-2} G_k,\qquad \lambda_k
 = k^{-2} \bar{\lambda}_k
\end{equation}
the resulting system of coupled RG equations is autonomous and has the 
structure
\begin{align}
 \partial_t g_k 
  &= 
  \beta_g(g_k,\lambda_k) \equiv \big[d-2 +\eta_N(g_k, \lambda_k)\big] g_k,\\
  \partial_t \lambda_k&=\beta_\lambda(g_k, \lambda_k)
\end{align}
with the anomalous dimension $\eta_N=-\partial_t \ln Z_{Nk}$. We shall employ 
the standard threshold functions $\Phi$, $\tilde{\Phi}$ of \cite{mr} along 
with a new type of threshold function, $\tilde{\tilde{\Phi}}$, defined 
according to
\begin{align}
 \Phi^p_n(w)
  &= \frac{1}{\Gamma(n)} \int_0^\infty \text{d}z\, z^{n-1} 
  \frac{R^{(0)}(z)-z R^{(0)}{}'(z)}{[z+R^{(0)}(z)+w]^p}\\
  \tilde{\Phi}^p_n(w)&= \frac{1}{\Gamma(n)} \int_0^\infty \text{d}z\, z^{n-1} 
  \frac{R^{(0)}(z)}{[z+R^{(0)}(z)+w]^p}\qquad\qquad \text{for}\quad n>0\\
  \tilde{\tilde{\Phi}}^p_n(w)&= \frac{1}{\Gamma(n)} \int_0^\infty \text{d}z\, 
  z^{n-1} \frac{R^{(0)}(z)-z R^{(0)}{}'(z)}{[R^{(0)}(z)+w]^p}
\end{align}
and $\tilde{\tilde{\Phi}}^p_0(w)=\tilde{\Phi}^p_0(w)=\Phi^p_0(w)= (1+w)^{-p}$. 
We can write down an explicit expression for $\eta_N$ in terms of the 
couplings $g,\,\lambda$ then:
\begin{equation}
 \eta_N(g,\lambda)= \frac{g \bar{B}_1(\lambda)}{1- g \bar{B}_2(\lambda)}
\end{equation}
The functions $\bar{B}_1$ and $\bar{B}_2$ are $\xi$-dependent generalizations 
of similar ones occurring in \cite{mr}:
\begin{equation}
\begin{split}
\bar{B}_1(\lambda) = &\frac{1}{3} (4\pi)^{1-d/2} 
 \bigg[ (d-1)(d+2)\, \Phi^1_{d/2-1}\big((\xi-2)\lambda\big)\\
 &+2 \Phi^1_{d/2-1}\bigg(\!-2\lambda\frac{d-2+\xi}{d-2}\bigg) -4 d \, 
 \Phi^1_{d/2-1}(0)-2 d(d-1) \,\tilde{\tilde{\Phi}}^1_{d/2-1}
 \bigg(\frac{2\mu^2}{\sqrt{\xi} }\bigg)\\
 &-6(d-1)(d+2)\, C_T(\xi)\, \Phi^2_{d/2}\big((\xi-2)\lambda\big)\\
 & -12\, C_S(\xi) \,\Phi^2_{d/2}\bigg(\!-2\lambda\frac{d-2+\xi}{d-2}\bigg) 
 -24 \Phi^2_{d/2}(0)\bigg]
\end{split}
\end{equation}
and
\begin{equation}
\begin{split}
\bar{B}_2(\lambda) 
 = 
 &-\frac{1}{6} (4\pi)^{1-d/2} \bigg[ (d-1)(d+2)\, 
 \tilde{\Phi}^1_{d/2-1}\big((\xi-2)\lambda\big)\\
 & +2 \tilde{\Phi}^1_{d/2-1}\bigg(\!-2\lambda\frac{d-2+\xi}{d-2}\bigg) 
 -6(d-1)(d+2)\, C_T(\xi)\, \tilde{\Phi}^2_{d/2}\big((\xi-2)\lambda\big)\\
 & -12 \,C_S(\xi)\, \tilde{\Phi}^2_{d/2}\bigg(\!-2\lambda\frac{d-2+\xi}{d-2}
 \bigg)\bigg].
\end{split}
\end{equation}
For the $\beta$-function of the cosmological constant we obtain
\begin{equation}
\begin{split}
 \beta_\lambda= 
 & -(2-\eta_N)\lambda\\ 
 &+ \frac{1}{2}g_k (4\pi)^{1-d/2}\bigg[2(d-1)(d+2)\,\Phi^1_{d/2} 
 \big((\xi-2)\lambda\big) + 4 \Phi^1_{d/2} 
 \bigg(\!\!-2\lambda\frac{d-2+\xi}{d-2}\bigg)\\
 & -8 d\, \Phi^1_{d/2}(0) -4d(d-1)\, 
 \tilde{\tilde{\Phi}}^1_{d/2}\bigg(\frac{2\mu^2}{\sqrt{\xi}}\bigg)-(d-1)(d+2)\,
  \eta_N \,\tilde{\Phi}^1_{d/2} \big((\xi-2)\lambda\big)\\
 & -2 \,\eta_N\, \tilde{\Phi}^1_{d/2} 
 \bigg(\!-2\lambda\frac{d-2+\xi}{d-2}\bigg)\bigg].
\end{split}
\end{equation}
These general $\xi$-dependent expressions are in exact correspondence to the 
eqns. (4.40) and (4.43) of ref. \cite{mr} for metric gravity. Analyzing the 
$\xi$-dependence of the RG flow they give rise to is a convenient way of 
exploring the field parametrization dependence of the flow.

An important observation is that for a constant, $\xi$-independent choice of 
$\mu$ (i.\,e. in the ghost adaptation schemes (i) and (iii)) {\it the above 
$\beta$-functions reduce precisely to those of the metric result in the limit 
$\xi \rightarrow 0$.} All prefactors and arguments of the threshold functions 
coincide and the function $\tilde{\tilde{\Phi}}$ vanishes in this limit. 
Although this result is far from obvious when considering the definition of 
$\xi$ in eq. \eqref{DefXi}, we can now regard this one parameter family of 
field redefinitions as an interpolation between the metric description (for 
$\xi \rightarrow 0$) and the usual vielbein decomposition (for $\xi=1$). 

In scheme (ii) however, the argument of $\tilde{\tilde{\Phi}}$ is constant, so 
that in the limit $\xi\rightarrow0$ the $\beta$-functions match the metric 
result except for the additional $\tilde{\tilde{\Phi}}$ contributions, which 
are precisely the terms due to the ${\sf O(}d)$ ghosts.

\section{Numerical analysis of the RG flow}
In this section we will analyze the RG flow in $d=4$ dimensions. We will 
compare results of different cutoff schemes, namely with the optimized shape 
function, $R^{(0)}(z)=(1-z) \theta(1-z)$, and the exponential one, $R^{(0)}(z) 
= sz/(e^{sz}-1)$, for shape parameters $s$ ranging from 2 to 20.
\subsection{The standard vielbein case $\xi = 1$}
To start with, let us consider the usual vielbein representation of the metric 
in \eqref{DefXi} and set $\xi =1$ for the time being. With $\xi$ fixed the 
flow continues to depend on the mass parameter $\mu\equiv \bar\mu / k$. We 
shall analyze this dependence in the following, highlighting especially the 
implications of those choices of $\mu$ that correspond to the three ghost 
adaptation schemes (i)--(iii).

A first encouraging result is that there exists a non-Gaussian fixed point 
(NGFP), for any value of the dimensionless constant $\mu\neq0$, and in all 
cutoff schemes we studied. 

\begin{figure}[htb]
\centering
 \includegraphics[width=.8 \linewidth]{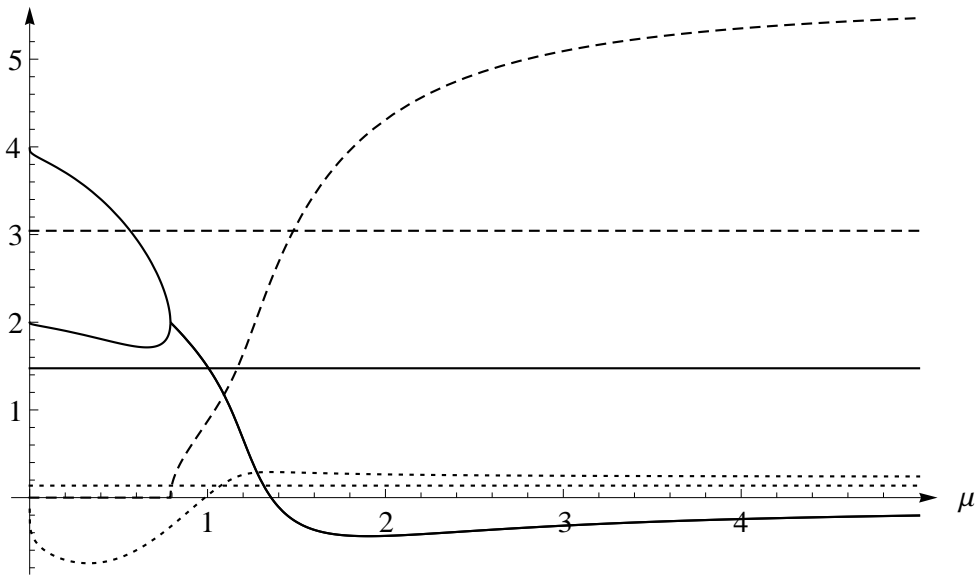}
\caption{The critical exponents $\theta_i=\theta_i'+i\theta_i''$ split into 
real and imaginary part (solid and dashed line, respectively) of the NGFP and 
the product $g^\ast\lambda^\ast$ (dotted line) as a function of the mass 
parameter $\mu$ for the optimized cutoff. The straight horizontal lines 
represent the values of the corresponding quantities in the metric 
calculation.}\label{Plot1}
\end{figure}

\noindent{\bf (A) Fixed point properties.} Figure \ref{Plot1} shows, for the 
case of the optimized cutoff, the $\mu$-dependence of three quantities one 
might expect to be universal, namely the critical exponents $\theta_i$ at the 
fixed point and the product $g^\ast\lambda^\ast$. We notice that, while the 
very existence of the fixed point is indeed universal, its properties heavily 
depend on the value of $\mu$: For $\mu\lesssim 0.8$ we find a UV attractive FP 
with two real critical exponents, which then turn into a complex conjugated 
pair. At $\mu\approx 1.35$ the FP changes its character and becomes UV 
repulsive in both directions. For large $\mu$-values the dependence on $\mu$ 
weakens for all three ``universal'' quantities. 

Employing the exponential cutoff (not shown here) essentially leads to the same 
picture: real critical exponents turn into a complex pair before the otherwise 
UV attractive FP gets UV repulsive for large $\mu$. In all cases the product 
$g^\ast \lambda^\ast$ changes its sign from negative to positive within the 
interval of $\mu$, in which the FP is attractive and has complex critical 
exponents.

It is important to stress that even in a much better truncation with many more 
invariants we would not expect these quantities to become independent of 
$\mu$: The parameter $\mu$ should not be considered a free parameter 
corresponding e.\,g. to different cutoff schemes, but it rather corresponds to 
an additional {\it coupling}. In principle its running is prescribed by an 
additional $\beta$-function which however is not determined by the present 
calculation. Therefore one should not worry too much about the 
$\mu$-dependence of the ``universal'' quantities.

In the {\bf ghost adaptation scheme (i)} the best we can do, as we did not 
calculate the running of $\mu$ in our truncation, is to sensibly choose a 
fixed value for the constant $\mu$. Most naturally we would choose a value of 
the order of 1 as any other choice would correspond to the introduction of an 
additional unmotivated physical scale other than $k$.

Strikingly, in all cutoff schemes studied there exists indeed a $\mu$-interval 
including, or at least close to $\mathbf{\boldsymbol{\mu}=1}$ in which the 
situation is similar to the metric theory: We find the NGFP, it is UV 
attractive, has $g^\ast \lambda^\ast>0$, and a pair of complex conjugate 
critical exponents. 

As an alternative to choosing $\mu=1$ it is therefore tempting to find the 
``best fit'' to the metric calculation by selecting a $\mu$-value such that 
there is also a quantitative agreement of the universal quantities.

In Fig. \ref{Plot1} the values corresponding to the metric calculation are 
given by the horizontal lines. We observe that the crossings of the lines of 
the same type are quite close to each other and are all located at a $\mu$ of 
the order of 1. For the optimized cutoff we find the crossing for the real 
part of the critical exponent $\theta_i'$ very much at $\mu \approx 1$ and for 
the product $g^\ast\lambda^\ast$ at about $\mu \approx 1.1$; the imaginary 
part $\theta_i''$ takes on its metric value at $\mu\approx 1.45$ in a region 
where the FP turned UV repulsive already. Taking the average of these values 
we arrive at $\mathbf{\boldsymbol{\mu}\approx 1.2}$, for which we expect the 
best agreement between the vielbein and the metric theory.

The fact that we find this agreement of metric and vielbein values in a 
relatively small $\mu$-interval close to the most natural value of $\mu=1$ can 
be interpreted as an indication that also the full quantum theories are 
similar to each other or perhaps equivalent.

The {\bf adaptation scheme (ii)} is expected to be the most reliable one. It 
yields the value $\mu=1/\sqrt{2}\approx0.7$ for $\xi=1$. However, in this 
scheme the results of scheme (i) are not confirmed: For the smaller value of 
the parameter $\mu$ we find a UV attractive NGFP at $g^\ast\lambda^\ast<0$, 
with two real critical exponents. 

The {\bf adaptation scheme (iii)} corresponds to large $\mu\rightarrow\infty$, 
so that we find a UV repulsive FP in this scheme.

\begin{figure}[phtb]
\centering
\subfigure[$\mu\!=\!1/\sqrt{2}$, ghost adaptation scheme (ii): The phase 
portrait resembles the one in QECG.
]{\includegraphics[width=0.45 
\linewidth]{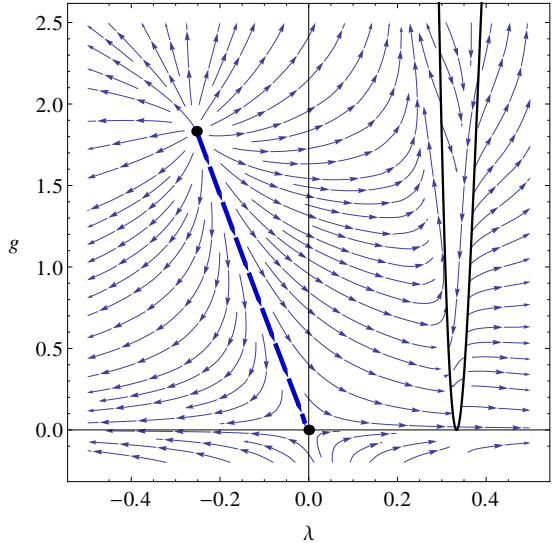}\label{Plot2a}}\quad
\subfigure[$\mu$=1: This value is the most natural choice when using ghost 
adaptation scheme (i).
]{\includegraphics[width=0.45 \linewidth]{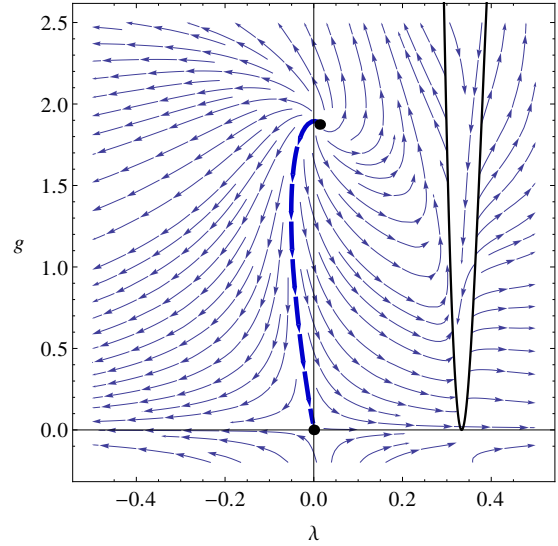}\label{Plot2b}}\\
\subfigure[$\mu$=1.2: For this value we obtain a situation most similar to the 
metric theory in scheme (i).
]{\includegraphics[width=0.45 
\linewidth]{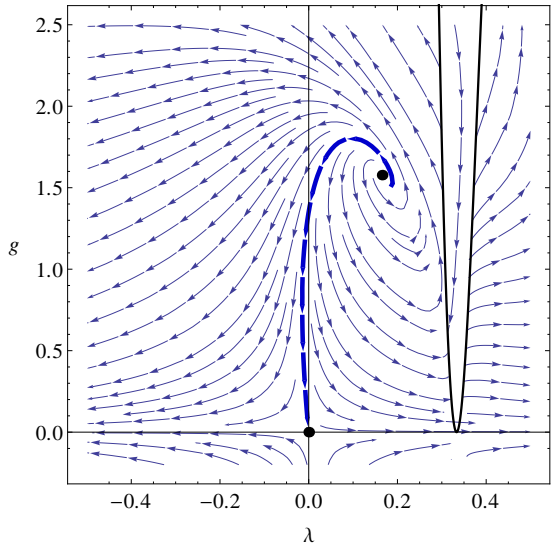}\label{Plot2c}}\quad
\subfigure[$\mu$=2: The limit cycle is a qualitatively new feature of the 
phase portrait in scheme (iii).
]{\includegraphics[width=0.45 
\linewidth]{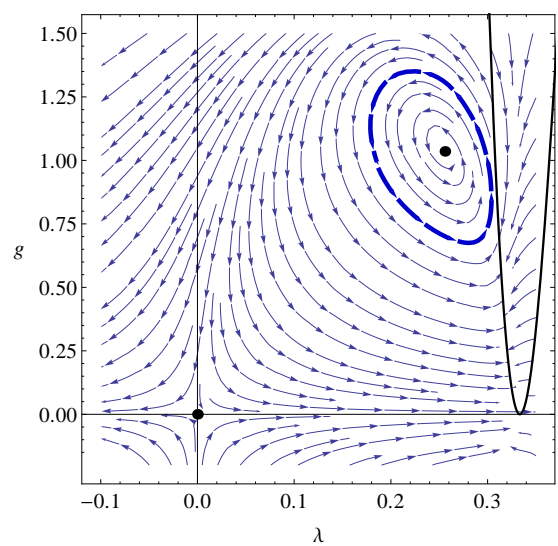}\label{Plot2d}}\caption{RG phase portraits for 
different values of the mass parameter $\mu$ at $\xi\!=\!1$. The figures show 
the impact of the ${\sf O(}d)$ ghost contribution: While for large $\mu$, when 
it is suppressed, we obtain a limit cycle, for smaller $\mu$ we find flow 
diagrams similar to the ones known from QEG and QECG, 
respectively.}\label{Plot2}
\end{figure}

\vspace{0.5cm}
\noindent{\bf (B) The phase portrait.} Let us now discuss the entire RG flow. 
In Fig. \ref{Plot2} we have plotted its phase portrait for different values of 
$\mu$. Figs. \ref{Plot2b} and \ref{Plot2c} correspond to the first adaptation 
scheme {\bf (i)}. We observe that the best fit case $\mu=1.2$ is indeed most 
similar to the metric flow known from Quantum Einstein Gravity (QEG) in the 
Einstein-Hilbert truncation \cite{mr,frank1,oliver}: We find a NGFP in the 
positive $(g,\lambda)$-quadrant with two attractive directions; the 
trajectories spiral into it due to the nonzero imaginary part of the critical 
exponents. It is in an interplay with the Gaussian fixed point (GFP) and there 
exists a ``separatrix'' that separates trajectories with positive and negative IR 
values for the cosmological constant $\lambda$. Also a major difference to the 
metric case is to be noted: The UV repulsive direction of the GFP has changed 
and points now into the negative $\lambda$-halfplane. Therefore the separatrix 
starts off with negative $\lambda$ before heading to the NGFP at 
$\lambda^\ast>0$. This effect can be traced back to be due to the ${\sf O(}d)$ 
ghost contributions.

For smaller $\mu$ (as e.\,g. $\mu=1/\sqrt{2}$ in adaptation scheme {\bf (ii)}) 
these contributions are enhanced in such a way, that the NGFP itself lies at 
$\lambda^\ast<0$ (cf. Fig. \ref{Plot2a}). Now the fixed point has real 
critical exponents, but is still UV attractive. Qualitatively this picture 
resembles much the RG flow of Quantum Einstein Cartan Gravity (QECG) in the 
planes of vanishing and infinite Immirzi parameter as found in \cite{e-omega}.

For large $\mu$ (scheme {\bf (iii)}), as exemplarily shown for the case of 
$\mu=2$ in Fig. \ref{Plot2d}, we find a rather different behavior. Although 
the flow looks similar to the metric case in large parts of the 
$(g,\lambda)$-plane, the NGFP is repulsive now; the critical exponents form a 
complex conjugate pair with a negative real part. These two circumstances lead 
to the formation of a {\it limit cycle} around the NGFP. This limit cycle is 
UV attractive for trajectories approaching it both from outside and from the 
interior.

Clearly such a limit cycle is an interesting and intriguing new possibility 
for the nonperturbative UV completion of a quantum field theory. It is 
``asymptotically safe'' in a novel sense. However, in this concrete case the 
picture of a limit cycle is hardly credible against the background of all RG 
flow studies of gravity to date. Nevertheless it is inspiring to see its 
formation for the first time in quantum gravity.

\vspace{0.5cm}
\noindent{\bf (C) Non-propagating ghosts.} The fact that we should not choose 
the parameter $\mu$ too large teaches us another important lesson: Consider 
the $\beta$-functions as given in the previous section. They involve the new 
threshold function $\tilde{\tilde{\Phi}}^1_{d/2}(w)$ that vanishes for 
$w\rightarrow \infty$ and diverges for $w\rightarrow0$. In both 
$\beta$-functions, the terms with $\tilde{\tilde{\Phi}}^1_{d/2}(w)$ are 
exactly the ghost contributions of the ${\sf O(}d)$ gauge group. Since the 
$\tilde{\tilde{\Phi}}$ argument is always $w= 2 \mu^2/\sqrt{\xi}$ we can 
control the magnitude of these contributions by changing $\mu$: We obtain a 
suppression for large $\mu$ and an infinite enhancement in the limit 
$\mu\rightarrow 0$. If we had not added a cutoff for the ${\sf O(}d)$ ghosts, 
the situation would correspond to the limit $\mu\rightarrow\infty$, i.\,e. 
adaptation scheme (iii). In this case we find a {\it UV repulsive} fixed 
point, quite different from all results known from metric calculations. We 
therefore conclude that contrary to the situation of perturbation theory 
\cite{Woodard} {\it it is crucial to include all modes of the non-background 
fields into the renormalization procedure, whether they are propagating or 
not, by introducing a cutoff-operator for all of them and retaining their 
contribution to the supertrace in the FRGE}. This implies that we should 
choose adaptation scheme (i) or (ii) but {\it not} (iii).

Similar remarks might also apply to perturbative calculations with regularization 
schemes which retain power divergences.\footnote{It might be interesting to 
reconsider the calculation \cite{bene-speziale} in this light since there a 
propertime regulator has been used.}

\subsection{Field parametrizations with $\xi \neq 1$}
When altering the value of $\xi$, we do not change theory space as both field 
content and symmetries remain the same. Therefore we expect to find the same 
fixed point properties in the RG flow for all values of $\xi$, resulting in 
universal quantities that, in case of a good approximation to the exact flow, 
are largely independent of $\xi$. We will use this criterion in order to test 
the reliability of the different ghost adaptation schemes in this section.

\begin{figure}[htb]
\centering
\subfigure[$\mu\!=\!1/\sqrt{2}$]{\includegraphics[width=0.45 
\linewidth]{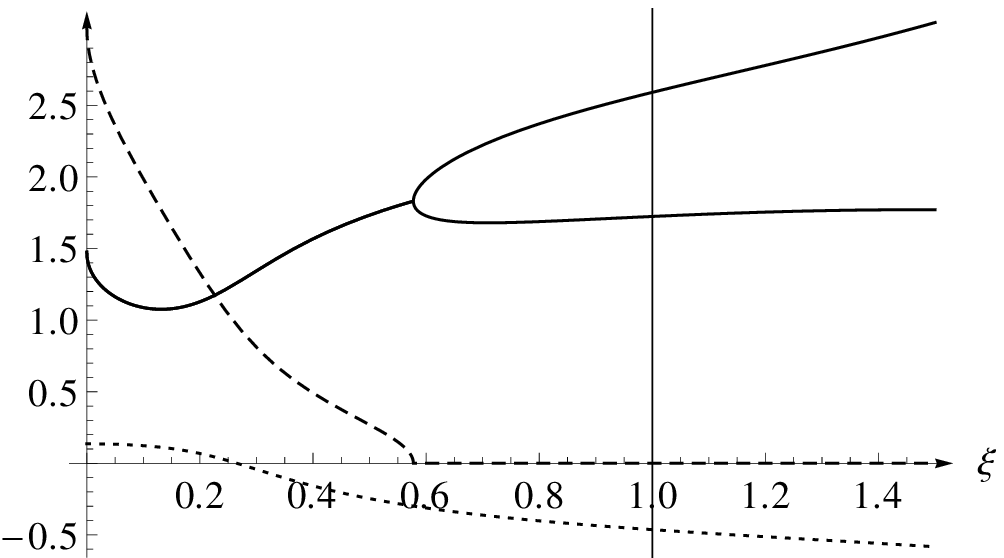}\label{Plot3a}}\quad
\subfigure[$\mu$=1.0]{\includegraphics[width=0.45 
\linewidth]{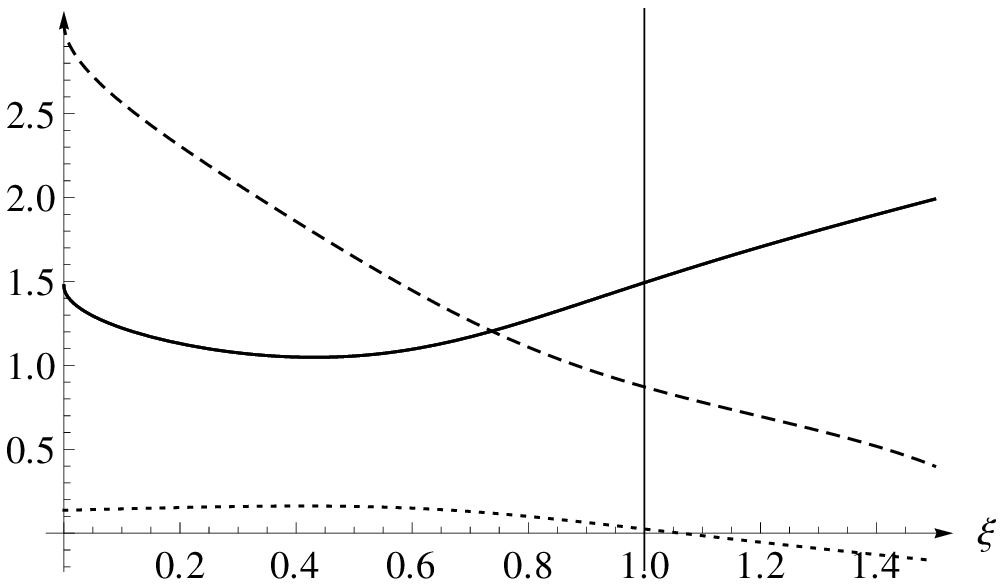}\label{Plot3b}}\quad
\subfigure[$\mu$=1.2]{\includegraphics[width=0.45 
\linewidth]{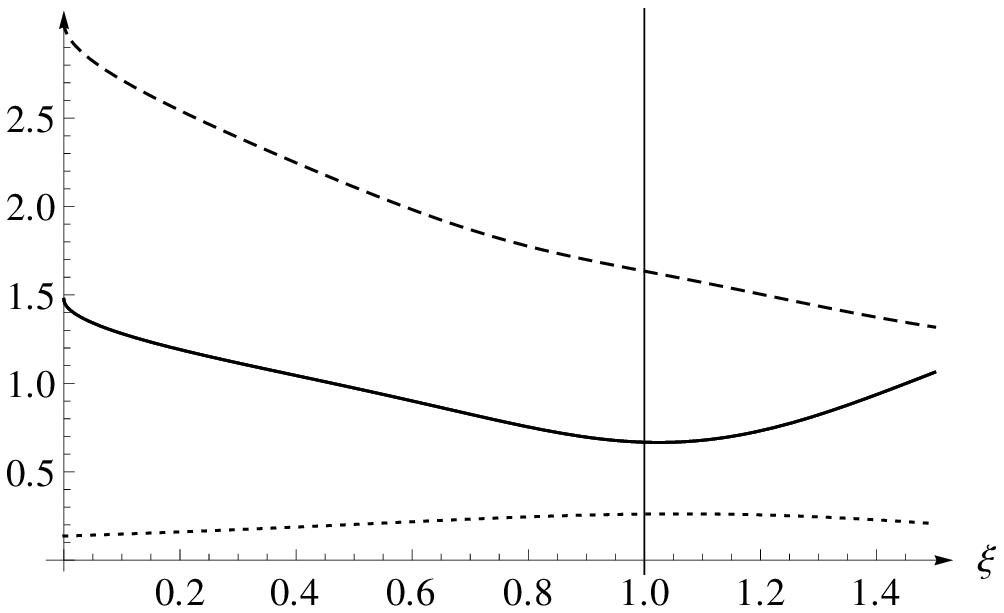}\label{Plot3c}}\quad
\subfigure[$\mu$=2]{\includegraphics[width=0.45 
\linewidth]{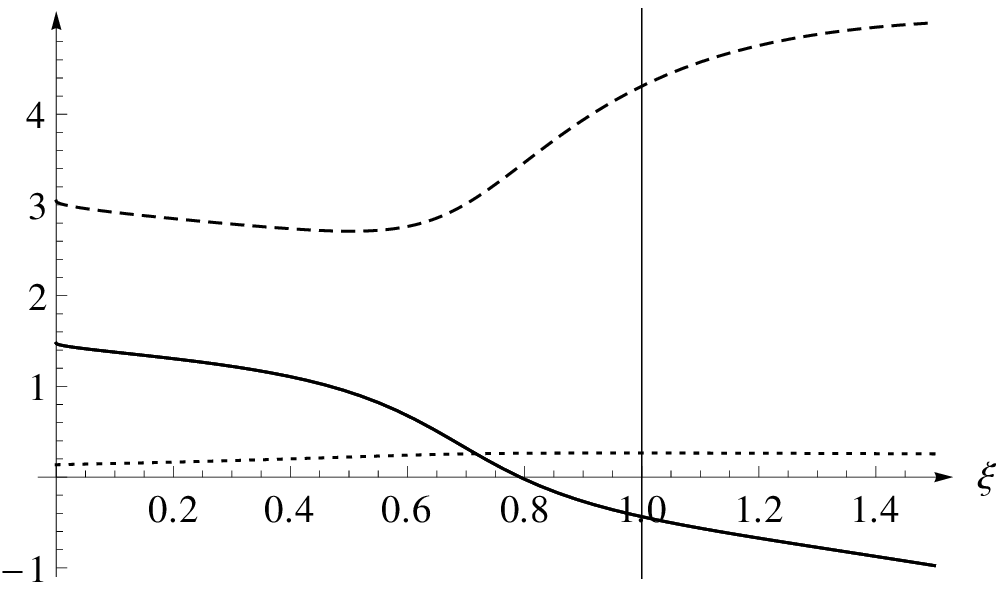}\label{Plot3d}}\caption{Critical exponents and 
$g^\ast \lambda^\ast$ calculated using the optimized cutoff, for different 
values of the mass parameter $\mu$, as functions of $\xi$: The real part of 
the critical exponents $\theta_i'$ (solid), its imaginary part $\theta_i''$ 
(dashed) and the product of the fixed point coordinates $g^\ast \lambda^\ast$ 
(dotted).}\label{Plot3}
\end{figure}

\begin{figure}[htb]
\centering
\subfigure[$\mu=(\xi/4)^{1/4}/\sqrt{2}$]{\includegraphics[width=0.45 
\linewidth]{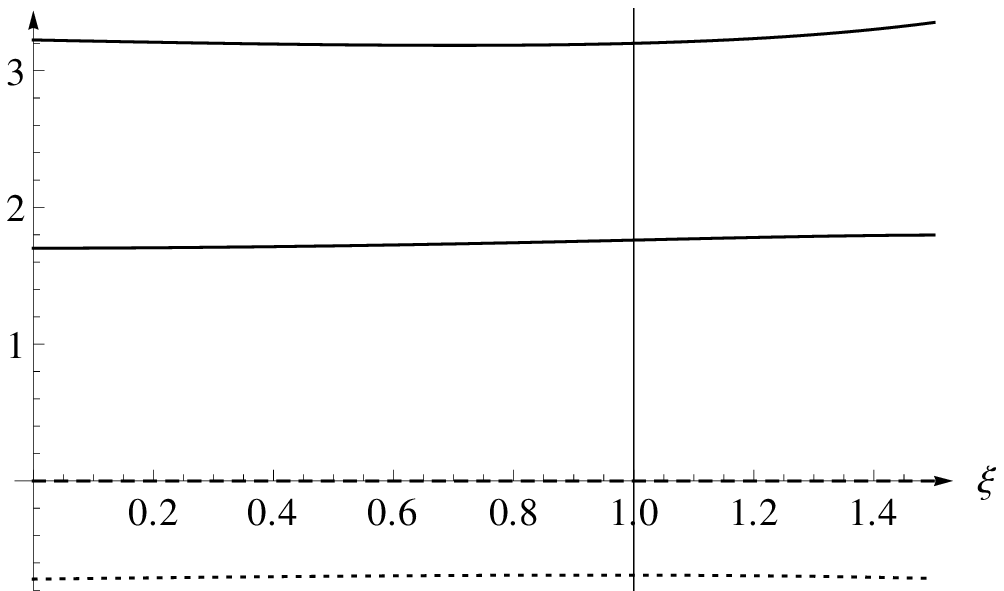}\label{Plot31a}}
\quad
\subfigure[$\mu=(\xi/4)^{1/4}$]{\includegraphics[width=0.45 
\linewidth]{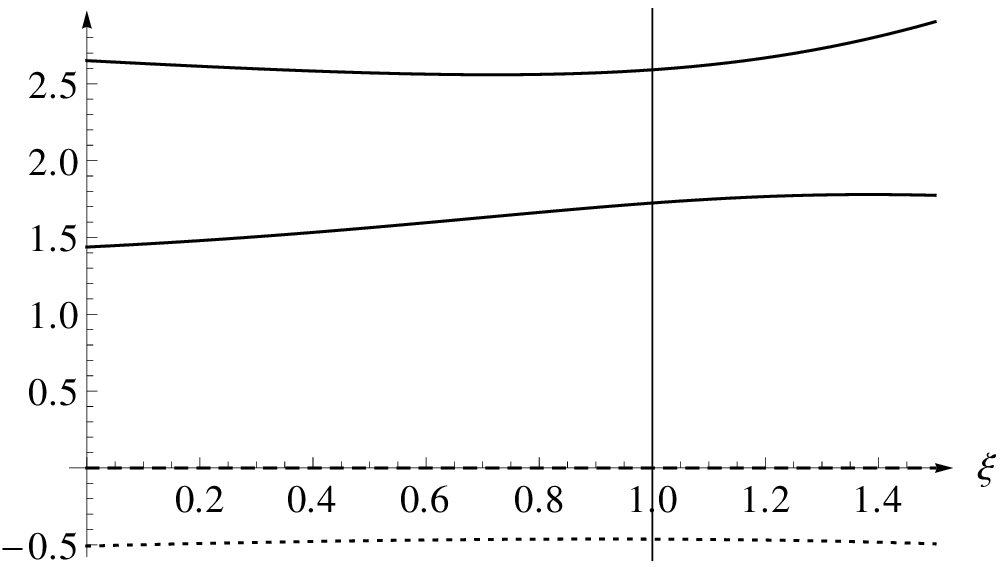}\label{Plot31b}}
\quad
\subfigure[$\mu=\sqrt{2}(\xi/4)^{1/4}$]{\includegraphics[width=0.45 
\linewidth]{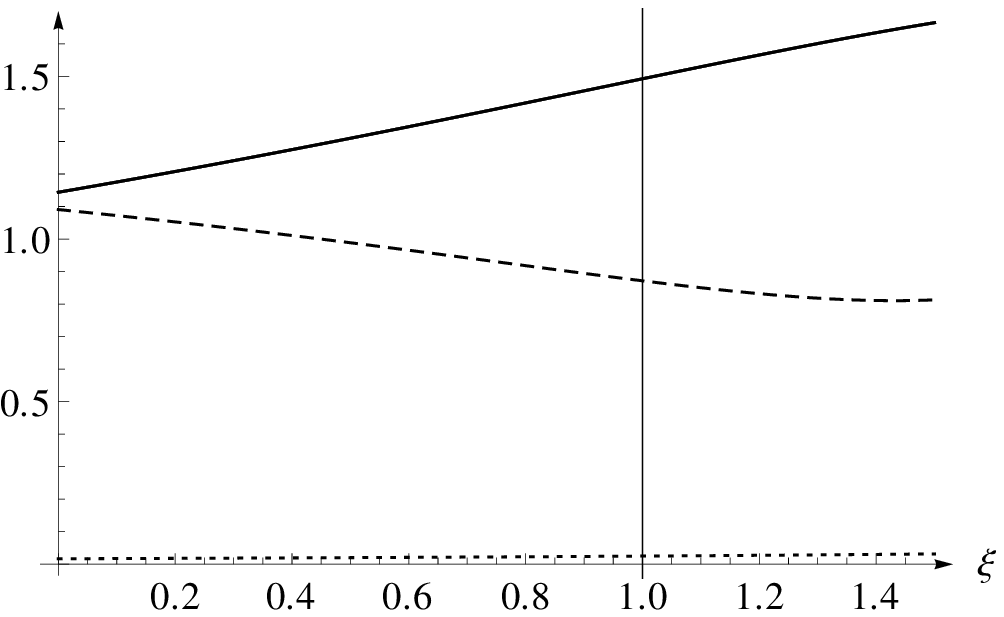}\label{Plot31c}}
\caption{Critical exponents and $g^\ast \lambda^\ast$ for different values of 
an adapted mass parameter $\mu$ as a function of $\xi$ ($\theta'$ solid, 
$\theta''$ dashed, $g^\ast \lambda^\ast$ dotted), calculated with the 
optimized cutoff.}\label{Plot31}
\end{figure}

\noindent{\bf (A) Adaptation scheme (i).} In Fig. \ref{Plot3} we have plotted 
the universal quantities (critical exponents and the product of the fixed 
point coordinates) for various values of the mass parameter $\mu$ as functions 
of $\xi$. As $\mu$ does not depend on $\xi$ in these examples, all of them 
correspond the adaptation scheme (i), although Fig. \ref{Plot3d} already shows 
typical characteristics of the large $\mu$ limit and can therefore also be 
seen as an example of scheme (iii). 

In all four cases we start with the values of the metric theory at $\xi=0$ and 
find for each of them a quite pronounced dependence on $\xi$. While for small 
$\mu$ as shown in Fig. \ref{Plot3a} the critical exponents turn from complex 
to real and $g^\ast \lambda^\ast$ turns negative as we move towards $\xi=1$, 
for large $\mu$ (Fig. \ref{Plot3d}) $g^\ast \lambda^\ast$ stays positive but 
the fixed point gets repulsive. Only in the region of $\mu\approx1$ the 
situation improves a little, as no quantity changes its sign in the interval 
of $\xi\in[0,1]$. However the quantities plotted are far from being constant 
with respect to $\xi$; furthermore if we compare the analogous results 
obtained with the family of $s$-dependent exponential cutoffs (as is done in 
the Appendix) we find that these results still show a substantial cutoff 
scheme dependence. Can we do better than this?

\noindent{\bf (B) Adaptation scheme (ii).} If we employ the optimally adapted 
cutoff (ii) instead (Fig. \ref{Plot31b}), the ${\sf O(}d)$ ghost contribution 
is now independent of $\xi$. Therefore $\xi=0$ does not correspond to the 
metric theory any more. In this case we find the universal quantities almost 
independent of $\xi$. 

Variants of this cutoff adaptation differing by a factor of $\sqrt{2}$ (Figs. 
\ref{Plot31a},\ref{Plot31c}) show that the universality can even be improved 
when choosing a smaller $\mu$. This effect, however, does not really improve 
the reliability of the flow.: In the limit $\mu\rightarrow0$ the constant 
${\sf O(}d)$ ghost contribution diverges and governs the RG flow, so that the 
effect of the physical field modes becomes negligible. Therefore it is evident 
that the $\xi$ dependence weakens when going to smaller values of $\mu$, but 
only at the cost of losing the physics content of the flow.

\noindent{\bf (C) Discussion.} The properties of the universal quantities 
calculated in this chapter show, that the influence of the ${\sf O(}d)$ ghosts 
on the fixed point properties is quite significant. While neglecting these 
contributions (adaptation scheme (iii)) leads to the implausible result that 
the fixed point changes its character and gets UV repulsive for some $\xi\in 
[0,1]$, the simple unadapted cutoff (scheme (i)) leads to universal quantities 
strongly dependent on $\xi$. 

Only the optimally adapted ghost cutoff (scheme (ii)) predicts relatively 
stable values for the universal quantities. These values indicate a fixed 
point at $\lambda^\ast<0$ with real critical exponents, that therefore may not 
be the one known from the metric theory. 

If this picture is correct, part of the $\xi$-dependence found in scheme (i) 
is clearly due to the fact, that in this scheme the quantities are forced to 
take on their metric values at $\xi=0$. This way we would have constructed an 
interpolation between theories of different universality class which obviously 
leads to a $\xi$-dependence of the ``universal quantities''.

Nevertheless, all results show a cutoff scheme, i.\,e. $R^{(0)}(\,\cdot\,)$\,- 
dependence that is more severe than in the metric case. It is analyzed further 
in the Appendix to which the reader might turn at this point. 

Apparently the truncation chosen is less reliable than the 
Einstein-Hilbert truncation of metric gravity, although it can be considered 
as its exact ``translation'' to the tetrad theory space. Together with the 
different FP properties this indicates that the quantum theories of metric and 
tetrad gravity (if both should turn out nonperturbatively renormalizable) are 
perhaps not similar to each other. For this reason it is 
crucial to use tetrads as fundamental field variables whenever an RG study of 
fermions coupled to gravity is performed even if only the pure gravity 
$\beta$-functions are investigated. Our results can be considered a warning 
that in a nonperturbative RG analysis the ${\sf O(}d)$ ghost sector cannot be 
ignored (as opposed to perturbation theory \cite{Woodard}). Seen in this light, 
the status of hybrid calculations which add fermionic contributions to metric 
QEG seems questionable.

\section{Summary and Conclusion}
In this paper we performed a first survey of the renormalization group flow on 
the ``tetrad only'' theory space $\mathcal{T}_{\text{tet}}=\{A[e^a_{\ 
\mu},\cdots]\}$. Its points are action functionals which, besides the 
indispensable background and ghost fields, depend on the vielbein $e^a_{\ 
\mu}$ only, and which are invariant under the semidirect product of spacetime 
diffeomorphisms and local Lorentz transformations. Contrary to the 
Einstein-Cartan theory the spin connection is not an independent field, but 
rather is identified with the Levi-Civita connection implied by $e^a_{\ \mu}$. 
This excludes the possibility of field configurations with torsion. We 
truncated $\mathcal{T}_{\text{tet}}$ so as to consist of a running 
Einstein-Hilbert term, along with the classical gauge fixing and ghost terms. 
As a result, the only difference in comparison to Quantum Einstein Gravity 
(QEG) in the Einstein-Hilbert truncation \cite{mr, oliver, frank1} is the use 
of $e^a_{\ \mu}$ rather than the metric $g_{\mu\nu}$ as the fundamental field 
variable and the larger group of gauge transformations ${\sf Diff}({\cal 
M})\ltimes{\sf O}(d)_{\text{loc}}$ replacing ${\sf Diff}({\cal M})$. In the 
present treatment the latter has the status of a composite field: $g_{\mu\nu}= 
\eta_{ab} e^a_{\ \mu} e^b_{\nu}$. Our main tool was the gravitational average 
action on $\mathcal{T}_{\text{tet}}$, and in particular the FRGE which governs 
its scale dependence. Since this framework is not covariant under field 
reparametrizations, and since the respective groups ${\bf G}$ are different, 
the RG flow on $\mathcal{T}_{\text{tet}}$ is likely to differ from the one of 
QEG, even at the exact level.

This expectation was confirmed by our explicit calculation. The details of the 
Einstein-Hilbert flows with $e^a_{\ \mu}$ and $g_{\mu\nu}$, respectively, as 
fundamental fields are indeed different in a significant way. However, their 
gross topological features are still similar, nevertheless. In particular we 
found on $\mathcal{T}_{\text{tet}}$ one, and only one, non-Gaussian fixed 
point, exactly as in QEG. Provided it is not a truncation artifact it seems 
suitable for taking a nonperturbative continuum limit there, thus defining an 
asymptotically safe field theory.

To assess the reliability of the approximations made we investigated the 
dependence of ``universal'' quantities such as critical exponents and the 
product $g^\ast \lambda^\ast$ on the cutoff shape function $R^{(0)}(\cdot)$ 
and the mass parameter $\mu$. The latter had to be introduced in order to give 
equal canonical dimensions to diffeomorphism and ${\sf O(}d)_{\text{loc}}$ 
ghosts. While in a more complete truncation it would be treated as a running 
coupling with its own $\beta$-function, we neglected its running in the 
present investigation. The upshot of the analysis is that the very existence 
of the NGFP indeed seems to be a universal feature, in the sense that it 
exists for all admissible cutoffs and values of $\mu$. 

However, further details, even the critical exponents show a variability with 
$R^{(0)}(\cdot)$ and $\mu$ which is significantly larger than in QEG with the 
same truncation. (In particular, in QEG there is no analog of the parameter 
$\mu$.) Thus we must conclude that, using the same type of flow equation and 
the same (Einstein-Hilbert) truncation, the use of the vielbein instead of the 
metric leads to a less robust RG flow.

Can we understand on general grounds why the flow of the metric theory might 
have better robustness properties than the one based upon the tetrad? A 
possible explanation is as follows.

The running couplings parametrizing a general functional $\Gamma_k$ are, per 
se, not measurable quantities, that is, typical observables are complicated 
combinations of these couplings, and in forming these combinations the scheme 
dependence which the individual couplings have (even in the exact theory!) 
cancels among them. Consider now a theory space whose actions are constrained 
to be invariant under a group ${\bf G}$ of gauge transformations which we make 
larger and larger. As a result, more and more excitations carried by the 
(fixed) set of fields considered are declared ``unphysical'' gauge modes. 
Nevertheless all those modes continue to contribute to the supertrace in the 
FRGE, but are counteracted by an increasing number of ghosts needed to 
gauge-fix ${\bf G}$. Loosely speaking, increasing the size of ${\bf G}$ 
reduces the amount of ``physical'' (in the sense of ``non-gauge'') or 
``observable'' contents encoded in the running couplings. In diagrammatic 
terms, the ratio of physical excitations relative to gauge excitations gets 
smaller when ${\bf G}$ grows. However, since those features of the RG flow 
which are due to the gauge modes have no reason to be scheme independent, one 
can expect that the larger is ${\bf G}$ the more scheme dependent is even the 
exact RG flow.

While, in $d=4$, metric gravity has 4 gauge parameters per spacetime point 
related to the diffeomorphisms, this number increases to 4+6 in tetrad gravity 
since local Lorentz invariance is demanded in addition. If we assume that both 
theories have the same number of physical degrees of freedom, it is clear that 
tetrad gravity has a smaller ratio of physical to unphysical field modes, and 
this might explain to some extent why its RG flow has the more delicate scheme 
dependence we observed.

To close with, let us come back to the issues raised in the Introduction which 
motivated the present work.

\noindent{\bf (A)} In ref. \cite{e-omega} the RG flow was computed for a 3D 
truncation of $\mathcal{T}_{\text{EC}}=\{A[e,\omega,\cdots]\}$ which has the 
same gauge transformations ${\sf Diff(\mathcal{M})\ltimes O(}d)_{\text{loc}}$ 
as $\mathcal{T}_{\text{tet}}=\{A[e,\cdots]\}$, but treats the spin connection 
as an independent field. There, too, the very existence of a NGFP is a robust 
feature which obtains for all cutoff and gauge choices, but the quantitative 
details are more scheme dependent then we are used to from QEG. In this 
respect the results of \cite{e-omega} are very reminiscent of what we saw in 
the present paper. In \cite{e-omega} both the truncated action and the 
fundamental variables are different from QEG (``Holst'' instead of 
``Einstein-Hilbert'', and $(e,\omega)$ instead of the metric). In the light of 
our present results we can say that the hitherto unexplained relatively strong 
scheme dependence seen in \cite{e-omega} could be entirely due to the 
different variables used and the related larger group of gauge 
transformations; even though the running actions used were quite different in 
the two cases (first vs. second order in derivatives, etc.) this is not 
necessarily the cause for the observed differences.

\noindent{\bf (B)} In the literature \cite{Eichhorn-Gies, 
vacca,percacci-perini} ``hybrid'' calculations were proposed in order to avoid 
re-calculating parts of the $\beta$-functions for the gravitational couplings 
in presence of fermionic matter. The idea is to use the tetrad formalism only 
when it comes to evaluating fermion loops, but to keep the metric as the 
fundamental variable for the gravity loops. While this can be legitimate in 
perturbation theory \cite{Woodard}, the present investigation revealed that 
the quantitative details of the flow of Newton's constant and the cosmological 
constant are significantly different in the metric and the vielbein formalism. 
Hence, adding the fermionic loops to the ``old'' metric $\beta$-functions does 
not seem a consistent procedure, even within the limited scope of a 
truncation. Thus we must conclude that one should refrain from such hybrid 
calculations when one aims at quantitative results.

\noindent{\bf (C)} In the symmetric vielbein gauge the ${\sf 
O(}d)_{\text{loc}}$ ghosts are non-propagating. It was therefore argued, in 
perturbation theory, that they simply may be ignored in practical calculations 
\cite{Woodard}. As we saw quite explicitly, the same is not true in the FRGE 
framework. The Lorentz ghosts do have a considerable impact on the RG flow we 
found, and moreover the arguments put forward in \cite{Woodard} are easily 
seen not to carry over to $\Gamma_k$ at $k>0$.

Several semi-quantitative calculations \cite{wet-shap, fine} have shown that 
the Standard Model coupled to asymptotically safe gravity may lead to a theory 
with enhanced predictivity, that is some of the perturbatively undetermined 
parameters of the Standard Model (like the mass of the Higgs boson 
\cite{wet-shap} or the fine-structure constant \cite{fine}) can be calculated 
in the coupled gravity + matter theory. The present paper has identified 
possible pitfalls in RG calculations of such coupled systems of gravity and 
fermions and indicated how to avoid them. This paves the way for a fully 
quantitative treatment of the considerations in refs. \cite{wet-shap} and 
\cite{fine}. Even though this might require more work than thought before, the 
chance to {\it compute} the Higgs mass or the fine-structure constant clearly 
will be worth the effort.

\noindent{\bf Acknowledgments:} We are grateful to R. Percacci for discussions and 
a careful reading of this paper.
\newpage
\appendix
\vspace{1.5 cm}
\noindent {\bf \Large Appendix}
\vspace{-0.5cm}
\section{The family of exponential cutoffs}
In this appendix we study the cutoff scheme dependence of $g^\ast 
\lambda^\ast$ and the critical exponents within the family of exponential 
cutoff functions $R^{(0)}(z)=sz/(e^{sz}-1)$ for shape parameters $s$ ranging 
from 2 to 20 in more detail.

\begin{figure}[phtb]
\centering
\subfigure[$s=2$]{\includegraphics[width=0.45 
\linewidth]{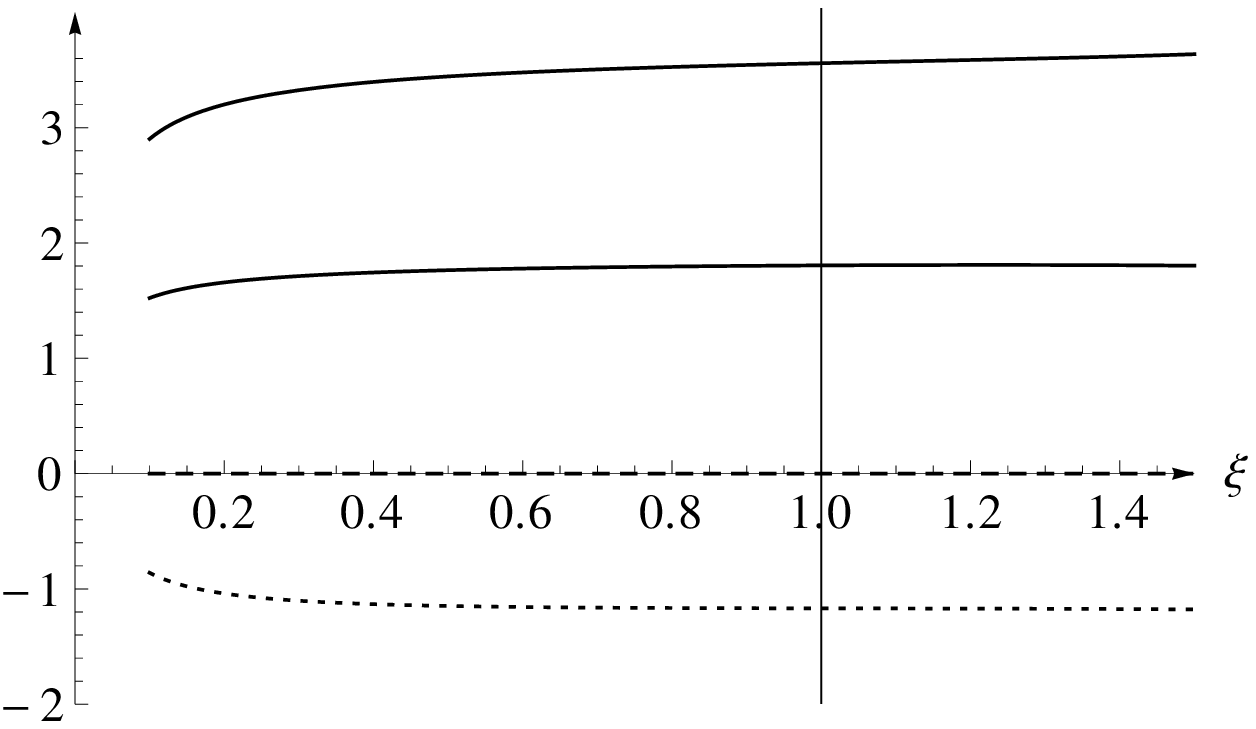}\label{Plot5a}}\quad
\subfigure[$s=3$]{\includegraphics[width=0.45 
\linewidth]{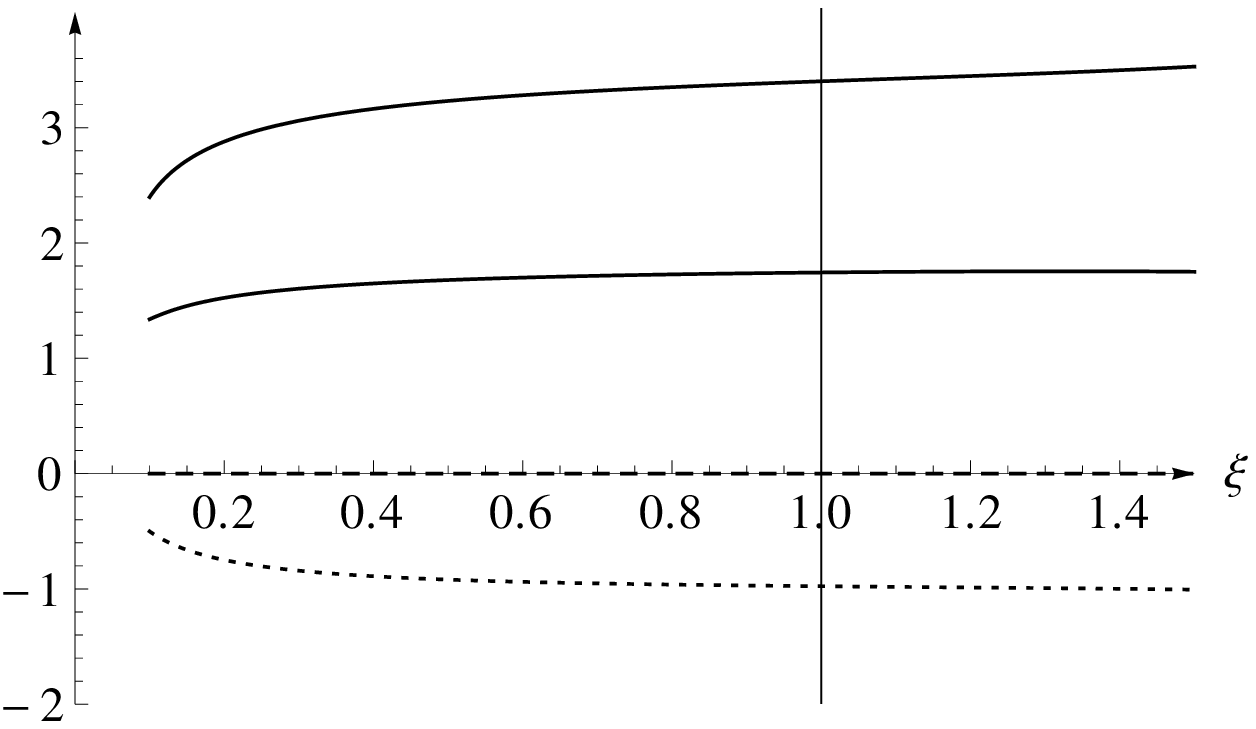}\label{Plot5b}}\quad
\subfigure[$s=4$]{\includegraphics[width=0.45 
\linewidth]{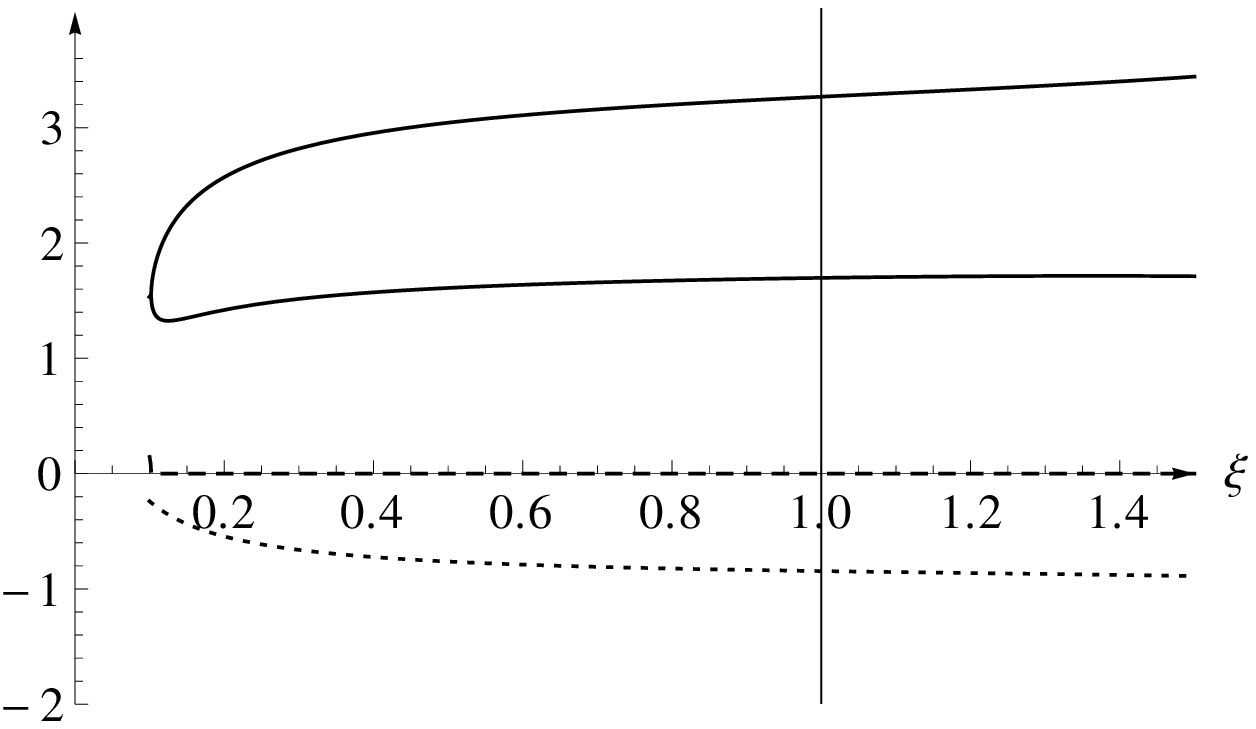}\label{Plot5c}}\quad
\subfigure[$s=7$]{\includegraphics[width=0.45 
\linewidth]{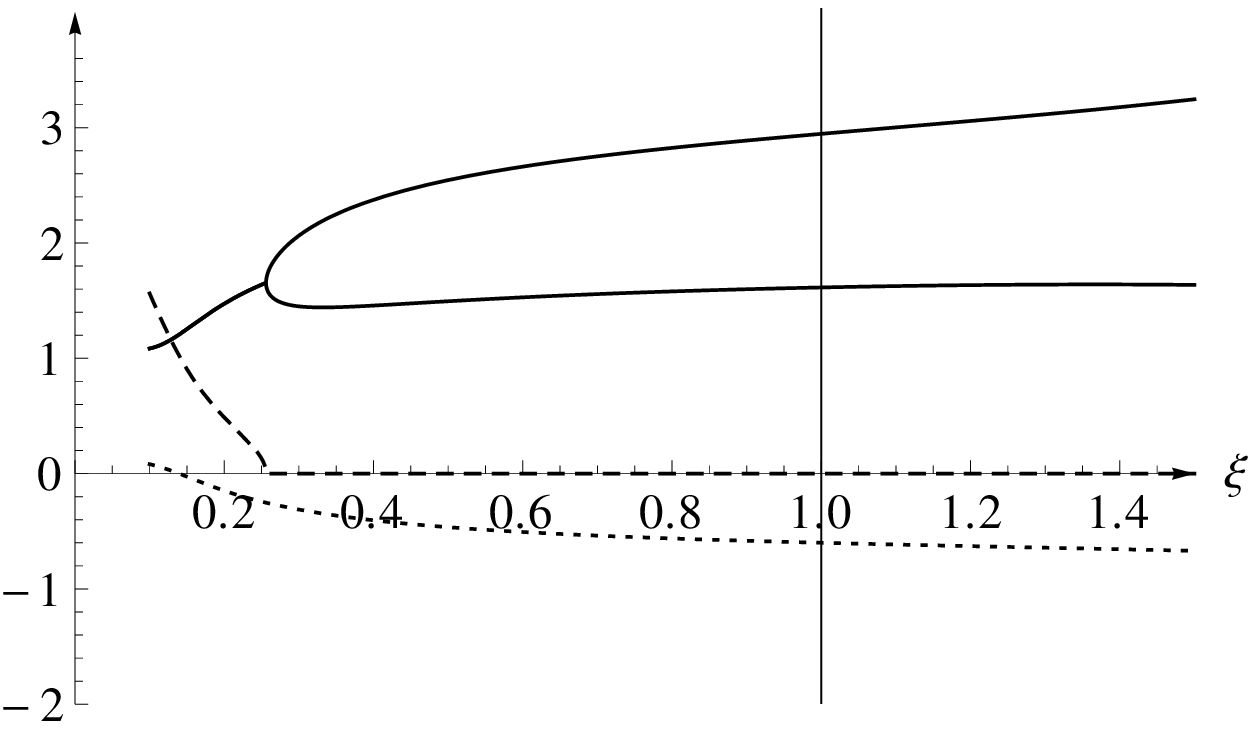}\label{Plot5d}}\\
\subfigure[$s=10$]{\includegraphics[width=0.45 
\linewidth]{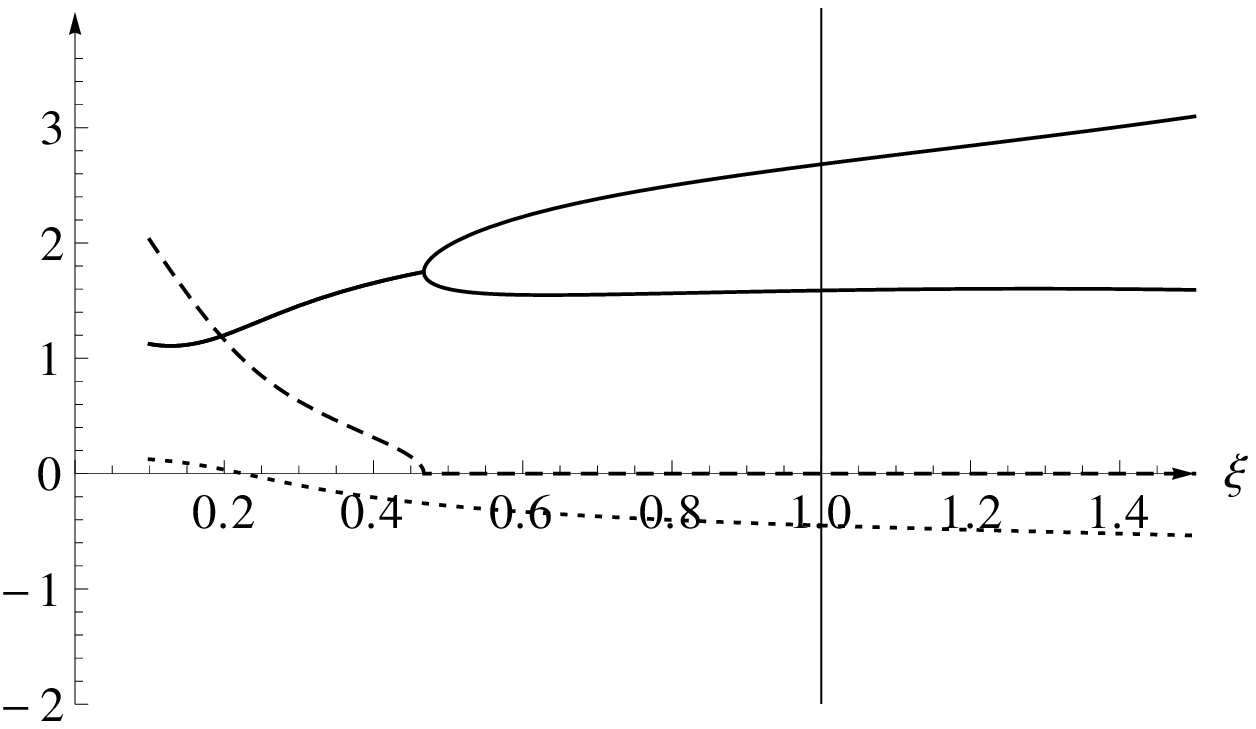}\label{Plot5e}}\quad
\subfigure[$s=20$]{\includegraphics[width=0.45 
\linewidth]{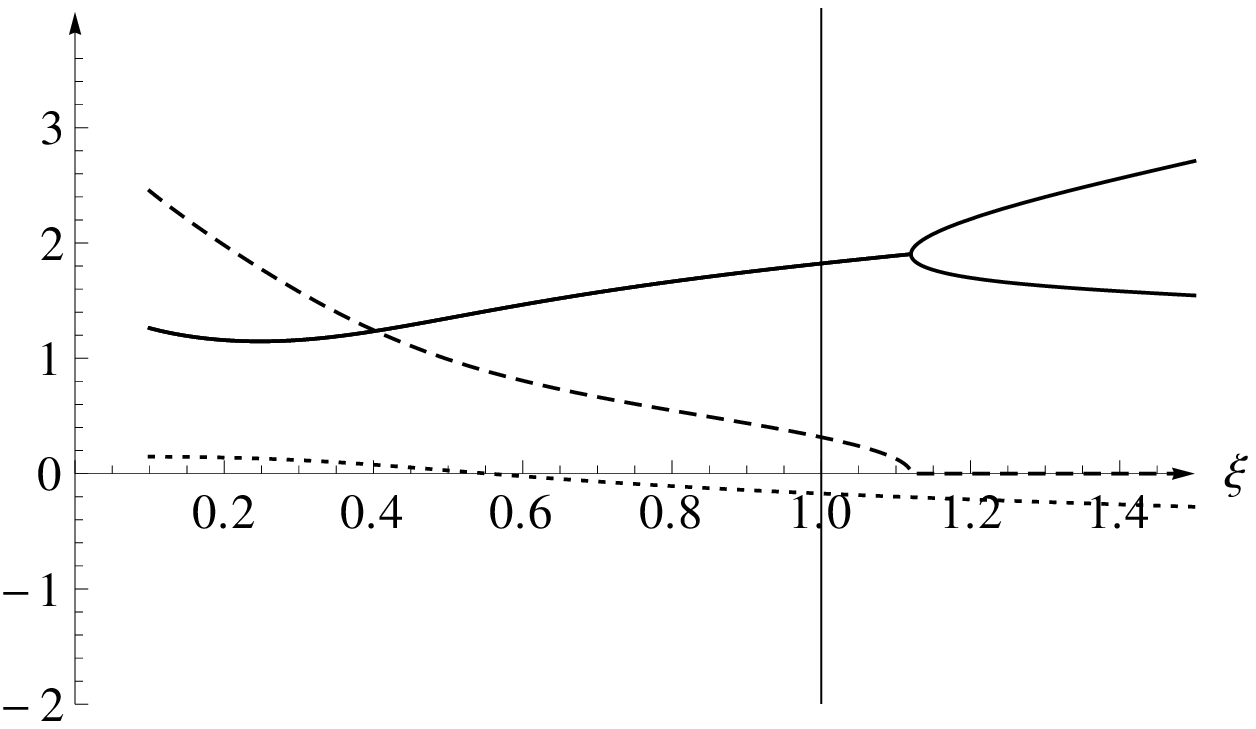}\label{Plot5f}}\caption{Critical 
exponents and $g^\ast \lambda^\ast$ for different shape parameters $s$ 
depending on $\xi$ with mass parameter $\mu=0.5$ ($\theta'$ solid, $\theta''$ 
dashed, $g^\ast \lambda^\ast$ dotted).}\label{Plot5}
\end{figure}

\begin{figure}[phtb]
\centering
\subfigure[$s=2$]{\includegraphics[width=0.45 
\linewidth]{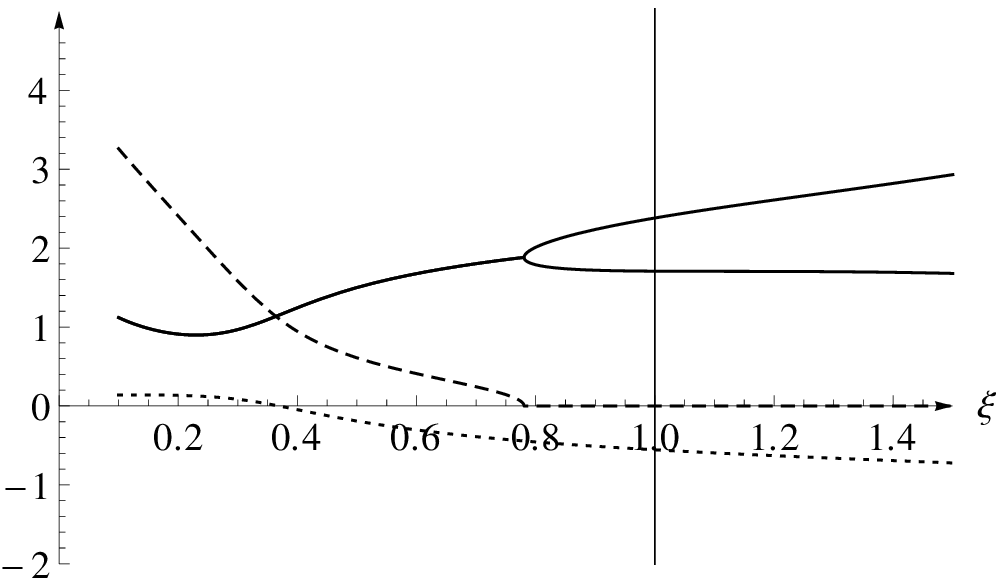}\label{Plot6a}}\quad
\subfigure[$s=3$]{\includegraphics[width=0.45 
\linewidth]{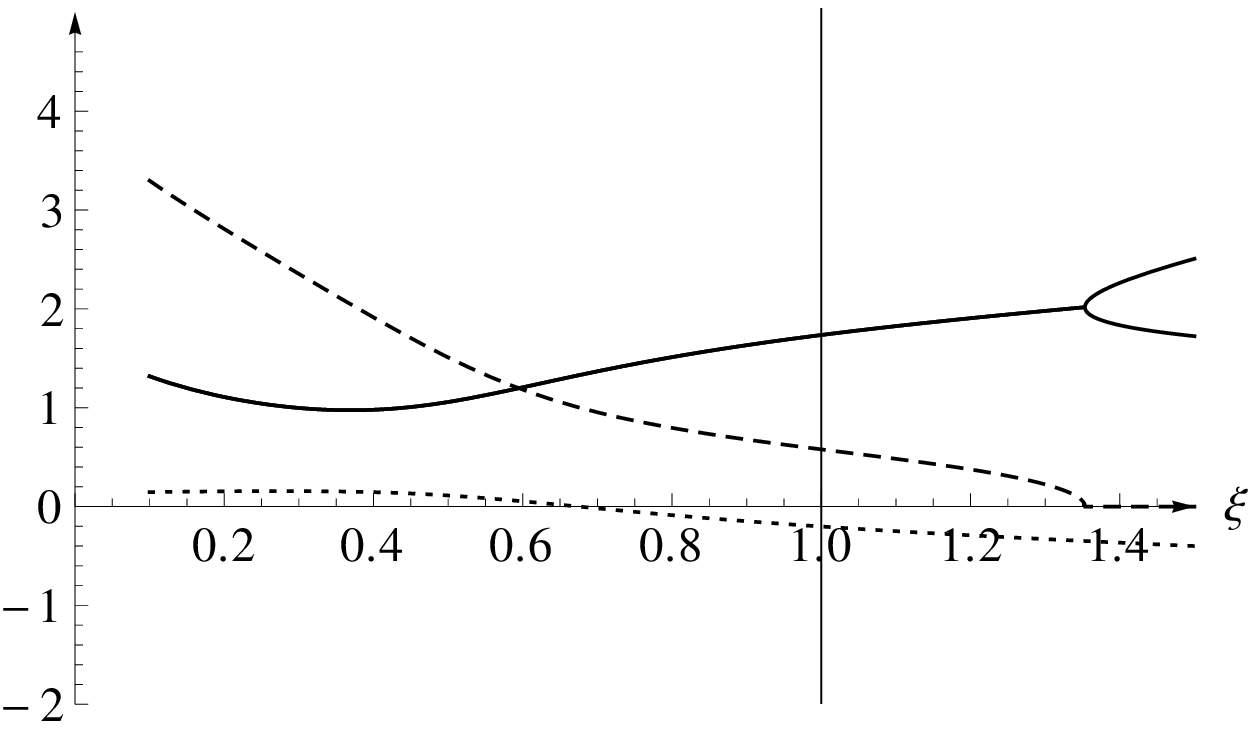}\label{Plot6b}}\quad
\subfigure[$s=5$]{\includegraphics[width=0.45 
\linewidth]{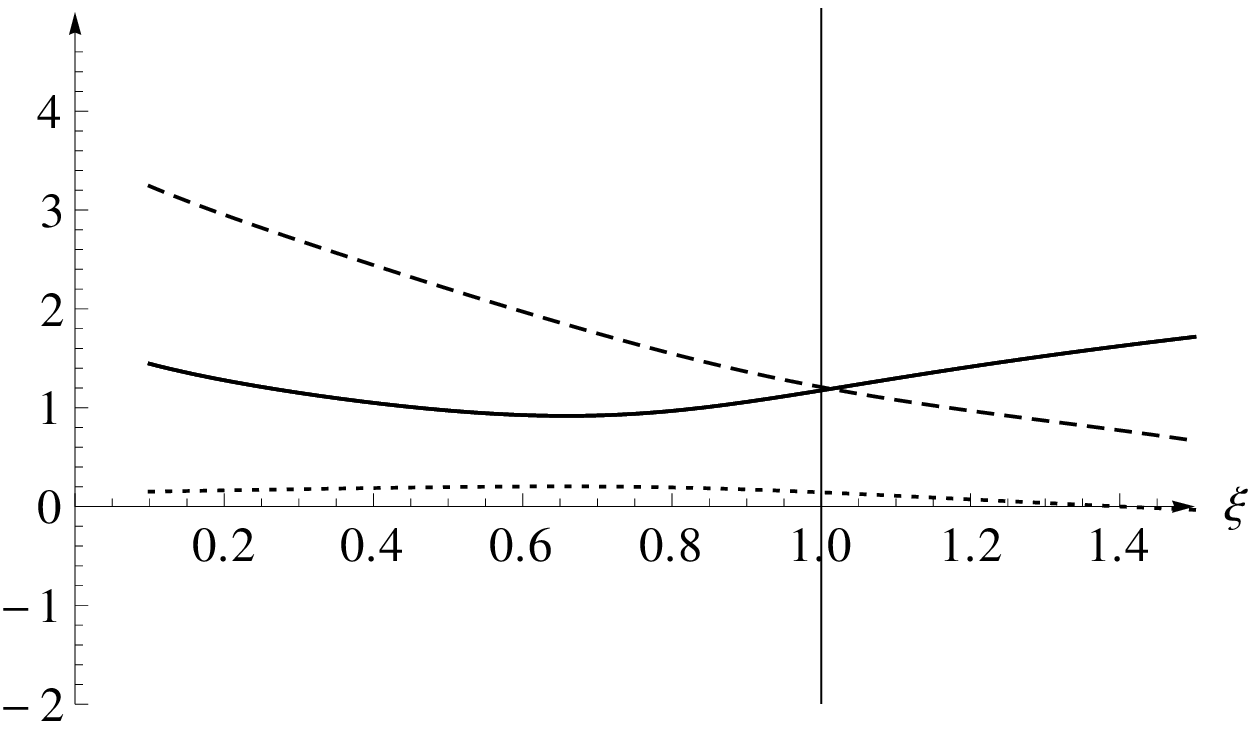}\label{Plot6c}}\quad
\subfigure[$s=7$]{\includegraphics[width=0.45 
\linewidth]{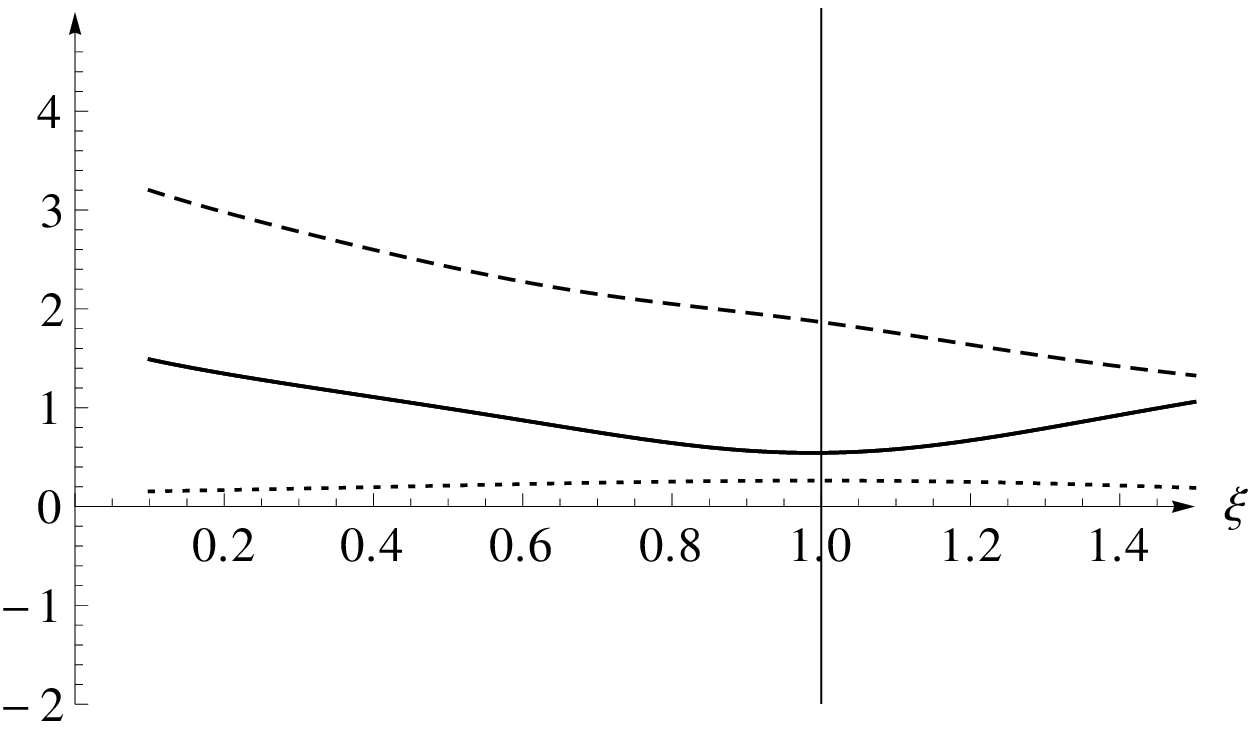}\label{Plot6d}}\\
\subfigure[$s=10$]{\includegraphics[width=0.45 
\linewidth]{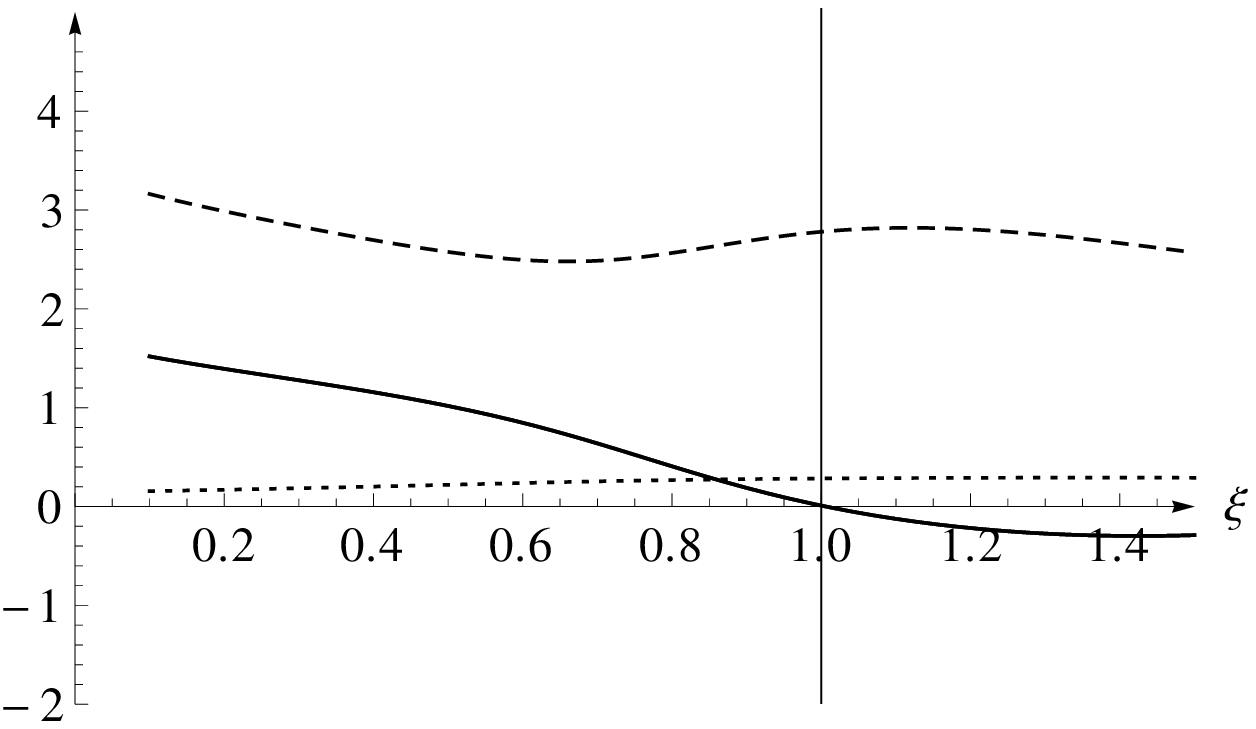}\label{Plot6e}}\quad
\subfigure[$s=20$]{\includegraphics[width=0.45 
\linewidth]{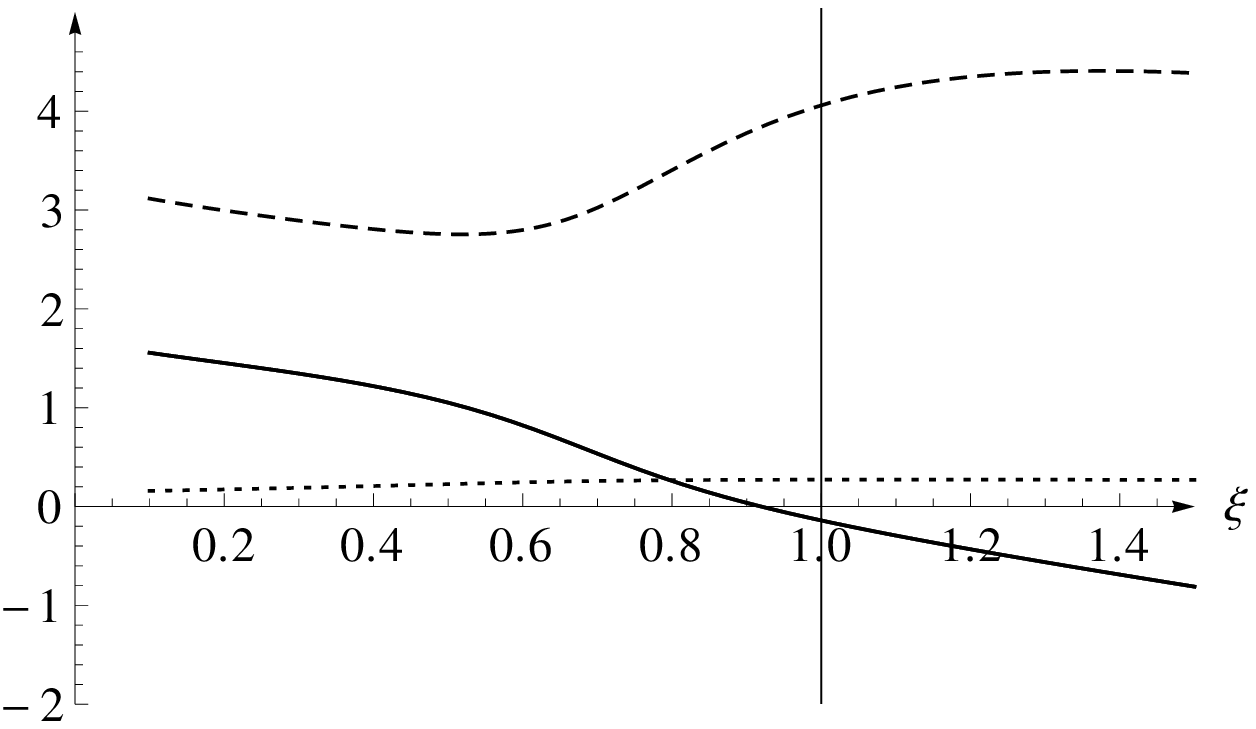}\label{Plot6f}}\caption{Critical 
exponents and $g^\ast \lambda^\ast$ for different shape parameters $s$ 
depending on $\xi$ with mass parameter $\mu=1$ ($\theta'$ solid, $\theta''$ 
dashed, $g^\ast \lambda^\ast$ dotted).}\label{Plot6}
\end{figure}

\begin{figure}[phtb]
\centering
\subfigure[$s=2$]{\includegraphics[width=0.45 
\linewidth]{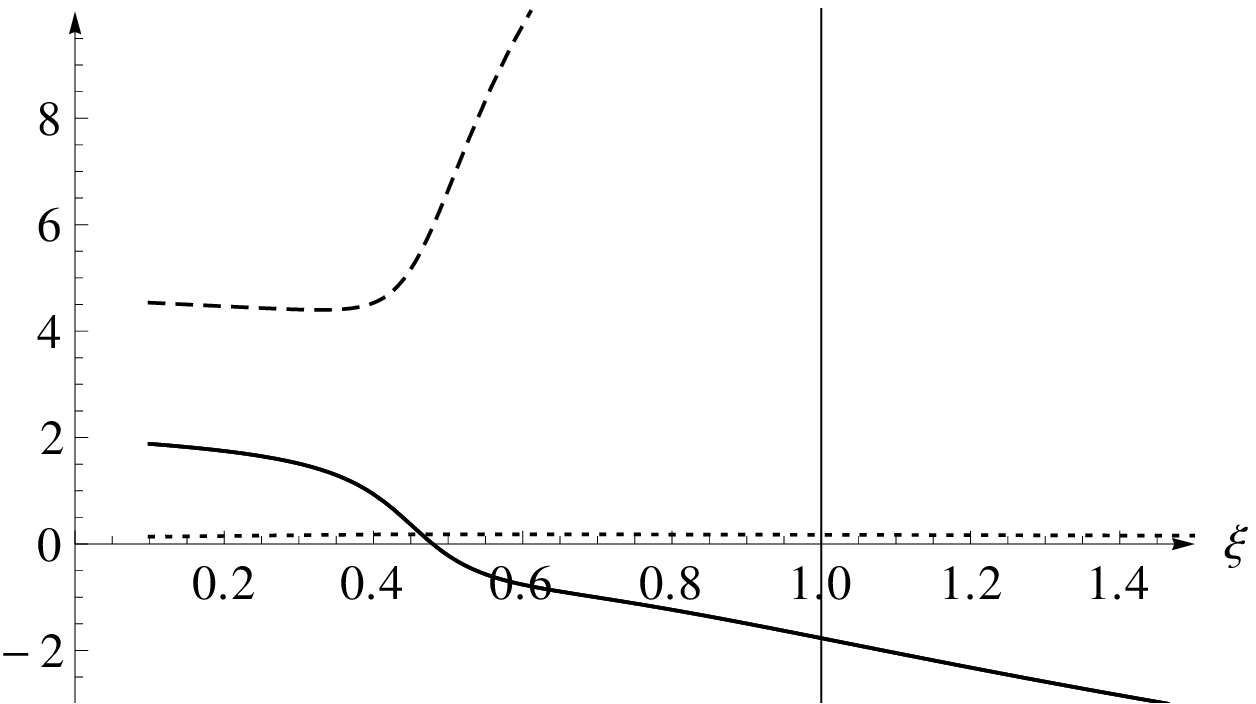}\label{Plot7a}}\quad
\subfigure[$s=3$]{\includegraphics[width=0.45 
\linewidth]{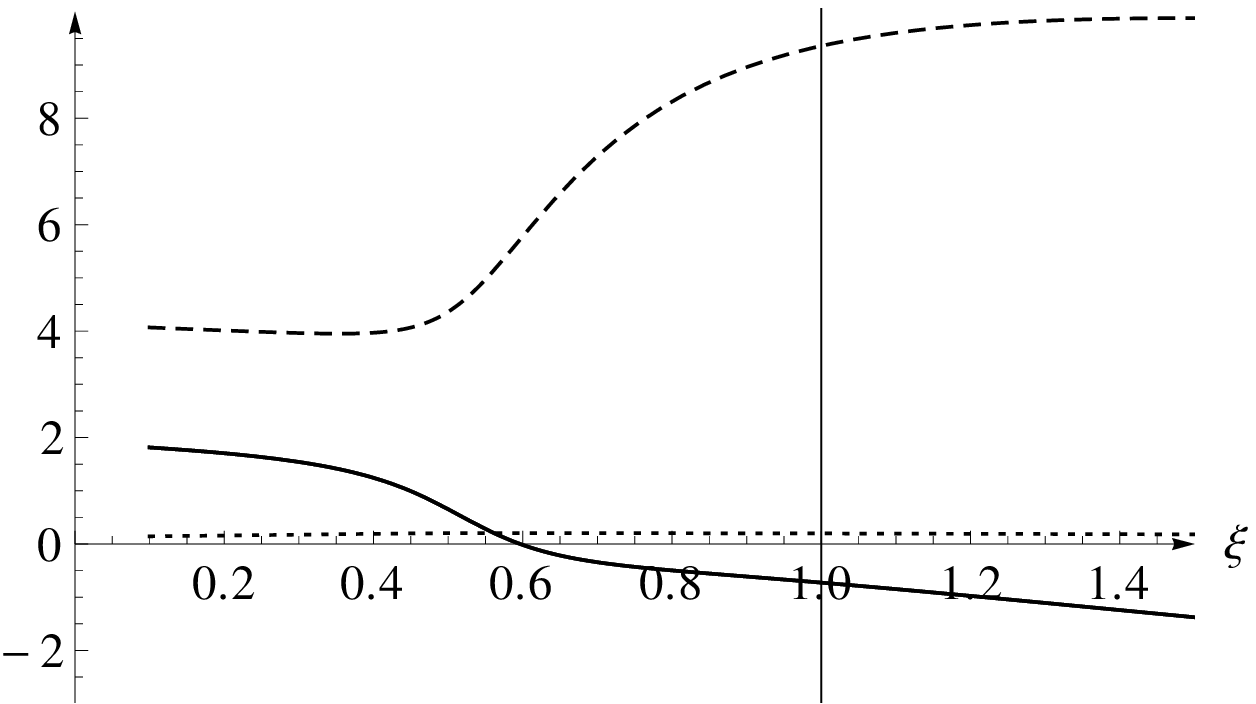}\label{Plot7b}}\quad
\subfigure[$s=5$]{\includegraphics[width=0.45 
\linewidth]{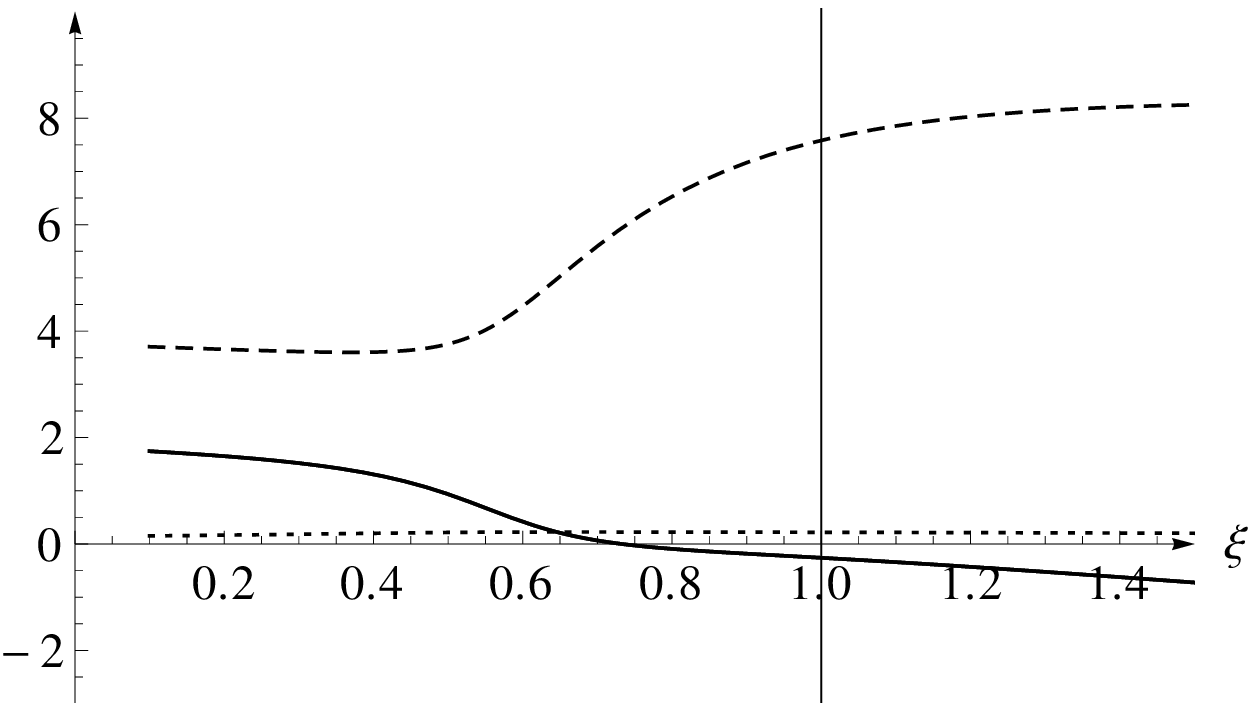}\label{Plot7c}}\quad
\subfigure[$s=7$]{\includegraphics[width=0.45 
\linewidth]{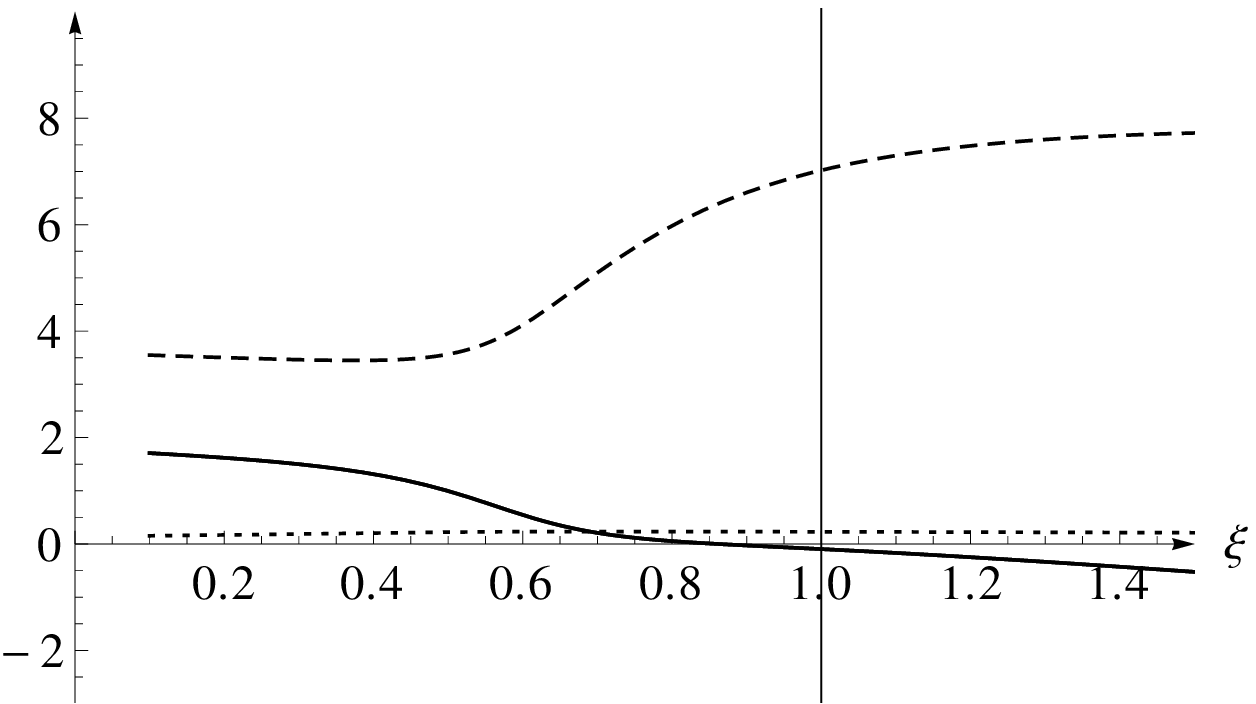}\label{Plot7d}}\\
\subfigure[$s=10$]{\includegraphics[width=0.45 
\linewidth]{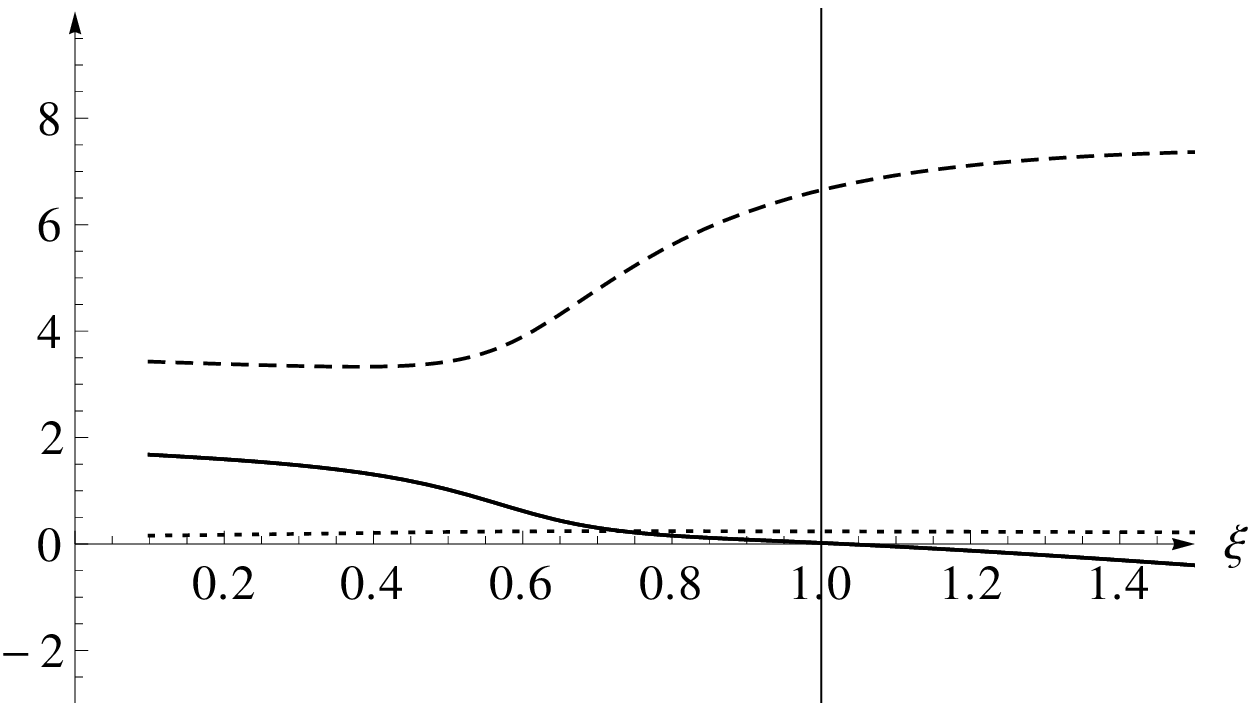}\label{Plot7e}}\quad
\subfigure[$s=20$]{\includegraphics[width=0.45 
\linewidth]{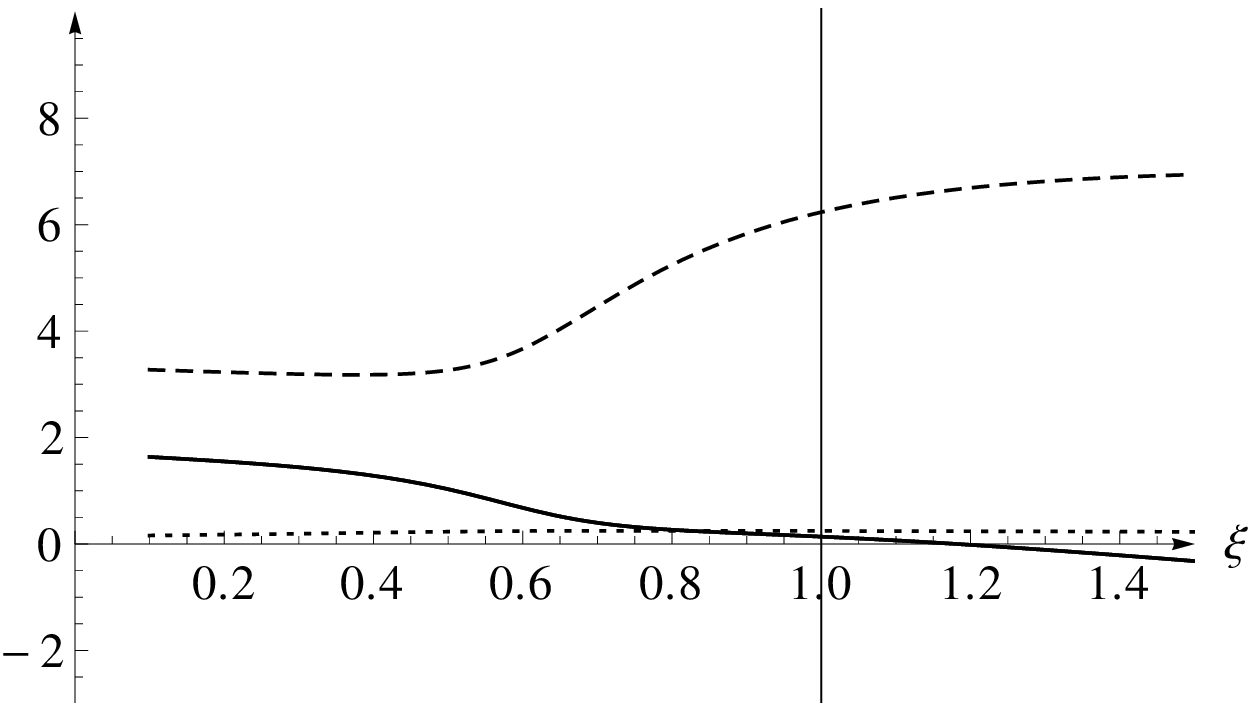}\label{Plot7f}}\caption{Critical 
exponents and $g^\ast \lambda^\ast$ for different shape parameters $s$ 
depending on $\xi$ with mass parameter $\mu=5$ ($\theta'$ solid, $\theta''$ 
dashed, $g^\ast \lambda^\ast$ dotted).}\label{Plot7}
\end{figure}

The first set of figures, Figs. \ref{Plot5} to \ref{Plot7}, corresponds to the 
application of ghost adaptation scheme (i), with each figure representing a 
different choice of $\mu$ ($\mu=0.5,\ 1,\ 5$ in Figs. \ref{Plot5}, 
\ref{Plot6}, \ref{Plot7}, respectively). Fig. \ref{Plot7} where $\mu$ is 
already rather large can also be seen as representing ghost adaptation scheme 
(iii). Each of the figures contains a series of plots ordered from small to 
large shape parameters employed in the exponential cutoff function. 

In a good approximation of the exact flow we would expect the plots to show 
only little $\xi$-dependence of the universal quantities, resulting in 
horizontal lines, as well as only small variations of the same picture for 
different shape parameters, i.\,e. almost equal plots within each of the 
figures. The dependence on $\mu$, on the other hand, could be more pronounced, 
as it should be seen as an additional coupling set to a fixed value.

We see, however, that there is a severe dependence on the parameter $\xi$ in 
all three figures; it leads to a change of sign of $g^\ast\lambda^\ast$, 
changing critical exponents from complex to real, and even a change of 
character of the fixed point from UV attractive to repulsive. Although we 
already made similar observations for the optimized cutoff function (cf. Fig. 
\ref{Plot3}), only the exponential cutoff functions reveal the full degree of 
scheme dependence in these results: While for the optimized cutoff we were 
able to choose a value of $\mu\approx1$ such that none of the above 
problematic changes occurred in the interval $\xi \in [0,1]$ (cf. Fig. 
\ref{Plot3} b, c), we now find that {\it changing $s$ has an effect similar to 
choosing a different $\mu$}. For that reason we find qualitatively the same 
plots (Fig. \ref{Plot3} a, b, c, d) obtained for the optimized cutoff function 
and distinguished choices of $\mu$ all within the family of exponential cutoff 
functions for the same value of $\mu=1$ (Fig. \ref{Plot6} a, c, d, f).

Only for large $\mu$ the variation of the plots within Fig. \ref{Plot7} is 
relatively weak. But here we find a large $\xi$-dependence of the critical 
exponents leading to a change of character of the fixed point in all plots, as 
we already found for the optimized cutoff in the same limit.

Taken together these observations show, that both adaptation schemes (i) and 
(iii) lead to severely scheme dependent results, that make it almost 
impossible draw any universally valid conclusion besides the existence of a 
NGFP.

\begin{figure}[phtb]
\centering
\subfigure[$s=2$]{\includegraphics[width=0.45 
\linewidth]{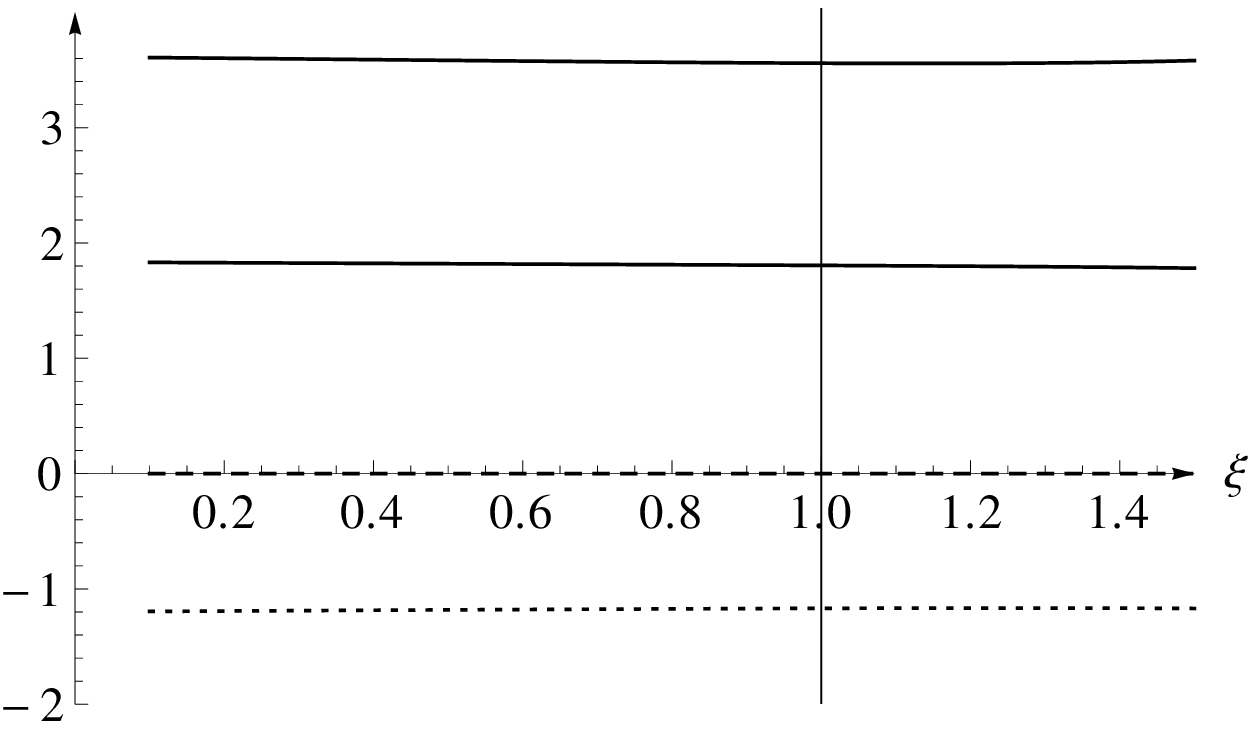}\label{Plot8a}}\quad
\subfigure[$s=3$]{\includegraphics[width=0.45 
\linewidth]{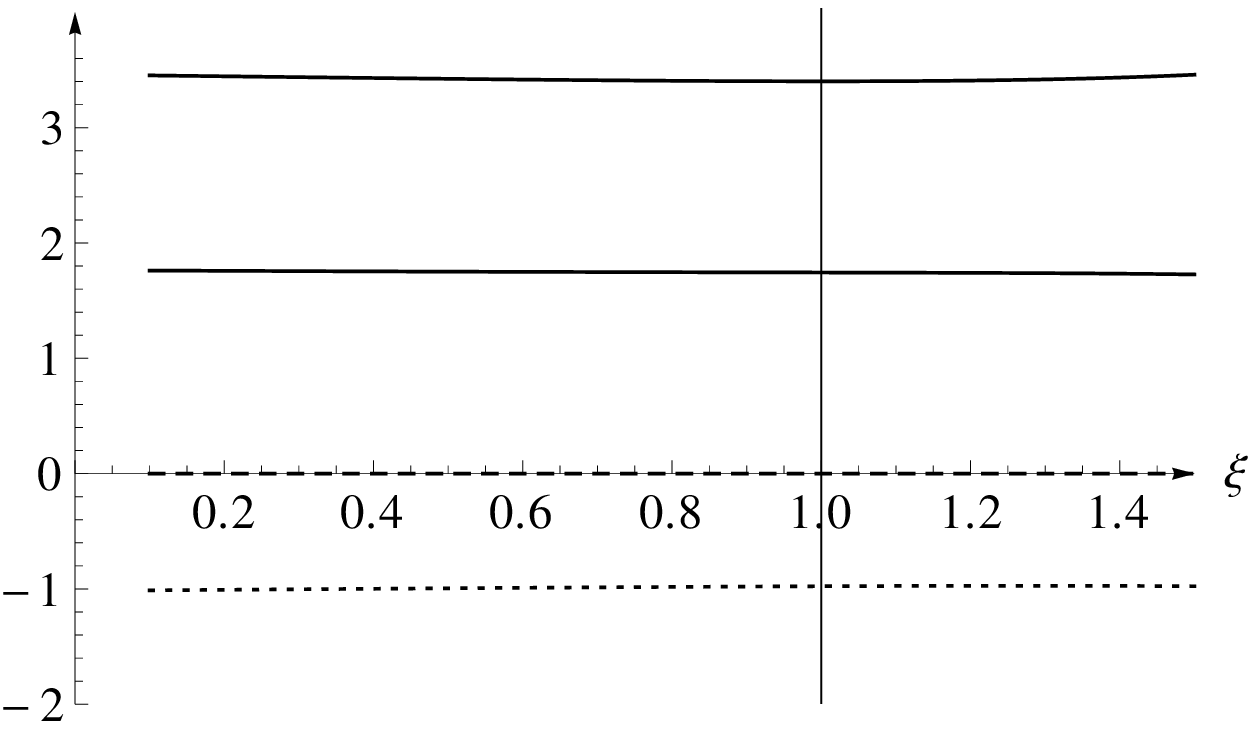}\label{Plot8b}}\quad
\subfigure[$s=4$]{\includegraphics[width=0.45 
\linewidth]{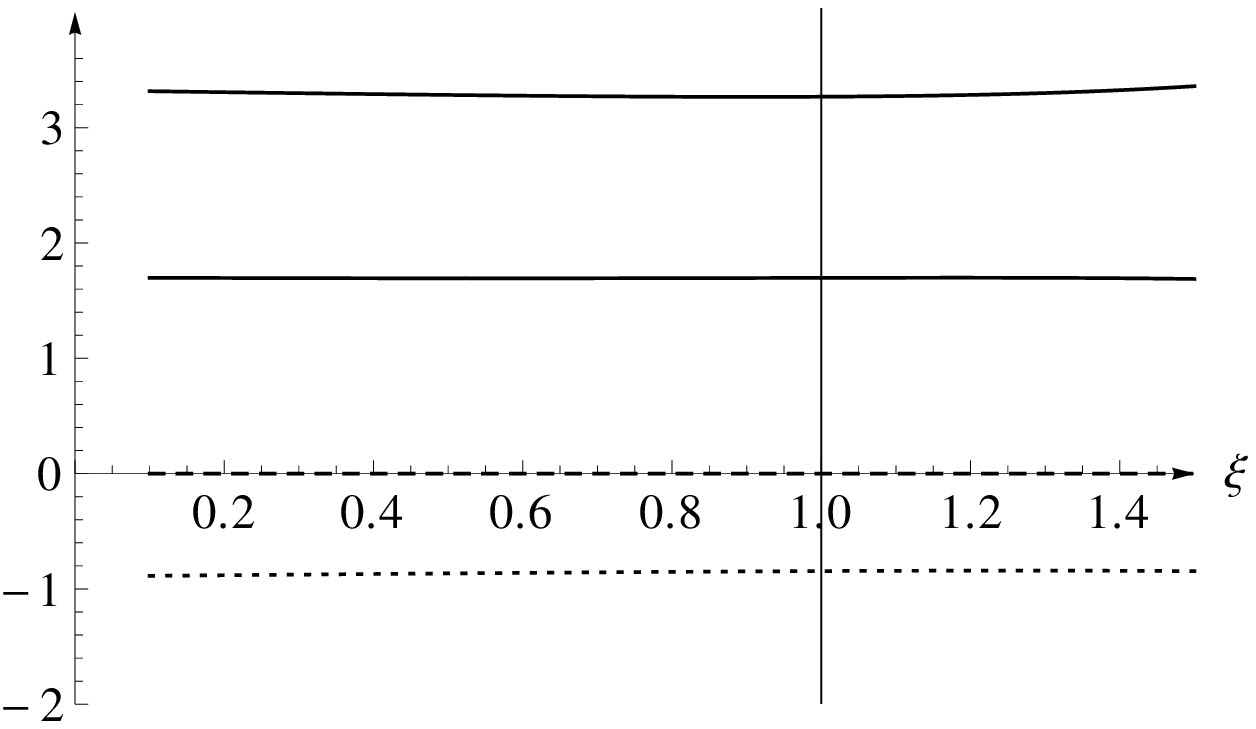}\label{Plot8c}}\quad
\subfigure[$s=5$]{\includegraphics[width=0.45 
\linewidth]{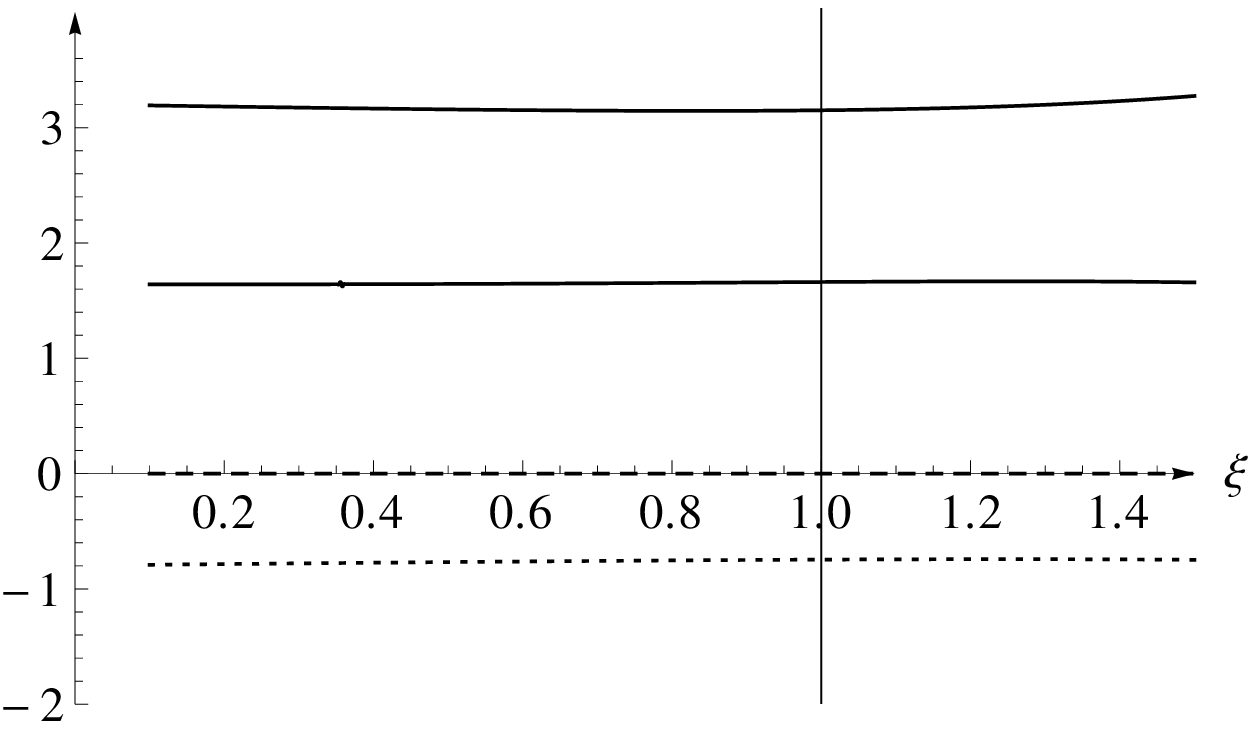}\label{Plot8d}}\\
\subfigure[$s=10$]{\includegraphics[width=0.45 
\linewidth]{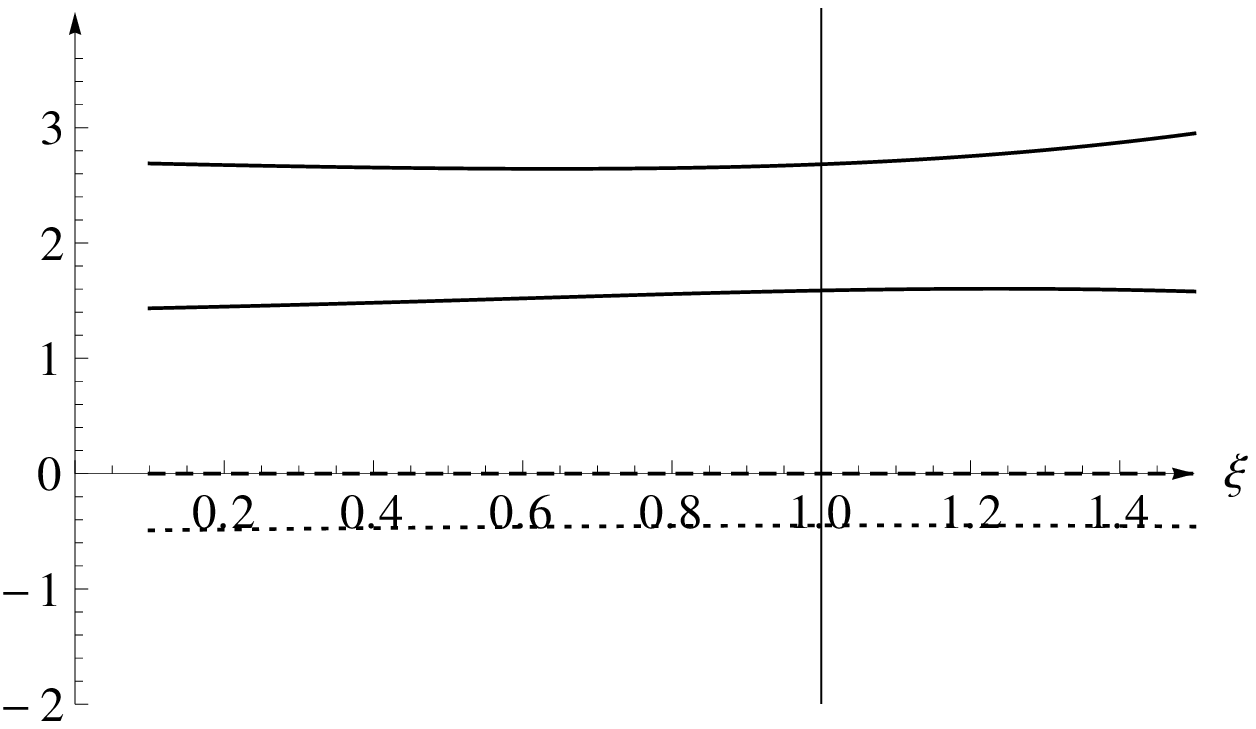}\label{Plot8e}}\quad
\subfigure[$s=20$]{\includegraphics[width=0.45 
\linewidth]{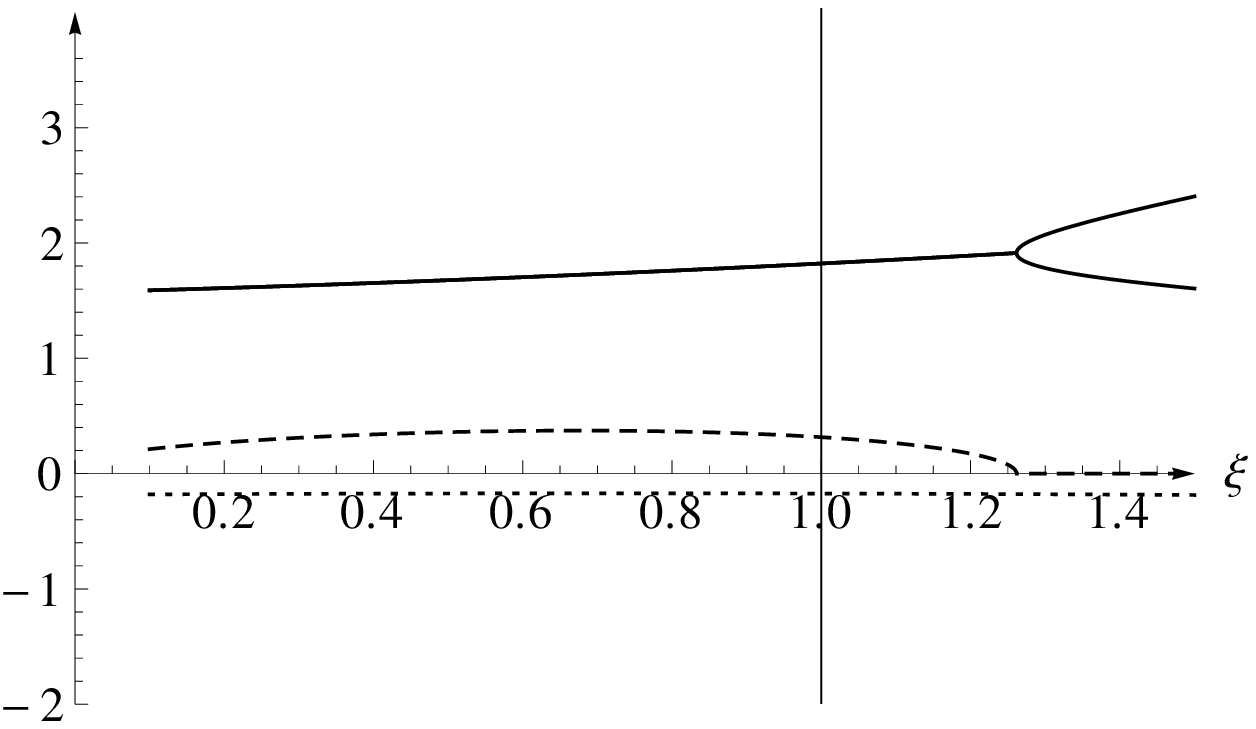}\label{Plot8f}}\caption{Critical 
exponents and $g^\ast \lambda^\ast$ for different shape parameters $s$ 
depending on $\xi$ with an adapted mass parameter $\mu=(\xi/4)^{1/4}/\sqrt{2}$ 
($\theta'$ solid, $\theta''$ dashed, $g^\ast \lambda^\ast$ 
dotted).}\label{Plot8}
\end{figure}

\begin{figure}[phtb]
\centering
\subfigure[$s=2$]{\includegraphics[width=0.45 
\linewidth]{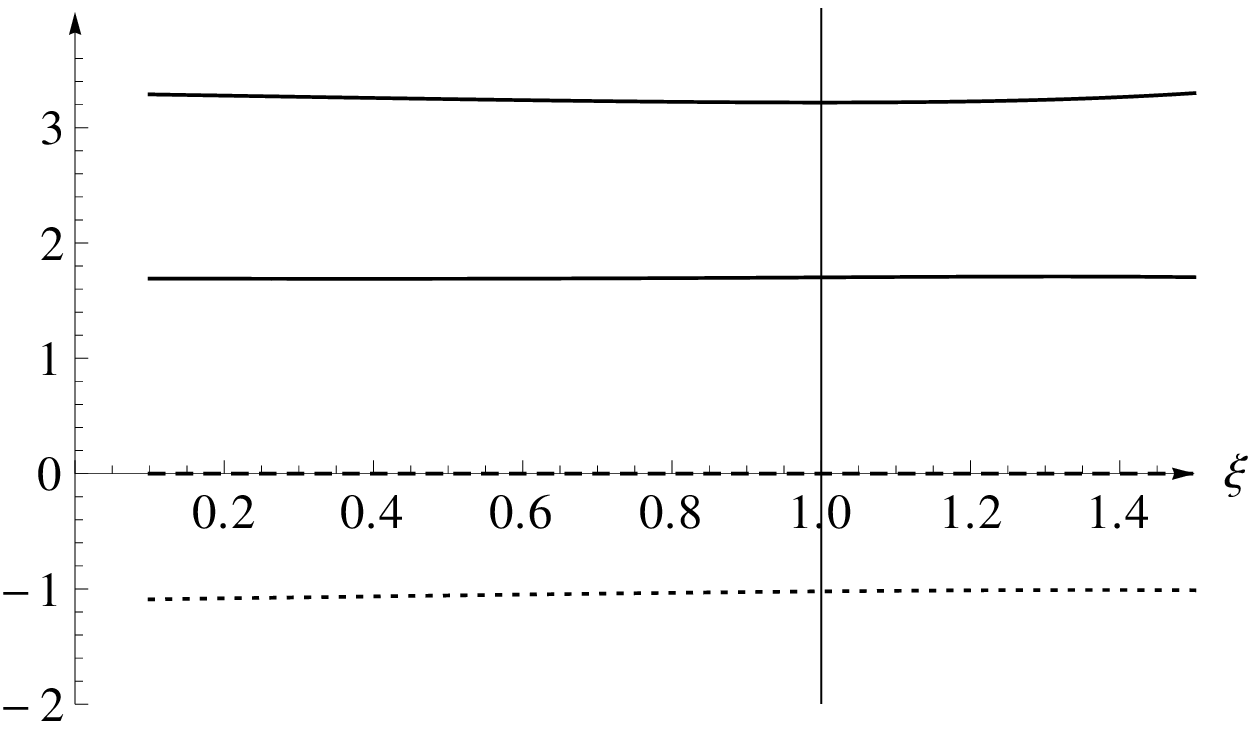}\label{Plot9a}}\quad
\subfigure[$s=3$]{\includegraphics[width=0.45 
\linewidth]{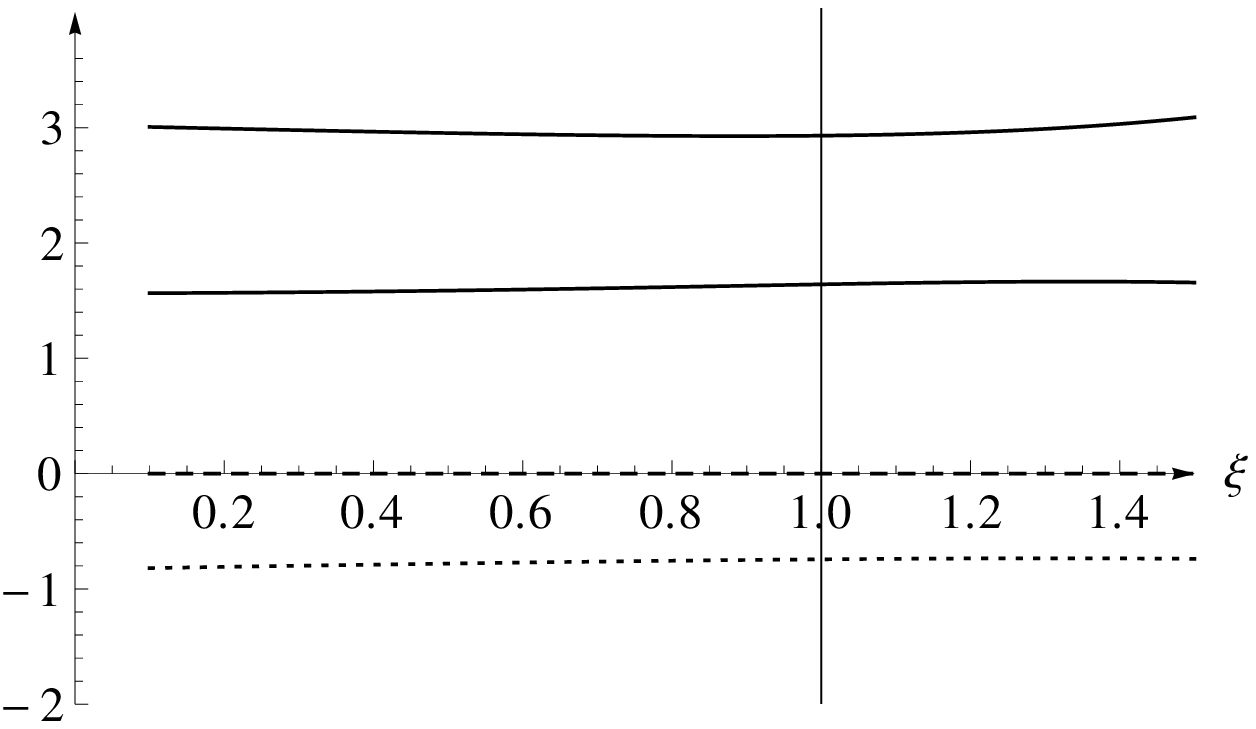}\label{Plot9b}}\quad
\subfigure[$s=5$]{\includegraphics[width=0.45 
\linewidth]{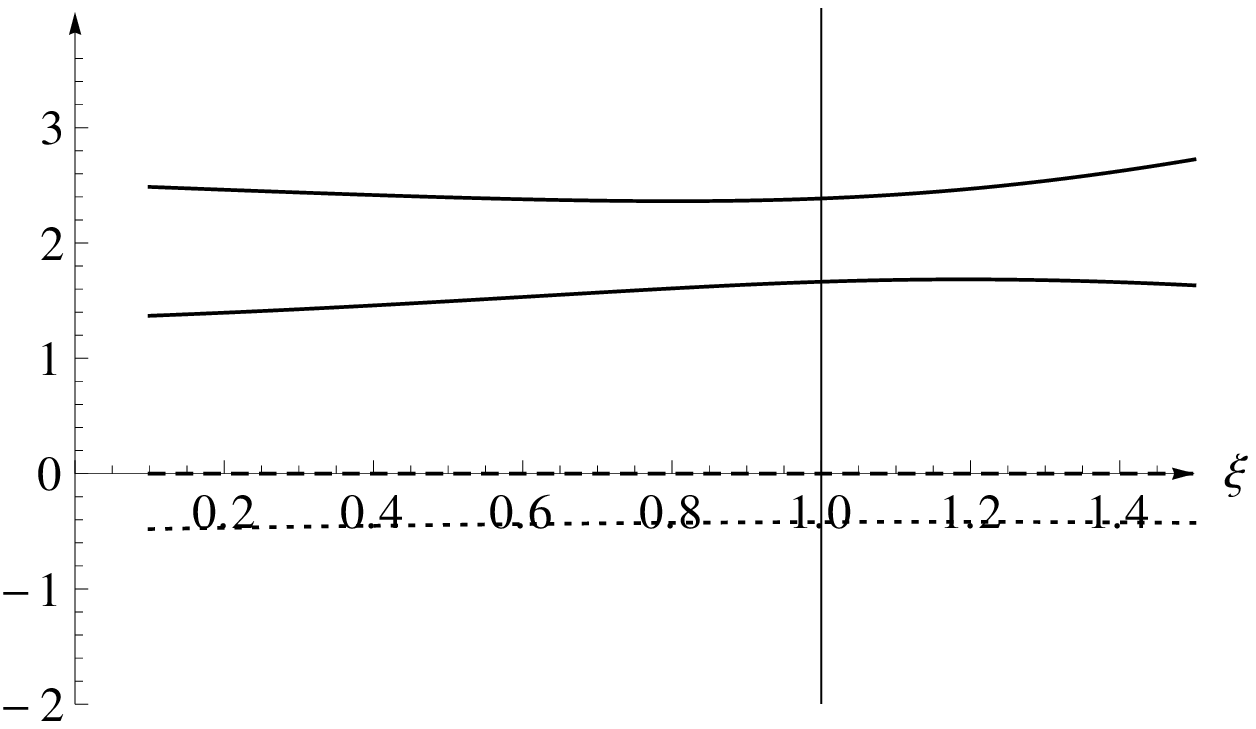}\label{Plot9c}}\quad
\subfigure[$s=7$]{\includegraphics[width=0.45 
\linewidth]{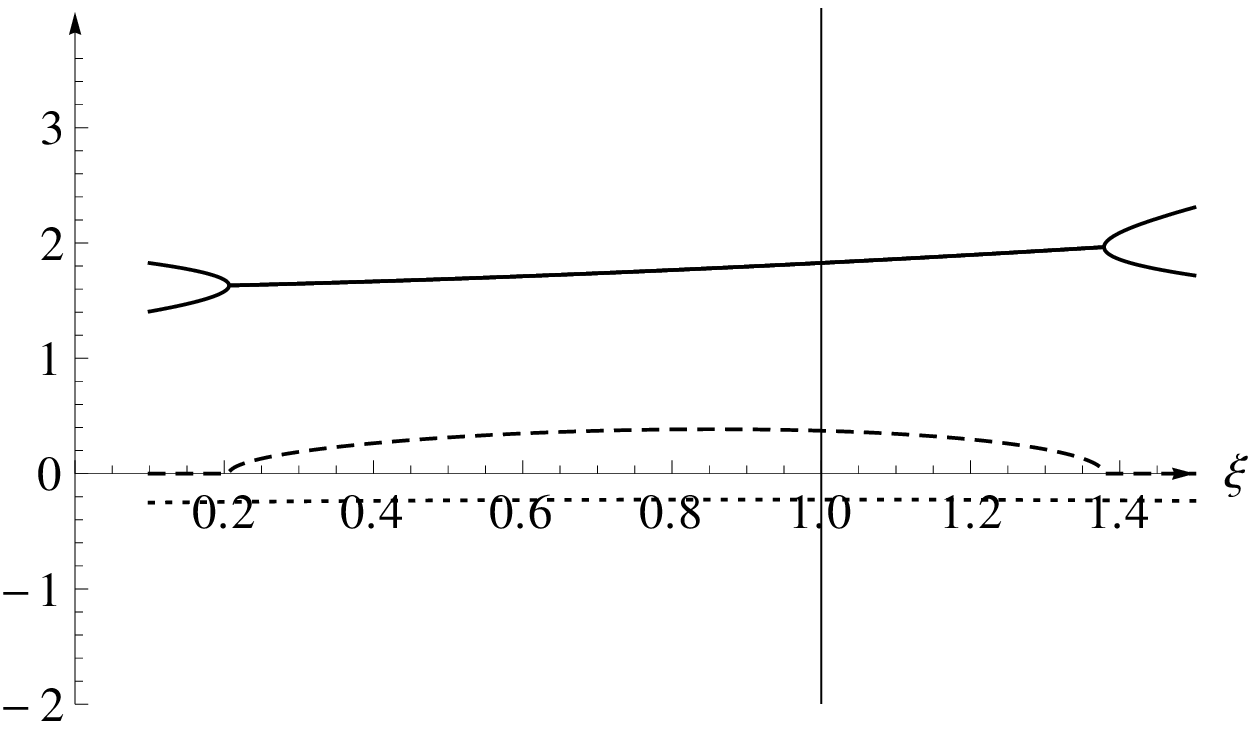}\label{Plot9d}}\\
\subfigure[$s=10$]{\includegraphics[width=0.45 
\linewidth]{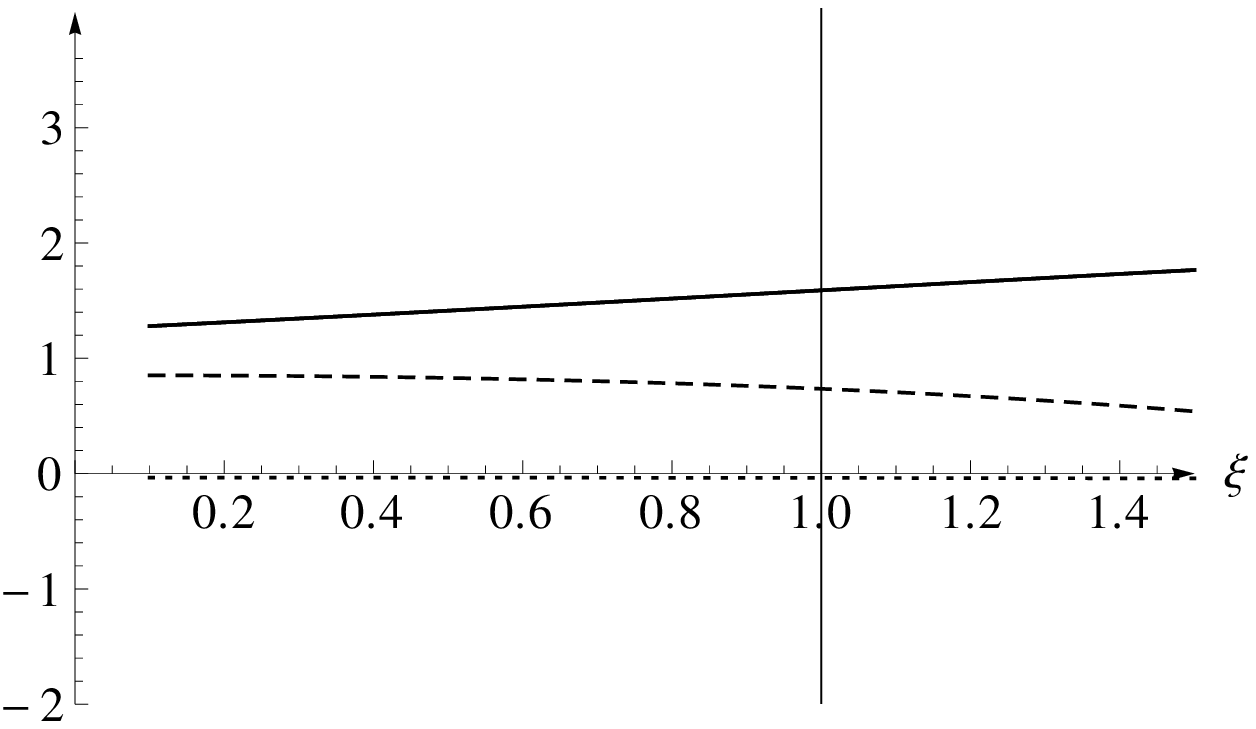}\label{Plot9e}}\quad
\subfigure[$s=20$]{\includegraphics[width=0.45 
\linewidth]{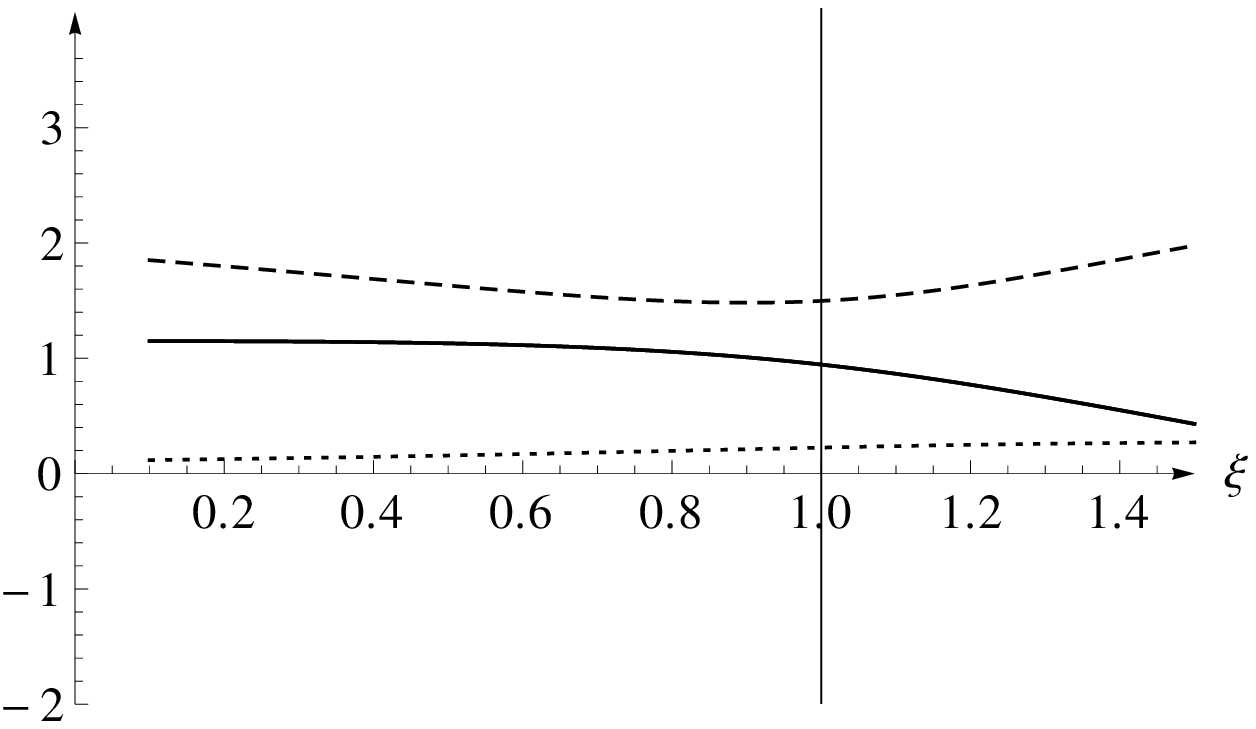}\label{Plot9f}}\caption{Critical
 exponents and $g^\ast \lambda^\ast$ for different shape parameters $s$ 
depending on $\xi$ with an adapted mass parameter $\mu=(\xi/4)^{1/4}$ 
($\theta'$ solid, $\theta''$ dashed, $g^\ast \lambda^\ast$ 
dotted).}\label{Plot9}
\end{figure}

\begin{figure}[phtb]
\centering
\subfigure[$s=2$]{\includegraphics[width=0.45 
\linewidth]{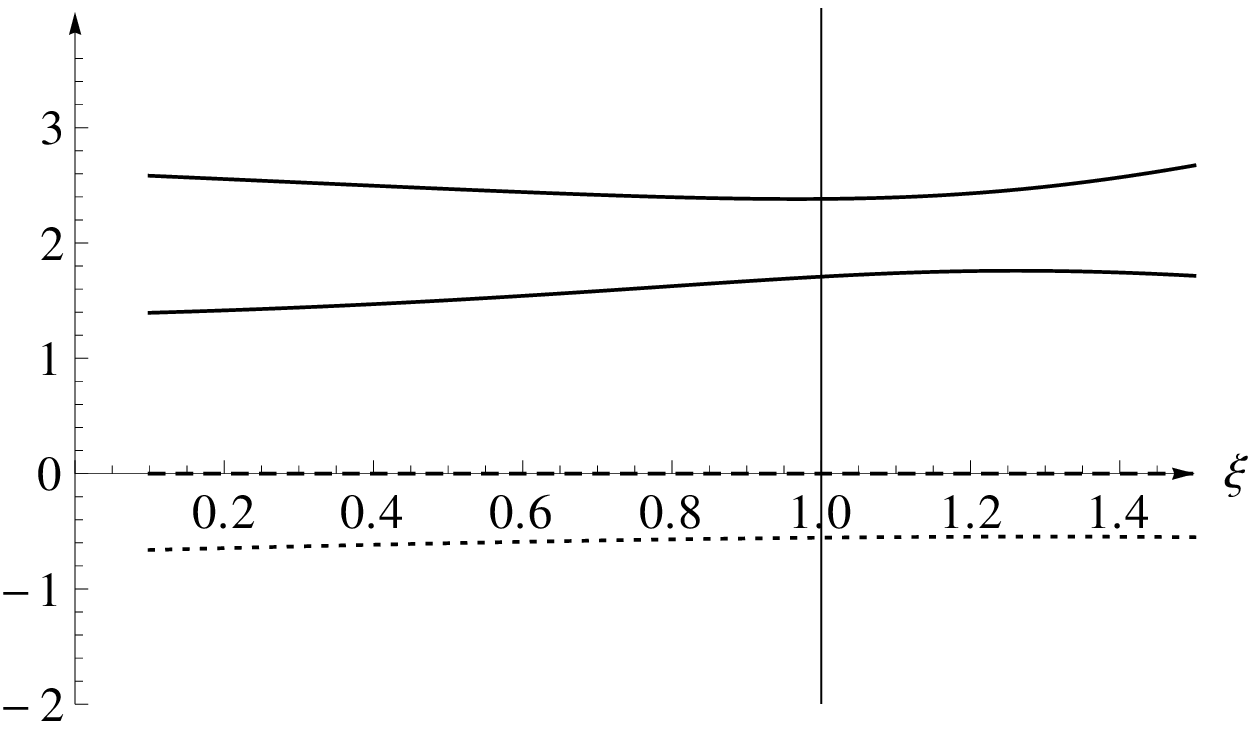}\label{Plot10a}}\quad
\subfigure[$s=3$]{\includegraphics[width=0.45 
\linewidth]{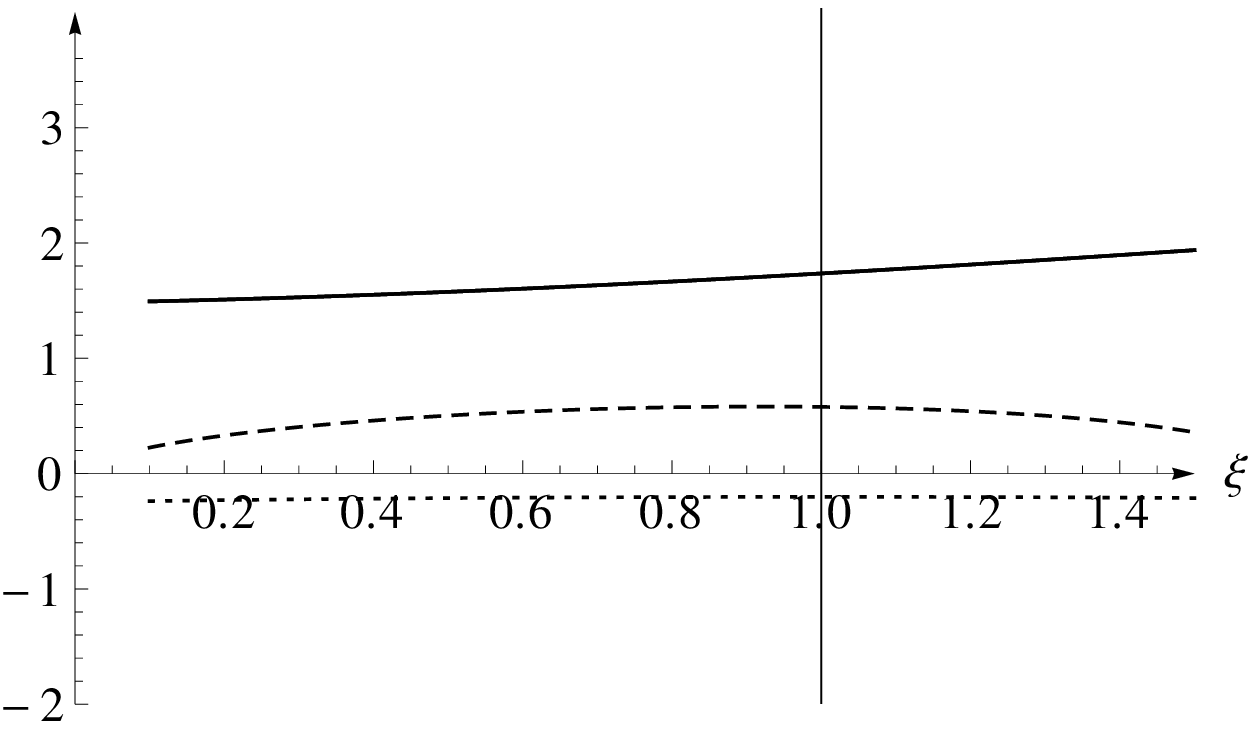}\label{Plot10b}}\quad
\subfigure[$s=4$]{\includegraphics[width=0.45 
\linewidth]{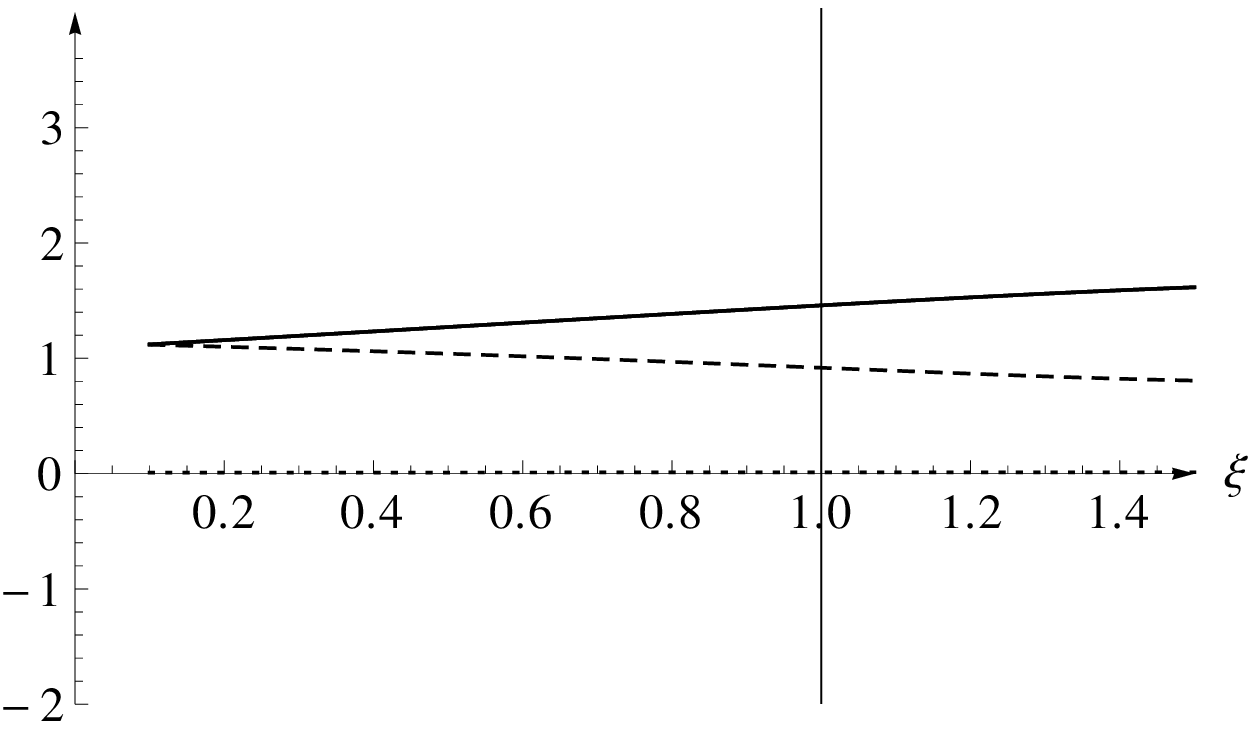}\label{Plot10c}}\quad
\subfigure[$s=5$]{\includegraphics[width=0.45 
\linewidth]{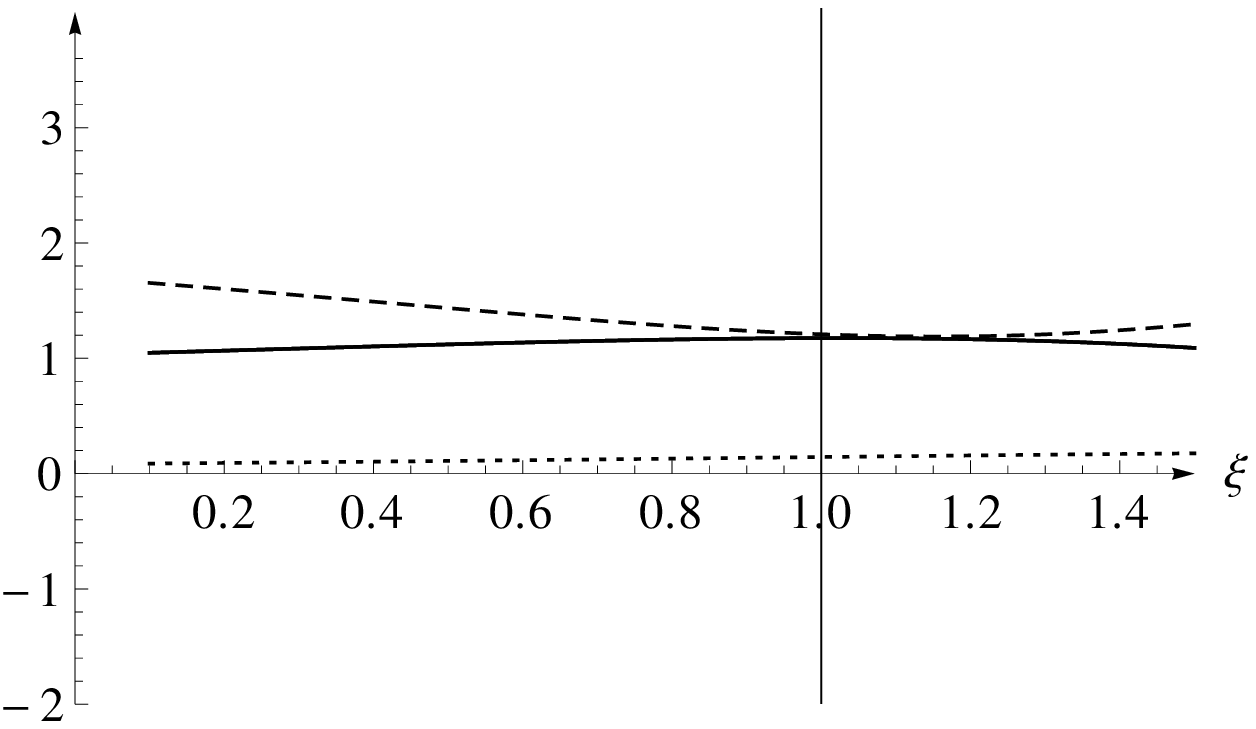}\label{Plot10d}}\\
\subfigure[$s=10$]{\includegraphics[width=0.45 
\linewidth]{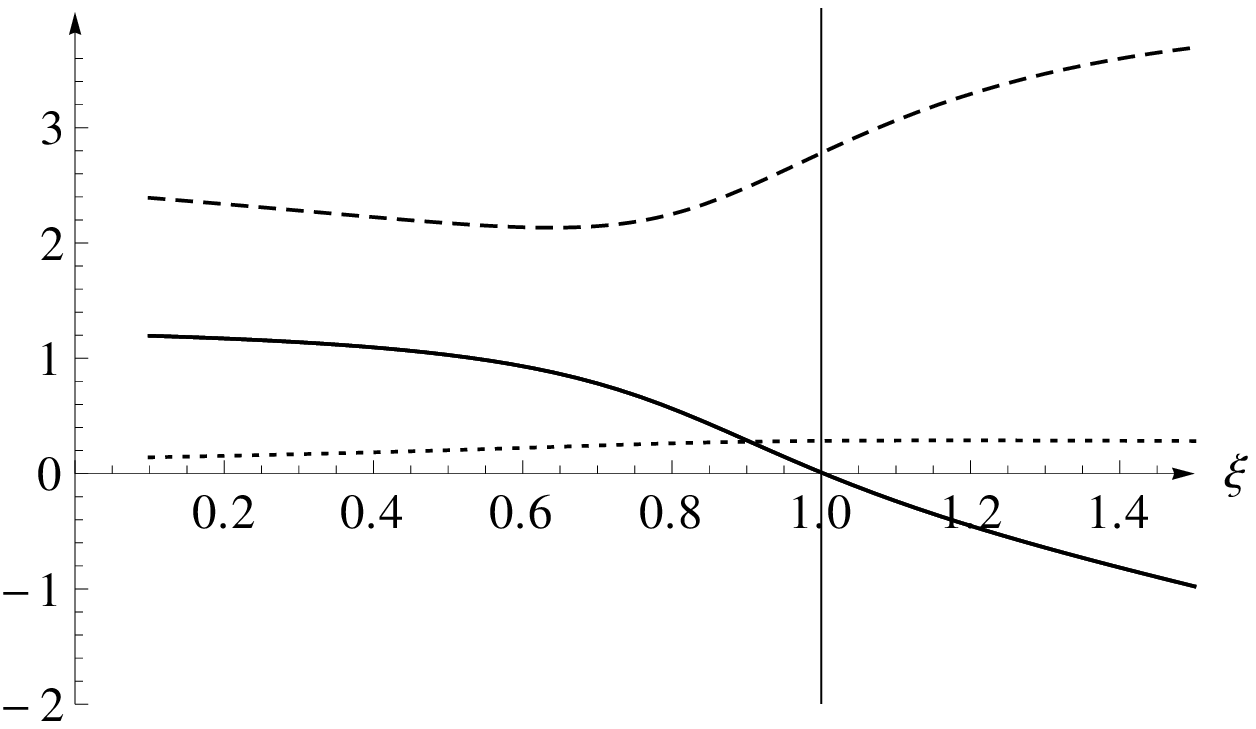}\label{Plot10e}}\quad
\subfigure[$s=20$]{\includegraphics[width=0.45 
\linewidth]{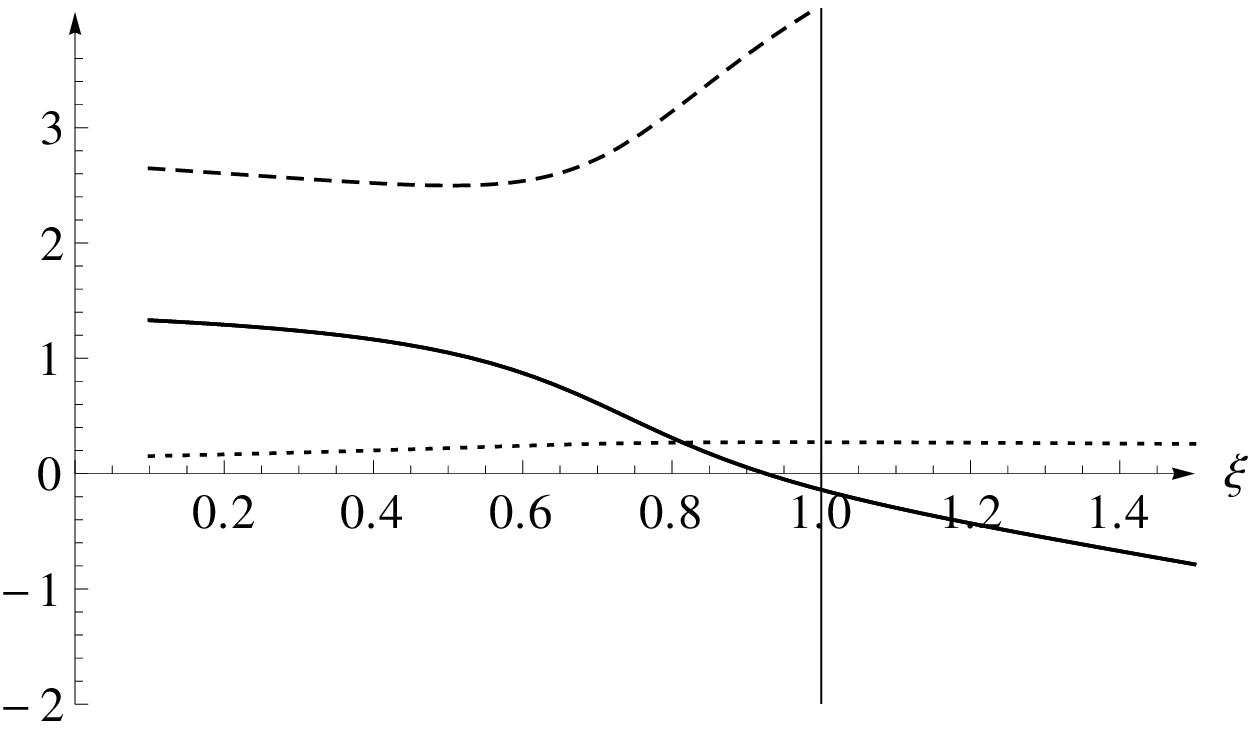}\label{Plot10f}}\caption{Critical 
exponents and $g^\ast \lambda^\ast$ for different shape parameters $s$ 
depending on $\xi$ with an adapted mass parameter $\mu=\sqrt{2} \cdot 
(\xi/4)^{1/4}$ ($\theta'$ solid, $\theta''$ dashed, $g^\ast \lambda^\ast$ 
dotted).}\label{Plot10}
\end{figure}

Let us therefore go on and discuss the second set of figures (Figs. 
\ref{Plot8} to \ref{Plot10}). Again each figure represents a certain choice of 
the parameter $\mu$ and contains a series of plots showing values of the same 
quantities obtained for different shape parameters $s$ of the exponential 
cutoff function. In this case however, we employed the three variants of the 
ghost adaptation scheme (ii) differing by a factor of $\sqrt{2}$, that we 
already introduced when we discussed this scheme for the optimized cutoff 
function in the main part of this paper (cf. Fig. \ref{Plot31}).

The first and most prominent observation is that {\it the $\xi$-dependence in 
the plots is considerably weaker for all three variants of scheme (ii) 
compared to schemes (i) and (iii).} While we still find some dependence on the 
shape parameter $s$ (in all the three figures there are real critical 
exponents for small $s$ turning complex for larger $s$), except for the plots 
\ref{Plot10e} and \ref{Plot10f}, all plots show almost horizontal lines, 
i.\,e. virtually no $\xi$-dependence of the universal quantities.

Secondly, as for the optimized cutoff, we find the weakest scheme dependence 
for the variant of adaptation scheme (ii) with the smallest value of $\mu$ 
(cf. Fig. \ref{Plot8}). This, however, is probably due to the effective 
suppression of the physical degrees of freedom in the limit of small $\mu$, as 
explained in the main part of the paper.

For these reasons we conclude that, {\it within the limits of the present 
truncation, the most reliable results are from the plots employing adaptation 
scheme (ii)} as shown in Fig. \ref{Plot9}. They suggest, 
in accordance with the optimized cutoff result in Fig. \ref{Plot31b}, a UV 
attractive FP in the $\lambda^\ast<0$ region, presumably with real critical 
exponents.

\newpage
\begin{thebibliography}{99}
%
\bibitem{A}
A.~Ashtekar, 
{\it Lectures on non-perturbative canonical gravity,} World Scientific,\\ Singapore (1991);\\ 
A.~Ashtekar and J.~Lewandowski, Class.\ Quant.\ Grav. 21 (2004) R53.
%
\bibitem{R}
C.~Rovelli, 
{\it Quantum Gravity,} Cambridge University Press, Cambridge (2004).
%
\bibitem{T}
Th.~Thiemann, 
{\it Modern Canonical Quantum General Relativity,}\\ Cambridge University Press, Cambridge (2007).
%
\bibitem{Kiefer}
C. Kiefer, {\it Quantum Gravity}, Second Edition, Oxford Science Publications,\\ Oxford, 2007.
%
\bibitem{Pleb}
J. F.~Plebanski, J. Math. Phys. 18 (1977) 2511.
%
\bibitem{CDJ}
R. Capovilla, T. Jacobson and J. Dell, Phys. Rev. Lett. 63 (1989) 2325;\\ Class. Quant. Grav. 8 (1991) 59 and 9 (1992) 1839.
%
\bibitem{Krasnov}
K. Krasnov, arXiv:1101.4788, arXiv:1103.4498.
%
\bibitem{ADM}
R.~Arnowitt, S.~Deser and C.~M.~Misner in {\it Gravitation: An Introduction to Current Research}, L.~Witten (Ed.), Wiley, New York, 1962.
%
\bibitem{Ash-Ham}
A.~Ashtekar, Phys. Rev. D 36 (1987) 1587.
%
\bibitem{wein}
S.~Weinberg 
in \textit{General Relativity, an Einstein Centenary Survey},\\
S.~W.~Hawking and W.~Israel (Eds.), 
Cambridge University Press (1979). 
%
\bibitem{mr}
M.~Reuter, Phys.\ Rev.\ D 57 (1998) 971 and
\mbox{hep-th/9605030}.
%
\bibitem{oliver}
O.~Lauscher and M.~Reuter, 
Phys.\ Rev.\ D 65 (2002) 025013 and 
\mbox{hep-th/0108040}; Phys. Rev. D 66 (2002) 025026 and \mbox{hep-th/0205062};\\ Class. Quant. Grav. 19 (2002) 483 and \mbox{hep-th/0110021}.
%
\bibitem{frank1}
M.~Reuter and F.~Saueressig, 
Phys.\ Rev.\ D 65 (2002) 065016 and 
\mbox{hep-th/0110054.}
%
\bibitem{NJP}
For a recent review on QEG and Asymptotic Safety and a comprehensive list of references see 
M.~Reuter and F.~Saueressig, arXiv:1202.2274;\\
Further details can be found in the {\it New Journal of Physics} special issue on QEG, http://iopscience.iop.org/1367-2630/focus/Focus on Quantum Einstein Gravity.
\bibitem{elisa1}
  E.~Manrique and M.~Reuter,
  Phys.\ Rev.\  D 79 (2009) 025008 and\\
 \mbox{arXiv:0811.3888}.
%
\bibitem{e-omega}
J.-E.~Daum and M.~Reuter, Phys. Lett. B doi:10.1016/j.physletb.2012.01.046 and \mbox{arXiv:1012.4280}; PoS (CNCFG 2010) 003 and \mbox{arXiv:1111.1000}.
%
\bibitem{bene-speziale}
D.~Benedetti and S.~Speziale, JHEP 1106 (2011) 107; arXiv:1111.0884.
%
\bibitem{percacci-perini}
R.~Percacci and D.~Perini,
Phys.\ Rev.\ D 67 (2003) 081503;\\
Phys.\ Rev.\ D 68 (2003) 044018;\\ D.~Dou and R.~Percacci, Class. Quant. Grav. 15 (1998) 3449.
%
\bibitem{vacca}
G.~P.~Vacca and O.~Zanusso, Phys. Rev. Lett. 105 (2010) 231601;\\
O.~Zanusso, L.~Zambelli, G.~P.~Vacca and R.~Percacci, Phys. Lett. B 689 (2010) 90.
%
\bibitem{Eichhorn-Gies}
A. Eichhorn and H. Gies, New J. Phys. 13 (2011) 125012.
%
\bibitem{DvN}
S. Deser and P. van Nieuwenhuizen, Phys. Rev. D 10 (1974) 411.
%
\bibitem{Joek}
J. Hoek, Lett. Math. Phys. 6 (1982) 49.
%
\bibitem{Woodard}
R. P. Woodard, Phys. Lett. B 148 (1984) 440.
%
\bibitem{avact}
C.~Wetterich, Phys. Lett. B 301 (1993) 90; M.~Reuter and C.~Wetterich, Nucl. Phys. B 417 (1994); Nucl. Phys. B 427 (1994) 291; Nucl. Phys. B 391 (1993) 147; Nucl. Phys. B 408 (1993) 91.
%
\bibitem{DeWitt-books}
B.~S.~DeWitt, {\it The Global Approach to  Quantum Field Theory}, Oxford University Press, Oxford, 2003.
%
\bibitem{elisa2}
E.~Manrique and M.~Reuter, Annals Phys. 325 (2010) 785 and \mbox{arXiv:0907.2617}.
%
\bibitem{je-uli}
J.-E.~Daum, U.~Harst and M.~Reuter,
JHEP 01 (2010) 084 and\\ \mbox{arXiv:0910.4938}.
%
\bibitem{wet-shap}
C.~Wetterich, \mbox{arXiv:1112.2910}.\\
M.~Shaposhnikov and C.~Wetterich, Phys. Lett. B 686 (2010) 196.
%
\bibitem{fine}
U.~Harst and M.~Reuter, JHEP 05 (2011) 119 and \mbox{arXiv:1101.6007}.
\end{thebibliography}
\end{document}